\documentclass[aps,psfig,epsfig,preprint] {revtex4}
\usepackage{epsfig}
\usepackage{color}

\begin{document}
\draft
\title{
{\normalsize \hskip4.2in USTC-ICTS-05-10} \\{\bf Spectroscopy of
$\rm{q}^3\overline{\rm{q}}^3$-States in Quark Model and
Baryon-Antibaryon Enhancements}}
\author{Gui-Jun Ding$^{a}$}
\author{Jialun Ping$^{b}$}
\author{ Mu-Lin Yan$^{a}$}

\affiliation{\centerline{$^a$ Interdisciplinary Center for
Theoretical Study,} \centerline{University of Science and Technology
of China,Hefei, Anhui 230026, China} \centerline{$^b$Department of
Physics, Nanjing Normal University, Nanjing, 210097, China }}

\begin{abstract}
We study the mass spectrum of the $\rm{q}^3\overline{\rm{q}}^3$
mesons both from common $S$ wave quark model with colormagnetic
interactions, and from the quark model with triquark correlations.
Two cluster configurations $(\rm{q}^3)-(\overline{\rm{q}}^3)$ and
$(\rm{q}^2\overline{\rm{q}})-(\rm{q}\overline{\rm{q}}^2)$ are
considered. In the spectrums of both models, we find rather stable
states which have the same quantum numbers as the particle
resonances which are corresponding to the $p\overline{p}$
enhancement, $p\overline{\Lambda}$ enhancement and
$\Lambda\overline{\Lambda}$ enhancement with spin-$\mathbf{0}$ or
$\mathbf{1}$. This implies these enhancements are NOT experimental
artifacts. The color-spin-flavor structures of $p\overline{p}$,
$p\overline{\Lambda}$, and $\Lambda\overline{\Lambda}$ enhancements
are revealed. It has been found that if there exists quark
correlation in the baryonium, the $p\overline{p}$ enhancement as a
baryonium is mainly a state of
$|\bf{120}_{cs},\overline{\bf{6}}_c,\bf{2},\bf{3}_f\oplus\overline{\bf{6}}_f\rangle
|\overline{\bf{120}}_{cs},\bf{6}_c,\bf{2},\overline{\bf{3}}_f\oplus\bf{6}_f\rangle$
in the configuration
$(\rm{q}^2\overline{\rm{q}})-(\rm{q}\overline{\rm{q}}^2)$, with few
mixture of $|\bf{70}_{cs},\bf{8}_c,\bf{2},\bf{8}_f\rangle
|\overline{\bf{70}}_{cs},\bf{8}_c,\bf{2},\bf{8}_f\rangle$ in the
configuration $(\rm{q}^3)-(\overline{\rm{q}}^3)$, where the
subscripts cs,~c and f indicate colorspin, color and flavor
respectively. The existence of spin-$\mathbf{1}$
$\Lambda\overline{\Lambda},\; p\overline{\Lambda},\; p\overline{p}$
enhancements is predicted.

\end{abstract}
\maketitle
\section{introduction}
There has recently been renewed theoretical interest in hadronic
states composed of more exotic quark configurations than the usual
q$\overline{\mbox{q}}$ for mesons and qqq for baryons. The interests
has been stimulated by the experimental discovery of many exotic
states, e.g. $\Theta^{+}$, $D_{s}$(2317), X(3872)\cite{exotic}.
Moreover, in the recent years, some baryon-antibaryon invariant mass
threshold enhancements have been observed in experiments[2-8]. In
order to interpret the $p\overline{p}$ enhancement
experiment\cite{Bes1}, the authors of Ref.\cite{Datta} pointed out
that the $\rm{q}^3\overline{\rm{q}}^3$ baryonium states should
exist. This claim comes from  analyzing the interaction among the
quarks(antiquarks) in terms of the colormagnetic hyperfine
interaction\cite{colormag}:
\begin{equation}
\label{0.1}H^{\prime}=-\sum_{i>j}C_{ij}\;\vec{\sigma}_{i}\cdot\vec{\sigma}_{j}\;
\vec{\lambda}_{i}\cdot\vec{\lambda}_{j}
\end{equation}
where $\vec{\sigma}$ and $\vec{\lambda}$ are the Pauli matrices and
Gell-Mann matrices, and the indices i and j run over all the
constituent quarks and antiquarks. This analysis has been further
studied by means of the chiral soliton model\cite{yan,ding}. And a
remarkable prediction on the baryonium's mesonic decays from this
picture in Ref.\cite{ding} has been supported by the recent BES
experiment\cite{Bes} significantly. In this case,  because of both
the experiment prompt and lack of a systemic study of the
$\rm{q}^3\overline{\rm{q}}^3$ system in the quark model with
colormagnetic interaction in the literature, we try to provide such
an investigation to the spectroscopy of the exotic
$(\rm{q}^3\overline{\rm{q}}^3)$ configurations in quark model in
this work.

The $p\overline{p}$ enhancement is observed in
$J/\psi\rightarrow\gamma p\overline{{p}}$ \cite{Bes1}\cite{Bes}, and
$\psi^{\prime}\rightarrow\pi^{0} p\overline{{p}},\eta
p\overline{{p}}$ \cite{Bes2}. The $p\overline{\Lambda}$ enhancement
has been observed in $J/\psi\rightarrow p \overline{\Lambda}K^{-}$
and $\psi^{\prime}\rightarrow p \overline{\Lambda}K^{-}$
\cite{Bes3}. In B decays, $p\overline{p}$, $p\overline{\Lambda}$ and
$\Lambda \overline{\Lambda}$ enhancements also were observed, such
as $B^{\pm}\rightarrow p\overline{p} K^{\pm}$\cite{bell1},
$B^{0}\rightarrow p\overline{\Lambda}\pi^{-}$\cite{bell2},
$B^{+}\rightarrow\Lambda\overline{\Lambda}K^{+}$\cite{bell3}. In
Ref.\cite{Bes1}, the data of $p\overline{p}$ invariant mass
threshold enhancement has been fitted with the $S$ and $P$ wave
Breit-Wigner resonance function, and the enhancement has naively
been attributed to existence of a $(p\overline{p})$-baryonium
resonance. Namely, the enhancement corresponds to a baryonium
particle. The theoretical studies on this picture are in
Refs.\cite{Datta,interpretation1,yan,ding}, and an experiment
recheck on it is in Ref.\cite{Bes}. Noting that in order to fix the
total angular momentum of the baryoniums the partial wave analysis
has to be done, and the $S$ wave or $P$ wave data fit can not fix
the spin of the baryonium. For instance, even though the baronium is
fitted out by $S$ wave, the total angular momentum of it can still
be non-zero because quark's intrinsic spin is $\mathbf{1/2}$ and the
particle is a multiquark system. For the sake of illustrating
convenience, we will  call the baryonium particle revealed by the
possible enhancement effect measurements as {\it
enhancement-baryonium}, or shortly {\it enhancement} hereafter.
There are many interesting phenomena and puzzles in the
baryon-antibaryon physics. In this paper, we mainly focus on these
exotic states observed in the baryon antibaryon final states.
Because the S wave fit is favored over the P wave fit , we do not
consider the radial excitations in enhancements, and the models used
in this paper are $S$ wave type quark models, i.e., the relative
angular momentum between any two quarks in the models is
$\mathbf{0}$.

Because the baryoniums finally decay into a baryon octet state plus
an antibaryon octet state and
$\bf{8}_{f}\otimes\bf{8}_{f}=\bf{1}_f\oplus\bf{8}_f\oplus\bf{8}_f\oplus\bf{10}_f
\oplus\overline{\bf{10}}_f\oplus\bf{27}_f$, these enhancements can
only possibly belong to flavor $\bf{1}_f$-plet,
$\bf{8}_f$-plet,$\bf{10}_{f}$-plet,$\overline{\bf{10}}_{f}$-plet or
$\bf{27}_{f}$-plet. And the quantum numbers of the $p\overline{p}$
enhancement most possibly are $J^{PC}=0^{-+}$, so it must consist of
$\rm{q}\overline{\rm{q}}$, $\rm{q}^{3}\overline{\rm{q}}^3$,
$\rm{q}^{5}\overline{\rm{q}}^5$ and so on, if the relative angular
momentum between quarks equals $\mathbf{0}$ . They are not possibly
the ordinary  mesonic states, then they at least are composed of
$\mathbf{3}$ quarks and $\mathbf{3}$ antiquarks.  The theoretical
interpretation of thess exotic state is a great challenge and many
proposals have been suggested
\cite{Datta,interpretation1,yan,ding,interpretation2}. Some of them
studied one particular state \cite{Datta,yan,ding}, others propose
the possible nature of these states or the possible classification
of these baryon-antibaryon enhancements\cite{interpretation2}. In
this work, we discuss the possible states of the baryon-antibaryon
system (or the $\rm{q}^3\overline{\rm{q}}^3$ system) and try to give
a unified interpretation of these baryon-antibaryon enhancements in
common quark model and in quark model with quark correlation. Since
all these states have been observed in charmonium decay and B decay,
it is possible that these states possess the same nature. It should
be interesting to interpret them in a uniform framework.
\par
Color plays important role in multiquark states. The configuration
of color is uniquely fixed in the usual q$\overline{\rm{q}} $ mesons
and baryons consisting of qqq. In the multiquark system, however,
the color degree of freedom is "unfrozen" and there exist a great
variety of color configurations to form color singlet states, e.g.,
in the system of the $\rm{q}^2\overline{\rm{q}}^2$, the two quark
can be either in  color $\mathbf{\overline{3}}_c$ or $\mathbf{6}_c$
state combining with the two antiquarks in color $\mathbf{3}_c$ or
$\mathbf{\overline{6}}_c$ state to give a singlet\cite{colorch}.
Many works also have been done in exploring this point of view in
pentaquark study\cite{petcolor}. This "color chemistry" idea is very
fascinating. Even though the question that how to cluster the quarks
into diquark, triquark or other structures in the systems by means
of the fundamental QCD dynamics remains to be open so far, the
quark-correlation has been proved to be a very useful and important
conception in the physics of multiquark
state\cite{correlation,ding2}.

\par
We will first discuss this system in common quark model where 3
quarks and 3 antiquarks are in relative S wave and there is not
correlation among the quarks(antiquarks). Under the QCD requirement
that the baryoniums have to be in color singlets, the spectrum of
enhancement has been calculated, and the eigenstates are linear
combinations of states which are in different SU(6)$_{cs}$
representations. For the sake of completeness, we then discuss the
effect of quark correlation in the baryinoim. There are many
different ways to form color singlet for the
$\rm{q}^3\overline{\rm{q}}^3$ system, and we expect the mesonic
states corresponding to the different color couplings between the
quark subsystems have different physical properties. We consider two
case in this work, the first case is the
$(\rm{q}^3)-(\overline{\rm{q}}^3)$ configuration and another
configuration is
$(\rm{q}^2\overline{\rm{q}})-(\rm{q}\overline{\rm{q}}^2)$. We assume
the mass splitting is due to the colormagnetic interaction
\cite{colormag}, we also list the possible states and the mass
splitting in the two cases. Since the S wave fit is favored, we
assume the relative angular momentum of the three quarks(antiquarks)
in each cluster is zero, and the two clusters are also in relative S
wave.
\par
In this work we take a close look at the
$\rm{q}^3\overline{\rm{q}}^3$ system in quark model with the mass
splitting due to colormagnetic hyperfine interaction, and we don't
consider the effect of $\rm{SU(3)}_f$ flavor symmetry breaking, and
the results are qualitatively correct.

The paper is organized as followings: in the  section II, we review
the description of colormagnetic hyperfine interaction between
quarks in hadrons and its application to the
$\rm{q}^2\overline{\rm{q}}^2$ system; Section III is devoted to
study the $\rm{q}^3\overline{\rm{q}}^3$ system in common quark
model; In Section IV, we study two interesting configurations with
quark correlation: $ (\rm{q}^3)-(\overline{\rm{q}}^3)$ and $
(\rm{q}^2\overline{\rm{q}})-(\overline{\rm{q}}^2\rm{q})$; Finally,
we briefly summary the results and give some discussions. In
Appendix A, we show an example to deal with the mixing of different
SU(6)$_{cs}$ multiplets under colormagnetic interaction, and we also
list the matrix form of the hamiltonian $H^{\prime}$ in the subspace
expanded by the states with definite spin in each case in Appendix
B.

\section{colormagnetic hyperfine interaction and its application to
the $\rm{q}^2\overline{\rm{q}}^2$ mesons}

The spectroscopically important interaction between quarks(aside
from the interaction which provides the confinement and sets the
overall scale) is the spin-spin force mediated by one gluon
exchange. The interaction hamiltonian $H'$ between quarks is given
by Eq(\ref{0.1}). Generally the coefficients $C_{ij}$ are dependent
on the quark masses of flavor i and j and the properties of the
spatial wave functions of the quarks and the antiquarks in the
system and parameterize the strength of the interaction. Following
the common colormagnetic hyperfine studies in literature
\cite{jaffe1,petcolor}, we take $C_{ij}=C$ neglecting the breaking
of the flavor $\rm{SU(3)}_f$ symmetry. It is straightforward to
express the colormagnetic interaction in terms of the quadratic
Casimir operators of SU(2)$_{s}$, SU(3)$_{c}$ and SU(6)$_{cs}$ of
the total system, the quarks and the antiquarks,
\begin{eqnarray}
\nonumber
H^{\prime}&=&C[8N+2C_{6}(tot)-\frac{4}{3}S_{tot}(S_{tot}+1)-2C_{3}(tot)+4C_{3}(Q)+\frac{8}{3}S_{Q}(S_{Q}+1)-4C_{6}(Q)\\
\label{1.2}&&+4C_{3}(\overline{Q})+\frac{8}{3}S_{\overline{Q}}(S_{\overline{Q}}+1)-4C_{6}(\overline{Q})]
\end{eqnarray}
Where C$_{3}$ and C$_{6}$ are the quadratic Casimir operator of the
SU(3)$_c$ and SU(6)$_{cs}$, $S$ is the spin, the labels Q,
$\overline{Q}$ and $tot$ refer to the quarks, antiquarks, and the
entire system respectively, and N is the total number of the quarks
and antiquarks in the system, for the $\rm{q}^3\overline{\rm{q}}^3$
system $N=6$. The Casimir operators are normalized conventionally,
e.g. $C_{3}(3)=\frac{4}{3},C_{6}(6)=\frac{35}{6}$. Eigenstates of
total SU(6)$_{cs}$ for $\rm{q}^{3}\overline{\rm{q}}^{3}$ are
generally mixtures of the states with different color and spin
representations of quarks and antiquarks respectively, and the
colormagnetic interaction usually mixes states of different total
colorspin.

So far various systems has been discussed in quark model with
colormagnetic interaction. It was first considered in
Ref.\cite{colormag}, where it was applied  to study the mass
splittings of the S wave color singlet q$\overline{\rm{q}}$ mesons
and $\rm{q}^3$ baryons, and it gives a good description of the
spectrums. And Jaffe studied the $\rm{q}^2\overline{\rm{q}}^2$
system and the $0^{+}$ mesonic states with mass bellow 1GeV, i.e.,
$a_0(980)$, $f_0(980)$ are identified as four quarks states
\cite{jaffe1}. Dibaryon( the $\rm{q}^6$ system ) also was discussed
\cite{q6}. Recently pentaquark (the $\rm{q}^4\overline{\rm{q}}$
system) has been widely studied in colormagnetic interaction models
\cite{petcolor}.

In order to clarify the notations and methods in the studies of
colomagnetic interaction, we briefly review Jaffe's works on
$\rm{q}^2\overline{\rm{q}}^2$\cite{jaffe1,jaffe2}. Because of the
Fermi statistics, the allowed states of the 2 quarks are
\begin{equation}
\label{aa1}|\mathbf{21}_{cs},\overline{\mathbf{3}}_f\rangle,
~~~~|\mathbf{15}_{cs},\mathbf{6}_f\rangle
\end{equation}
The 2 antiquarks are in conjugate representations,
\begin{equation}
\label{aa2}|\overline{\mathbf{21}}_{cs},\mathbf{3}_f\rangle,
~~~~|\overline{\mathbf{15}}_{cs},\overline{\mathbf{6}}_f\rangle
\end{equation}
where the first number in the bracket is the dimension of the
irreducible representation of $\rm{SU(6)}_{cs}$ and the second
number is the dimension of the irreducible representation of
$\rm{SU(3)}_{f}$. The physical $\rm{q}^2\overline{\rm{q}}^2$ states
are color singlets, and there are four combinations:
\begin{eqnarray}
\nonumber&&\mathbf{21}_{cs}\otimes\overline{\mathbf{21}}_{cs}=\mathbf{1}_{cs}\oplus\mathbf{35}_{cs}\oplus\mathbf{405}_{cs},~~
\rm{with\;\;flavor\;\;in\;}
\overline{\mathbf{3}}_f\otimes\mathbf{3}_f=\mathbf{1}_f\oplus\mathbf{8}_f\\
\nonumber&&\mathbf{15}_{cs}\otimes\overline{\mathbf{15}}_{cs}=\mathbf{1}_{cs}\oplus\mathbf{35}_{cs}\oplus\mathbf{189}_{cs},~~
\rm{with\;\;flavor\;\;in\;}
\mathbf{6}_f\otimes\overline{\mathbf{6}}_f=\mathbf{1}_f\oplus\mathbf{8}_{f}\oplus\mathbf{27}_{f}\\
\nonumber&&\mathbf{21}_{cs}\otimes\overline{\mathbf{15}}_{cs}=\mathbf{35}_{cs}\oplus\mathbf{280}_{cs},~~
\rm{with\;\;flavor\;\;in\;}
\overline{\mathbf{3}}_f\otimes\overline{\mathbf{6}}_f=\mathbf{8}_f\oplus\overline{\mathbf{10}}_f\\
\label{aa3}&&\mathbf{15}_{cs}\otimes\overline{\mathbf{21}}_{cs}=\overline{\mathbf{35}}_{cs}\oplus\overline{\mathbf{280}}_{cs},~~
\rm{with\;\;flavor\;\;in\;}
\mathbf{6}_f\otimes\mathbf{3}_f=\mathbf{8}_f\oplus\mathbf{10}_{f}
\end{eqnarray}
The four sector's multiplets are not mixed by the colormagnetic
interaction.

For the $\mathbf{21}_{cs}\otimes\overline{\mathbf{21}}_{cs}$ case,
the $SU(6)_{cs}$ decomposition with respect to $SU(3)_{c}\otimes
SU(2)_{s}$ reveals the following color singlets:
\begin{eqnarray}
\nonumber&&(\mathbf{1}_c,\mathbf{1})\subset\mathbf{1}_{cs}\\
\nonumber&&(\mathbf{1}_c,\mathbf{3})\subset\mathbf{35}_{cs}\\
\label{aa4}&&(\mathbf{1}_c,\mathbf{1})~\rm{and}~(\mathbf{1}_c,\mathbf{5})\subset\mathbf{405}_{cs}
\end{eqnarray}
The wavefunctions of the spin-$\mathbf{2}$ states
$|\mathbf{405}_{cs},\mathbf{1}_c,\mathbf{5};\mathbf{3}_f\otimes\overline{\mathbf{3}}_f\rangle$
and spin-$\mathbf{1}$
$|\mathbf{35}_{cs},\mathbf{1}_c,\mathbf{3};\mathbf{3}_f\otimes\overline{\mathbf{3}}_f\rangle$
respectively are,
\begin{eqnarray}
\nonumber&&|\mathbf{405}_{cs},\mathbf{1}_c,\mathbf{5};\mathbf{3}_f\otimes\overline{\mathbf{3}}_f\rangle
=|(\mathbf{21}_{cs},\mathbf{6}_c,\mathbf{3};\overline{\mathbf{3}}_f,)(\overline{\mathbf{21}}_{cs},
\overline{\mathbf{6}}_c,\mathbf{3};\mathbf{3}_f),\mathbf{1}_c,\mathbf{5};\mathbf{3}_f\otimes\overline{\mathbf{3}}_f\rangle\\
\nonumber&&|\mathbf{35}_{cs},\mathbf{1}_c,\mathbf{3};\mathbf{3}_f\otimes\overline{\mathbf{3}}_f\rangle
=|(\mathbf{21}_{cs},\mathbf{6}_c,\mathbf{3};\overline{\mathbf{3}}_f,)(\overline{\mathbf{21}}_{cs},
\overline{\mathbf{6}}_c,\mathbf{3};\mathbf{3}_f),\mathbf{1}_c,\mathbf{3};\mathbf{3}_f\otimes\overline{\mathbf{3}}_f\rangle\\
\end{eqnarray}
The corresponding colormagnetic interaction energy are
$\frac{32}{3}$ and $-16$ respectively. Here the wave function is
expressed in terms of the conventional notations\cite{notation}:
\begin{eqnarray*}
&&|D_{cs}(tot),D_c(tot),2S(tot)+1;D_{f}(tot)\rangle \\
&=&|(D_{cs}(Q),D_c(Q),2S(Q)+1;D_{f}(Q)),(D_{cs}(\overline{Q}),
D_c(\overline{Q}),2S(\overline{Q})+1;D_{f}(\overline{Q})),\\
&&D_c(tot),2S(tot)+1;D_f(tot)\rangle,
\end{eqnarray*}
where $D_{cs}(tot),D_{cs}(Q),D_{cs}(\overline{Q})$ denote the
dimension of the irreducible $SU(6)_{cs}$ representation of the
total system, the quarks, and the antiquarks respectively. And
similarly for the notations $D_c(id),D_f(id)$, the $id$ represents
$tot, Q ~\rm{or}~\overline{Q}$. $S(tot),S(Q),S(\overline{Q})$ are
respectively the spin of the total system, the quarks and the
antiquarks.

The two spin-$\mathbf{0}$ states are linear combinations of
$(\overline{\mathbf{3}}_c,\mathbf{1})\otimes(\mathbf{3}_c,\mathbf{1})$
with
$(\mathbf{6}_c,\mathbf{3})\otimes(\overline{\mathbf{6}}_c,\mathbf{3})$,
\begin{eqnarray}
\nonumber&&|\mathbf{1}_{cs},\mathbf{1}_c,\mathbf{1};\mathbf{3}_f\otimes\overline{\mathbf{3}}_f\rangle=\sqrt{\frac{6}{7}}|(\mathbf{21}_{cs},\mathbf{6}_c,\mathbf{3};\overline{\mathbf{3}}_f,)(\overline{\mathbf{21}}_{cs},\overline{\mathbf{6}}_c,\mathbf{3};\mathbf{3}_f),\mathbf{1}_c,\mathbf{1};\mathbf{3}_f\otimes\overline{\mathbf{3}}_f\rangle\\
\nonumber&&+\sqrt{\frac{1}{7}}|(\mathbf{21}_{cs},\overline{\mathbf{3}}_c,\mathbf{1};\overline{\mathbf{3}}_f,)(\overline{\mathbf{21}}_{cs},\mathbf{3}_c,\mathbf{1};\mathbf{3}_f),\mathbf{1}_c,\mathbf{1};\mathbf{3}_f\otimes\overline{\mathbf{3}}_f\rangle\\
\nonumber&&|\mathbf{405}_{cs},\mathbf{1}_c,\mathbf{1};\mathbf{3}_f\otimes\overline{\mathbf{3}}_f\rangle=\sqrt{\frac{1}{7}}|(\mathbf{21}_{cs},\mathbf{6}_c,\mathbf{3};\overline{\mathbf{3}}_f,)(\overline{\mathbf{21}}_{cs},\overline{\mathbf{6}}_c,\mathbf{3};\mathbf{3}_f),\mathbf{1}_c,\mathbf{1};\mathbf{3}_f\otimes\overline{\mathbf{3}}_f\rangle\\
\label{aa5}&&-\sqrt{\frac{6}{7}}|(\mathbf{21}_{cs},\overline{\mathbf{3}}_c,\mathbf{1};\overline{\mathbf{3}}_f,)(\overline{\mathbf{21}}_{cs},\mathbf{3}_c,\mathbf{1};\mathbf{3}_f),\mathbf{1}_c,\mathbf{1};\mathbf{3}_f\otimes\overline{\mathbf{3}}_f\rangle
\end{eqnarray}
The colormagnetic interaction mixes these two states, and the
eigenstates of $H'$ (Eq.(\ref{1.2})) are
\begin{eqnarray}
\nonumber&&|\mathbf{1},\mathbf{3}_f\otimes\overline{\mathbf{3}}_f\rangle_1=0.972|\mathbf{1}_{cs},\mathbf{1}_c,\mathbf{1};\mathbf{3}_f\otimes\overline{\mathbf{3}}_f\rangle+0.233|\mathbf{405}_{cs},\mathbf{1}_c,\mathbf{1};\mathbf{3}_f\otimes\overline{\mathbf{3}}_f\rangle\\
\label{aa6}&&|\mathbf{1},\mathbf{3}_f\otimes\overline{\mathbf{3}}_f\rangle_2=0.233|\mathbf{1}_{cs},\mathbf{1}_c,\mathbf{1};\mathbf{3}_f\otimes\overline{\mathbf{3}}_f\rangle-0.972|\mathbf{405}_{cs},\mathbf{1}_c,\mathbf{1};\mathbf{3}_f\otimes\overline{\mathbf{3}}_f\rangle
\end{eqnarray}
The colormagnetic interaction energies ( i.e. the eigenvalues of
$H'$ )  are -43.26 and -19.37 respectively. The color singlet states
and the corresponding colormagnetic energies for the other three
sectors in Eq.(\ref{aa3}) have also been derived in the same way by
Jaffe\cite{jaffe1}. Thus, the color singlet spectroscopy of
$\rm{q}^2\overline{\rm{q}}^2$ states in quark model has been
revealed. Among all of these multiplets, the nonet multiplet
$|\mathbf{1},\mathbf{3}_f\otimes\overline{\mathbf{3}}_f\rangle_1$
has the largest colormagnetic binding energy. Therefore, they are
the lightest states, and one can physically regard them  as the
known $0^{+}$ nonet mesons below 1 GeV. All of these have been done
and be well known. An interesting question and task is how to extent
Jaffe's analysis on $\rm{q}^2\overline{\rm{q}}^2$-system in quark
model into the $\rm{q}^3\overline{\rm{q}}^3$-system, and to explore
the corresponding physics. In the following of this paper, we
perform such an investigation.

\section{the $\rm{q}^3\overline{\rm{q}}^3 $ system and the enhancement from the common quark
model}
Since the experimental discovery of the enhancements which can be
regarded as baryoniums, it is necessary to explore the
$\rm{q}^3\overline{\rm{q}}^3$ system in the framework of
colormagnetic interaction. Now let us to extent the Jaffe's studies
on $\rm{q}^2\overline{\rm{q}}^2$ system to
$\rm{q}^3\overline{\rm{q}}^3$ one. Due to the increase of a pair of
quark and antiquark, the dimension of the irreducible representation
will become rather large, and more techniques and skills of group
theory are needed for dealing with this multiquark system. The
calculations are rather lengthy. In order to ensure the correctness
of calculation results, we have checked each analytical group theory
derivations by the computer program calculations.
The results obtained by the two approaches are exactly the same.

Similar to studying the usual $\rm{q}\overline{\rm{q}}$ meson and
$\rm{q}^3$ baryon from the common quark model, we study the
$\rm{q}^3\overline{\rm{q}}^3 $ system in the quark model in order to
understand the observed baryon antibaryon enhancements, and all
quarks are in relative S wave. The configuration is schematically
shown in Fig.3.

Because of the Fermi statistics, the three quarks's state must be
antisymmetric in the combined wavefunctions of
flavor$\otimes$color$\otimes$spin$\otimes$space. Because the quarks
are in a relative S wave, the flavor$\otimes$color$\otimes$spin
wavefunction must be antisymmetric. Under the group
$\rm{SU(18)}\supset \rm{SU(6)}_{cs}\otimes \rm{SU(3)}_{f}\supset
\rm{SU(3)}_{c}\otimes \rm{SU(2)}_{s}\otimes \rm{SU(3)}_{f}$, the
three quarks must be in a completely antisymmetric state and it
therefore transform under the
$\frac{18\cdot17\cdot16}{3!}=\mathbf{816}$ dimensional irreducible
representation of SU(18). The reduction of the SU(18) irreducible
representation $\mathbf{816}$ with respect to
$\rm{SU(6)}_{cs}\otimes \rm{SU(3)}_{f}$ is \cite{group}
\begin{equation}
\label{2.1}\mathbf{816}=(\mathbf{70}_{cs},\mathbf{8}_f)+(\mathbf{56}_{cs},
\mathbf{1}_f)+(\mathbf{20}_{cs},\mathbf{10}_f),
\end{equation}
Here the notation is the same as Eq.(\ref{aa1}), and the allowed
states of the 3 quarks are
\begin{equation}
\label{2.11}|\mathbf{70}_{cs},\mathbf{8}_f\rangle,\;\;
|\mathbf{56}_{cs},\mathbf{1}_f\rangle,\;\;|\mathbf{20}_{cs},\mathbf{10}_f\rangle,\;\;
\end{equation}
And the 3 antiquarks are in conjugate representations,
\begin{equation}
\label{2.12}|\overline{\mathbf{70}}_{cs},\mathbf{8}_f\rangle,\;\;
|\overline{\mathbf{56}}_{cs},
\mathbf{1}_f\rangle,\;\;|\overline{\mathbf{20}}_{cs},\overline{\mathbf{10}}_f\rangle.\;\;
\end{equation}
Next we combine the states of the 3 quarks in Eq.(\ref{2.11}) and
that of the 3 antiquarks in Eq.(\ref{2.12}) to obtain the states of
the total systems which is color singlet, in the following we
consider the allowed states of the total system and the
corresponding colormagnetic energy case by case. In the text, we
mainly  list the results and illustrate calculation skills, and in
Appendix A, we take the mixing of the three spin-$\mathbf{1}$ states
in the case of $\mathbf{70}_{cs}\otimes\mathbf{20}_{cs}$ as an
example in order to show how analytical calculations are performed
in details. In Appendix B, we write the explicit matrix form of the
interaction hamiltonian $H^{\prime}$ in the subspace expanded by the
states of definite spin in each case, from which we can obtain the
eigenstates and the corresponding colormagnetic energies.
\begin{enumerate}
\item $\mathbf{20}_{cs}\otimes\overline{\mathbf{20}}_{cs}$

From Eq.(\ref{2.11}) and Eq.(\ref{2.12}), the multiplets must be  in
the flavor representation
$\mathbf{10}_f\otimes\overline{\mathbf{10}}_f=\mathbf{64}_f\oplus\mathbf{27}_f\oplus\mathbf{8}_f\oplus\mathbf{1}_f$,
from Fig.1(VI) we can see
\begin{equation}
\label{2.60}\mathbf{20}_{cs}\otimes\overline{\mathbf{20}}_{cs}=\mathbf{175}_{cs}\oplus\mathbf{189}_{cs}\oplus\mathbf{35}_{cs}\oplus\mathbf{1}_{cs}
\end{equation}
The $\rm{SU}(6)_{cs}$ decomposition with respect to
$\rm{SU(3)}_c\otimes \rm{SU(2)}_s$ which are listed in Table VII and
Table VIII shows the following color singlet,
\begin{eqnarray}
\nonumber&&(\mathbf{1}_c,\mathbf{1})\subset\mathbf{1}_{cs}\\
\nonumber&&(\mathbf{1}_c,\mathbf{3})\subset\mathbf{35}_{cs}\\
\nonumber&&(\mathbf{1}_c,\mathbf{3})\;\; \rm{and} \;\;(\mathbf{1}_c,\mathbf{7})\subset\mathbf{175}_{cs}\\
\label{2.61}&&(\mathbf{1}_c,\mathbf{1})\;\; \rm{and}
\;\;(\mathbf{1}_c,\mathbf{5})\subset\mathbf{189}_{cs}
\end{eqnarray}
In order to derive the mass splitting due to $H^{\prime}$ by using
Eq.(\ref{1.2}), we must know which representations of the
$\rm{SU(3)}_{c}$ and of $\rm{SU}(2)_{s}$  the quarks and antiquarks
in the system belong to. In the followings, we study the $H'$
eigenstates with spin-$\mathbf{3}$, -$\mathbf{1}$ and -$\mathbf{0}$
respectively, and their corresponding eigenvalues:

$\mathbf{A)}$, The spin-$\mathbf{3}$ states
$|\mathbf{175}_{cs},\mathbf{1}_c,\mathbf{7};\mathbf{10}_f\otimes\mathbf{\overline{10}}_f\rangle$
and the spin-$\mathbf{2}$ states
$|\mathbf{189}_{cs},\mathbf{1}_c,\mathbf{5};\mathbf{10}_f\otimes\mathbf{\overline{10}}_f\rangle$
are trivial. The  wavefunction of the former is
\begin{equation}
\label{2.62}|\mathbf{175}_{cs},\mathbf{1}_c,\mathbf{7};\mathbf{10}_f\otimes\overline{\mathbf{10}}_f\rangle=|(\mathbf{20}_{cs},\mathbf{1}_c,\mathbf{4};\mathbf{10}_f),
(\overline{\mathbf{20}}_{cs},\mathbf{1}_c,\mathbf{4};\overline{\mathbf{10}}_f),\mathbf{1}_c,\mathbf{7};\mathbf{10}_f\otimes\overline{\mathbf{10}}_f\rangle
\end{equation}
Since
\begin{eqnarray*}
&&C_6(tot)=C_6(\mathbf{175_{cs}})=24,\;S_{tot}=\mathbf{3},\;C_{3}(tot)=C_3(\mathbf{1_{c}})=0,\;\\
&&C_3(Q)=C_3(\mathbf{1_c})=0,\;S_{Q}=\frac{3}{2},\;C_6(Q)=C_6(\mathbf{20_{cs}})=\frac{21}{2},\;\\
&&C_3(\overline{Q})=C_3(\mathbf{1_{c}})=0,\;S_{\overline{Q}}=\frac{3}{2},\;C_6(\overline{Q})=C_6(\mathbf{\overline{20}_{cs}})=\frac{21}{2}.
\end{eqnarray*}
Substituting the above quantities into Eq.(\ref{1.2}), then we get
the corresponding colormagnetic energy $16C$. And the latter is
\begin{equation}
\label{2.63}|\mathbf{189}_{cs},\mathbf{1}_c,\mathbf{5};\mathbf{10}_f\otimes\overline{\mathbf{10}}_f\rangle=|(\mathbf{20}_{cs},\mathbf{1}_c,\mathbf{4};\mathbf{10}_f),
(\overline{\mathbf{20}}_{cs},\mathbf{1}_c,\mathbf{4};\overline{\mathbf{10}}_f),\mathbf{1}_c,\mathbf{5};\mathbf{10}_f\otimes\overline{\mathbf{10}}_f\rangle
\end{equation}
The corresponding SU(6) Casimir operator is
$C_6(tot)=C_6(\mathbf{189_{cs}})=20$, and the colormagnetic energy
is also 16C.

$\mathbf{B)}$, The two spin-$\mathbf{1}$ states
$|\mathbf{35}_{cs},\mathbf{1}_c,\mathbf{3};\mathbf{10}_f\otimes\overline{\mathbf{10}}_f\rangle$
and
$|\mathbf{175}_{cs},\mathbf{1}_c,\mathbf{3};\mathbf{10}_f\otimes\overline{\mathbf{10}}_f\rangle$
are linear combinations of the states
$|(\mathbf{20}_{cs},\mathbf{8}_c,\mathbf{2};\mathbf{10}_{f}),(\overline{\mathbf{20}}_{cs},\mathbf{8}_c,\mathbf{2};\overline{\mathbf{10}}_{f}),\mathbf{1}_c,\mathbf{3};\mathbf{10}_f\otimes\overline{\mathbf{10}}_f\rangle$
and
$|(\mathbf{20}_{cs},\mathbf{1}_c,\mathbf{4};\mathbf{10}_{f}),(\overline{\mathbf{20}}_{cs},
\mathbf{1}_c,\mathbf{4};\overline{\mathbf{10}}_{f}),\mathbf{1}_c,
\mathbf{3};\mathbf{10}_f\otimes\overline{\mathbf{10}}_f\rangle$.
Following the method of the Appendix A, we can obtain
\begin{eqnarray}
\nonumber&&|\mathbf{175}_{cs},\mathbf{1}_c,\mathbf{3};\mathbf{10}_f\otimes\overline{\mathbf{10}}_f\rangle=\sqrt{\frac{5}{9}}|(\mathbf{20}_{cs},\mathbf{8}_c,\mathbf{2};\mathbf{10}_{f}),
(\overline{\mathbf{20}}_{cs},\mathbf{8}_c,\mathbf{2};\overline{\mathbf{10}}_{f}),\mathbf{1}_c,\mathbf{3};\mathbf{10}_f\otimes\overline{\mathbf{10}}_f\rangle\\
\nonumber&&-\sqrt{\frac{4}{9}}|(\mathbf{20}_{cs},\mathbf{1}_c,\mathbf{4};\mathbf{10}_{f}),(\overline{\mathbf{20}}_{cs},\mathbf{1}_c,\mathbf{4};\overline{\mathbf{10}}_{f}),\mathbf{1}_c,\mathbf{3};\mathbf{10}_f\otimes\overline{\mathbf{10}}_f\rangle,\\
\nonumber&&|\mathbf{35}_{cs},\mathbf{1}_c,\mathbf{3};\mathbf{10}_f\otimes\overline{\mathbf{10}}_f\rangle=-\sqrt{\frac{4}{9}}|(\mathbf{20}_{cs},\mathbf{8}_c,\mathbf{2};\mathbf{10}_{f}),
(\overline{\mathbf{20}}_{cs},\mathbf{8}_c,\mathbf{2};\overline{\mathbf{10}}_{f}),\mathbf{1}_c,\mathbf{3};\mathbf{10}_f\otimes\overline{\mathbf{10}}_f\rangle\\
\label{2.64}&&-\sqrt{\frac{5}{9}}|(\mathbf{20}_{cs},\mathbf{1}_c,\mathbf{4};\mathbf{10}_{f}),(\overline{\mathbf{20}}_{cs},\mathbf{1}_c,\mathbf{4};\overline{\mathbf{10}}_{f}),\mathbf{1}_c,\mathbf{3};\mathbf{10}_f\otimes\overline{\mathbf{10}}_f\rangle
\end{eqnarray}
These two states are mixed by the colormagnetic interaction
hamiltonian $H^{\prime}$,
\begin{eqnarray*}
&&H^{\prime}|\mathbf{175}_{cs},\mathbf{1}_c,\mathbf{3};\mathbf{10}_f\otimes\overline{\mathbf{10}}_f\rangle=\frac{16C}{9}(19|\mathbf{175}_{cs},\mathbf{1}_c,\mathbf{3};\mathbf{10}_f\otimes\overline{\mathbf{10}}_f\rangle-\sqrt{5}|\mathbf{35}_{cs},\mathbf{1}_c,\mathbf{3};\mathbf{10}_f\otimes\overline{\mathbf{10}}_f\rangle)\\
&&H^{\prime}|\mathbf{35}_{cs},\mathbf{1}_c,\mathbf{3};\mathbf{10}_f\otimes\overline{\mathbf{10}}_f\rangle=\frac{16C}{9}(-\sqrt{5}|\mathbf{175}_{cs},\mathbf{1}_c,\mathbf{3};\mathbf{10}_f\otimes\overline{\mathbf{10}}_f\rangle+5|\mathbf{35}_{cs},\mathbf{1}_c,\mathbf{3};\mathbf{10}_f\otimes\overline{\mathbf{10}}_f\rangle)
\end{eqnarray*}
From above, we can learn that the hamiltonian $H^{\prime}$ can be
written explicitly as the following matrix form in subspace expanded
by the two states,
\begin{equation}
\frac{16C}{9}\left(\begin{array}{cc}
19&-\sqrt{5}\\
-\sqrt{5}&5
\end{array}\right)
\end{equation}
The eigenstates are
\begin{eqnarray}
\nonumber&&|\mathbf{3};\mathbf{10}_f\otimes\overline{\mathbf{10}}_f\rangle_{1}=0.988|\mathbf{175}_{cs},\mathbf{1}_c,\mathbf{3};\mathbf{10}_f\otimes\overline{\mathbf{10}}_f\rangle-0.154|\mathbf{35}_{cs},\mathbf{1}_c,\mathbf{3};\mathbf{10}_f\otimes\overline{\mathbf{10}}_f\rangle\\
\nonumber&&=0.839|(\mathbf{20}_{cs},\mathbf{8}_c,\mathbf{2};\mathbf{10}_{f}),(\overline{\mathbf{20}}_{cs},\mathbf{8}_c,\mathbf{2};\overline{\mathbf{10}}_{f}),\mathbf{1}_c,\mathbf{3};\mathbf{10}_f\otimes\overline{\mathbf{10}}_f\rangle-0.544|(\mathbf{20}_{cs},\mathbf{1}_c,\mathbf{4};\mathbf{10}_{f}),\\
\label{2.65}&&(\overline{\mathbf{20}}_{cs},\mathbf{1}_c,\mathbf{4};\overline{\mathbf{10}}_{f}),\mathbf{1}_c,\mathbf{3};\mathbf{10}_f\otimes\overline{\mathbf{10}}_f\rangle,\\
\nonumber&& \\
\nonumber&&|\mathbf{3};\mathbf{10}_f\otimes\overline{\mathbf{10}}_f\rangle_{2}=0.154|\mathbf{175}_{cs},\mathbf{1}_c,\mathbf{3};\mathbf{10}_f\otimes\overline{\mathbf{10}}_f\rangle+0.988|\mathbf{35}_{cs},\mathbf{1}_c,\mathbf{3};\mathbf{10}_f\otimes\overline{\mathbf{10}}_f\rangle\\
\nonumber&&=-0.544|(\mathbf{20}_{cs},\mathbf{8}_c,\mathbf{2};\mathbf{10}_{f}),(\overline{\mathbf{20}}_{cs},\mathbf{8}_c,\mathbf{2};\overline{\mathbf{10}}_{f}),\mathbf{1}_c,\mathbf{3};\mathbf{10}_f\otimes\overline{\mathbf{10}}_f\rangle-0.839|(\mathbf{20}_{cs},\mathbf{1}_c,\mathbf{4};\mathbf{10}_{f}),\\
\label{2.66}&&(\overline{\mathbf{20}}_{cs},\mathbf{1}_c,\mathbf{4};\overline{\mathbf{10}}_{f}),\mathbf{1}_c,\mathbf{3};\mathbf{10}_f\otimes\overline{\mathbf{10}}_f\rangle
\end{eqnarray}
The corresponding colormagnetic energy are 34.397C and 8.269C.

$\mathbf{C)}$, In the same way, we can calculate the mixing of two
spin-$\mathbf{0}$ states
$|\mathbf{189}_{cs},\mathbf{1}_c,\mathbf{1};\mathbf{10}_f\otimes\overline{\mathbf{10}}_f\rangle$
and
$|\mathbf{1}_{cs},\mathbf{1}_c,\mathbf{1};\mathbf{10}_f\otimes\overline{\mathbf{10}}_f\rangle$.
The eigenstates are
\begin{eqnarray}
\nonumber&&|\mathbf{1};\mathbf{10}_f\otimes\overline{\mathbf{10}}_f\rangle_{1}=0.996|\mathbf{189}_{cs},\mathbf{1}_c,\mathbf{1};\mathbf{10}_f\otimes\overline{\mathbf{10}}_f\rangle-0.090|\mathbf{1}_{cs},\mathbf{1}_c,\mathbf{1};\mathbf{10}_f\otimes\overline{\mathbf{10}}_f\rangle\\
\nonumber&&=0.526|(\mathbf{20}_{cs},\mathbf{8}_c,\mathbf{2};\mathbf{10}_{f}),(\overline{\mathbf{20}}_{cs},\mathbf{8}_c,\mathbf{2};\overline{\mathbf{10}}_{f}),\mathbf{1}_c,\mathbf{1};\mathbf{10}_f\otimes\overline{\mathbf{10}}_f\rangle-0.851|(\mathbf{20}_{cs},\mathbf{1}_c,\mathbf{4};\mathbf{10}_{f}),\\
\label{2.67}&&(\overline{\mathbf{20}}_{cs},\mathbf{1}_c,\mathbf{4};\overline{\mathbf{10}}_{f}),\mathbf{1}_c,\mathbf{1};\mathbf{10}_f\otimes\overline{\mathbf{10}}_f\rangle,\\
\nonumber&& \\
\nonumber&&|\mathbf{1};\mathbf{10}_f\otimes\overline{\mathbf{10}}_f\rangle_{2}=0.090|\mathbf{189}_{cs},\mathbf{1}_c,\mathbf{1};\mathbf{10}_f\otimes\overline{\mathbf{10}}_f\rangle+0.996|\mathbf{1}_{cs},\mathbf{1}_c,\mathbf{1};\mathbf{10}_f\otimes\overline{\mathbf{10}}_f\rangle\\
\nonumber&&=-0.851|(\mathbf{20}_{cs},\mathbf{8}_c,\mathbf{2};\mathbf{10}_{f}),(\overline{\mathbf{20}}_{cs},\mathbf{8}_c,\mathbf{2};\overline{\mathbf{10}}_{f}),\mathbf{1}_c,\mathbf{1};\mathbf{10}_f\otimes\overline{\mathbf{10}}_f\rangle-0.526|(\mathbf{20}_{cs},\mathbf{1}_c,\mathbf{4};\mathbf{10}_{f}),\\
\label{2.68}&&(\overline{\mathbf{20}}_{cs},\mathbf{1}_c,\mathbf{4};\overline{\mathbf{10}}_{f}),\mathbf{1}_c,\mathbf{1};\mathbf{10}_f\otimes\overline{\mathbf{10}}_f\rangle
\end{eqnarray}
The colormagnetic energy separately are 25.889C and -9.889C.

\item $\mathbf{56}_{cs}\otimes\overline{\mathbf{20}}_{cs}$ and  $\mathbf{20}_{cs}\otimes\overline{\mathbf{56}}_{cs}$

The two multiplets are related by charge conjugation, and we discuss
the case of $\mathbf{56}_{cs}\otimes\overline{\mathbf{20}}_{cs}$,
the flavor represntation is
$\mathbf{1}_f\otimes\overline{\mathbf{10}}_f=\overline{\mathbf{10}}_f$,
and
\begin{equation}
\label{2.48}\mathbf{56}_{cs}\otimes\overline{\mathbf{20}}_{cs}=\mathbf{840}_{cs}\oplus\mathbf{280}_{cs}
\end{equation}
The color singlets are as followings
\begin{eqnarray}
\nonumber&&(\mathbf{1}_c,\mathbf{3})\subset\mathbf{280}_{cs}\\
\label{2.49}&&(\mathbf{1}_c,\mathbf{1})\subset\mathbf{840}_{cs}
\end{eqnarray}
The two wavefunctions
$|\mathbf{840}_{cs},\mathbf{1}_c,\mathbf{1};\mathbf{1}_f\otimes\overline{\mathbf{10}}_f\rangle$
and
$|\mathbf{280}_{cs},\mathbf{1}_c,\mathbf{3};\mathbf{1}_f\otimes\overline{\mathbf{10}}_f\rangle$
are trivial, and they are as  follows
\begin{equation}
\label{2.50}|\mathbf{840}_{cs},\mathbf{1}_c,\mathbf{1};\mathbf{1}_f\otimes\overline{\mathbf{10}}_f\rangle=|(\mathbf{56}_{cs},\mathbf{8}_c,\mathbf{2};\mathbf{1}_f),(\overline{\mathbf{20}}_{cs},\mathbf{8}_c,\mathbf{2};\overline{\mathbf{10}}_f),
\mathbf{1}_c,\mathbf{1};\mathbf{1}_f\otimes\overline{\mathbf{10}}_f\rangle
\end{equation}
\begin{equation}
\label{2.51}|\mathbf{280}_{cs},\mathbf{1}_c,\mathbf{3};\mathbf{1}_f\otimes\overline{\mathbf{10}}_f\rangle=|(\mathbf{56}_{cs},\mathbf{8}_c,\mathbf{2};\mathbf{1}_f),(\overline{\mathbf{20}}_{cs},\mathbf{8}_c,\mathbf{2};\overline{\mathbf{10}}_f),
\mathbf{1}_c,\mathbf{3};\mathbf{1}_f\otimes\overline{\mathbf{10}}_f\rangle
\end{equation}
The colormagnetic energy respectively are 16C and $-\frac{32}{3}$C.
In the same way we can obtain the wavefunctions in the case of
$\mathbf{20}_{cs}\otimes\overline{\mathbf{56}}_{cs}$ by interchange
quarks and antiquarks in Eq.(\ref{2.50}) and Eq.(\ref{2.51}), and
the corresponding colormagnetic energy is invariant.

\item $\mathbf{56}_{cs}\otimes\overline{\mathbf{56}}_{cs}$

The multiplets belong to the flavor representation
$\mathbf{1}_f\otimes\mathbf{1}_f=\mathbf{1}_f$, and
\begin{equation}
\label{2.52}\mathbf{56}_{cs}\otimes\overline{\mathbf{56}}_{cs}=\mathbf{2695}_{cs}\oplus\mathbf{405}_{cs}\oplus\mathbf{35}_{cs}\oplus\mathbf{1}_{cs}
\end{equation}
The overall color singlets are as followings,
\begin{eqnarray}
\nonumber&&(\mathbf{1}_c,\mathbf{1})\subset\mathbf{1}_{cs}\\
\nonumber&&(\mathbf{1}_c,\mathbf{3})\subset\mathbf{35}_{cs}\\
\nonumber&&(\mathbf{1}_c,\mathbf{1})\;\;\rm{and} \;\;(\mathbf{1}_c,\mathbf{5})\subset\mathbf{405}_{cs} \\
\label{2.53}&& (\mathbf{1}_c,\mathbf{3})\;\;\rm{and}
\;\;(\mathbf{1}_c,\mathbf{7})\subset\mathbf{2695}_{cs}
\end{eqnarray}
$\mathbf{A)}$, The spin-$\mathbf{2}$ and spin-$\mathbf{3}$ states
are trivial, their wavefunctions separately are
\begin{equation}
\label{2.54}|\mathbf{2695}_{cs},\mathbf{1}_c,\mathbf{7};\mathbf{1}_f\otimes\mathbf{1}_f\rangle=|(\mathbf{56}_{cs},\mathbf{10}_c,\mathbf{4};\mathbf{1}_f),
(\overline{\mathbf{56}}_{cs},\overline{\mathbf{10}}_c,\mathbf{4};\mathbf{1}_f),\mathbf{1}_c,\mathbf{7};\mathbf{1}_f\otimes\mathbf{1}_f\rangle
\end{equation}
\begin{equation}
\label{2.55}|\mathbf{405}_{cs},\mathbf{1}_c,\mathbf{5};\mathbf{1}_f\otimes\mathbf{1}_f\rangle=|(\mathbf{56}_{cs},\mathbf{10}_c,\mathbf{4};\mathbf{1}_f),
(\overline{\mathbf{56}}_{cs},\overline{\mathbf{10}}_c,\mathbf{4};\mathbf{1}_f),\mathbf{1}_c,\mathbf{5};\mathbf{1}_f\otimes\mathbf{1}_f\rangle
\end{equation}
The corresponding colormagnetic energy separately are 16C and -16C.

$\mathbf{B)}$, The two spin-$\mathbf{0}$ states
$|\mathbf{1}_{cs},\mathbf{1}_c,\mathbf{1};\mathbf{1}_f\otimes\mathbf{1}_f\rangle$
and
$|\mathbf{405}_{cs},\mathbf{1}_c,\mathbf{1};\mathbf{1}_f\otimes\mathbf{1}_f\rangle$
are linear combinations of the states
$|(\mathbf{56}_{cs},\mathbf{10}_c,\mathbf{4};\mathbf{1}_f),
(\overline{\mathbf{56}}_{cs},\overline{\mathbf{10}}_c,\mathbf{4};\mathbf{1}_f),\mathbf{1}_c,\mathbf{1};\mathbf{1}_f\otimes\mathbf{1}_f\rangle$
and $|(\mathbf{56}_{cs},\mathbf{8}_c,\mathbf{2};\mathbf{1}_f),
(\overline{\mathbf{56}}_{cs},\mathbf{8}_c,\mathbf{2};\mathbf{1}_f),\mathbf{1}_c,\mathbf{1};\mathbf{1}_f\otimes\mathbf{1}_f\rangle$.
By using the latter two states  as base vectors, the matrix
representation of $H'$ can be derived explicitly (please see
Appendix B, Eq.(\ref{B7})). Thus, $H'$'s  eigenstates read
\begin{eqnarray}
\nonumber&&|\mathbf{1},\mathbf{}1_f\otimes\mathbf{1}_f\rangle_1=0.591|(\mathbf{56}_{cs},\mathbf{10}_c,\mathbf{4};\mathbf{1}_f),(\overline{\mathbf{56}}_{cs},\overline{\mathbf{10}}_c,\mathbf{4};\mathbf{1}_f),\mathbf{1}_c,\mathbf{1};\mathbf{1}_f\otimes\mathbf{1}_f\rangle\\
\label{2.56}&&+0.807|(\mathbf{56}_{cs},\mathbf{8}_c,\mathbf{2};\mathbf{1}_f),(\overline{\mathbf{56}}_{cs},\mathbf{8}_c,\mathbf{2};\mathbf{1}_f),\mathbf{1}_c,\mathbf{1};\mathbf{1}_f\otimes\mathbf{1}_f\rangle,\\
\nonumber&& \\
\nonumber&&|\mathbf{1},\mathbf{}1_f\otimes\mathbf{1}_f\rangle_2=0.807|(\mathbf{56}_{cs},\mathbf{10}_c,\mathbf{4};\mathbf{1}_f),(\overline{\mathbf{56}}_{cs},\overline{\mathbf{10}}_c,\mathbf{4};\mathbf{1}_f),\mathbf{1}_c,\mathbf{1};\mathbf{1}_f\otimes\mathbf{1}_f\rangle\\
\label{2.57}&&-0.591|(\mathbf{56}_{cs},\mathbf{8}_c,\mathbf{2};\mathbf{1}_f),(\overline{\mathbf{56}}_{cs},\mathbf{8}_c,\mathbf{2};\mathbf{1}_f),\mathbf{1}_c,\mathbf{1};\mathbf{1}_f\otimes\mathbf{1}_f\rangle
\end{eqnarray}
The corresponding colormagnetic energy are -82.533C and -29.467C.

$\mathbf{C)}$, The two spin-$\mathbf{1}$ states
$|\mathbf{35}_{cs},\mathbf{1}_c,\mathbf{3};\mathbf{1}_f\otimes\mathbf{1}_f\rangle$
and
$|\mathbf{2695}_{cs},\mathbf{1}_c,\mathbf{3};\mathbf{1}_f\otimes\mathbf{1}_f\rangle$(or
$|(\mathbf{56}_{cs},\mathbf{10}_c,\mathbf{4};\mathbf{1}_f),
(\overline{\mathbf{56}}_{cs},\overline{\mathbf{10}}_c,\mathbf{4};\mathbf{1}_f),\mathbf{1}_c,\mathbf{3};\mathbf{1}_f\otimes\mathbf{1}_f\rangle$,
$|(\mathbf{56}_{cs},\mathbf{8}_c,\mathbf{2};\mathbf{1}_f),
(\overline{\mathbf{56}}_{cs},\mathbf{8}_c,\mathbf{2};\mathbf{1}_f),\mathbf{1}_c,\mathbf{3};\mathbf{1}_f\otimes\mathbf{1}_f\rangle$)
are mixed by the hamiltonian $H^{\prime}$. The matrix representation
of $H^{\prime}$ in the space expanded by the latter two states
vector have been given by Eq.(\ref{B8}) in Appendix B, then the
eigenstates are
\begin{eqnarray}
\nonumber&&|\mathbf{3},\mathbf{1}_f\otimes\mathbf{1}_f\rangle_1=0.864|(\mathbf{56}_{cs},\mathbf{10}_c,\mathbf{4};\mathbf{1}_f),(\overline{\mathbf{56}}_{cs},\overline{\mathbf{10}}_c,\mathbf{4};\mathbf{1}_f),\mathbf{1}_c,\mathbf{3};\mathbf{1}_f\otimes\mathbf{1}_f\rangle\\
\label{2.58}&&+0.504|(\mathbf{56}_{cs},\mathbf{8}_c,\mathbf{2};\mathbf{1}_f),(\overline{\mathbf{56}}_{cs},\mathbf{8}_c,\mathbf{2};\mathbf{1}_f),\mathbf{1}_c,\mathbf{3};\mathbf{1}_f\otimes\mathbf{1}_f\rangle,\\
\nonumber&& \\
\nonumber&&|\mathbf{3},\mathbf{1}_f\otimes\mathbf{1}_f\rangle_2=-0.504|(\mathbf{56}_{cs},\mathbf{10}_c,\mathbf{4};\mathbf{1}_f),(\overline{\mathbf{56}}_{cs},\overline{\mathbf{10}}_c,\mathbf{4};\mathbf{1}_f),\mathbf{1}_c,\mathbf{3};\mathbf{1}_f\otimes\mathbf{1}_f\rangle\\
\label{2.59}&&+0.864|(\mathbf{56}_{cs},\mathbf{8}_c,\mathbf{2};\mathbf{1}_f),(\overline{\mathbf{56}}_{cs},\mathbf{8}_c,\mathbf{2};\mathbf{1}_f),\mathbf{1}_c,\mathbf{3};\mathbf{1}_f\otimes\mathbf{1}_f\rangle
\end{eqnarray}
the corresponding eigenvalue are -48.331C, -5.003C.

\item $\mathbf{70}_{cs}\otimes\overline{\mathbf{20}}_{cs}\;\;and\;\;\mathbf{20}_{cs}\otimes\overline{\mathbf{70}}_{cs}.$

Similar to the case 2., the multiplets are related together by
charge conjugation, then we discuss
$\mathbf{70}_{cs}\otimes\overline{\mathbf{20}}_{cs}$, and we can
easily obtain the results for
$\mathbf{20}_{cs}\otimes\overline{\mathbf{70}}_{cs}$. The states can
belong to the flavor representation
$\mathbf{8}_f\otimes\overline{\mathbf{10}}_f=\overline{\mathbf{35}}_f\oplus\mathbf{27}_f\oplus\overline{\mathbf{10}}_f\oplus\mathbf{8}_f$,
and
\begin{equation}
\label{2.39}\mathbf{70}_{cs}\oplus\overline{\mathbf{20}}_{cs}=\mathbf{896}_{cs}\oplus\mathbf{280}_{cs}\oplus\mathbf{189}_{cs}\oplus\mathbf{35}_{cs}
\end{equation}
The overall color singlets are
\begin{eqnarray}
\nonumber&& (\mathbf{1}_c,\mathbf{3})\subset\mathbf{35}_{cs}\\
\nonumber&& (\mathbf{1}_c,\mathbf{1})\;\;\rm{and}\;\;
(\mathbf{1}_c,\mathbf{5})\subset\mathbf{189}_{cs}\\
\nonumber &&(\mathbf{1}_c,\mathbf{3})\subset\mathbf{280}_{cs}\\
\label{2.40}&& (\mathbf{1}_c,\mathbf{3})\;\;\rm{and}\;\;
(\mathbf{1}_c,\mathbf{5})\subset\mathbf{896}_{cs}
\end{eqnarray}
$\mathbf{A)}$, The spin-$\mathbf{0}$ states are trivial, its
wavefunctions are
\begin{equation}
\label{2.41}|\mathbf{189}_{cs},\mathbf{1}_c,\mathbf{1};\mathbf{8}_f\otimes\overline{\mathbf{10}}_f\rangle=|(\mathbf{70}_{cs},\mathbf{8}_c,\mathbf{2};\mathbf{8}_f),
(\overline{\mathbf{20}}_{cs},\mathbf{8}_c,\mathbf{2};\overline{\mathbf{10}}_f),\mathbf{1}_c,\mathbf{1};\mathbf{8}_f\otimes\overline{\mathbf{10}}_f\rangle
\end{equation}
The colormagnetic energy is 8C.

$\mathbf{B)}$, The three spin-$\mathbf{1}$ states
$|\mathbf{35}_{cs},\mathbf{1}_c,\mathbf{3};\mathbf{8}_f\otimes\overline{\mathbf{10}}_f\rangle$,
$|\mathbf{280}_{cs},\mathbf{1}_c,\mathbf{3};\mathbf{8}_f\otimes\overline{\mathbf{10}}_f\rangle$
and
$|\mathbf{896}_{cs},\mathbf{1}_c,\mathbf{3};\mathbf{8}_f\otimes\overline{\mathbf{10}}_f\rangle$
are linear combinations of the states
$|(\mathbf{70}_{cs},\mathbf{8}_c,\mathbf{4};\mathbf{8}_f),(\overline{\mathbf{20}}_{cs},\mathbf{8}_c,\mathbf{2};\overline{\mathbf{10}}_f),\mathbf{1}_c,\mathbf{3};\mathbf{8}_f\otimes\overline{\mathbf{10}}_f\rangle$,
$|(\mathbf{70}_{cs},\mathbf{8}_c,\mathbf{2};\mathbf{8}_f),(\overline{\mathbf{20}}_{cs},\mathbf{8}_c,\mathbf{2};\overline{\mathbf{10}}_f),\mathbf{1}_c,\mathbf{3};\mathbf{8}_f\otimes\overline{\mathbf{10}}_f\rangle$
and
$|(\mathbf{70}_{cs},\mathbf{1}_c,\mathbf{2};\mathbf{8}_f),(\overline{\mathbf{20}}_{cs},\mathbf{1}_c,\mathbf{4};\overline{\mathbf{10}}_f),\mathbf{1}_c,\mathbf{3};\mathbf{8}_f\otimes\overline{\mathbf{10}}_f\rangle$
(please see Appendix A for details),
\begin{eqnarray}
\nonumber&&|\mathbf{896}_{cs},\mathbf{1}_c,\mathbf{3};\mathbf{8}_f\otimes\overline{\mathbf{10}}_f\rangle=-\sqrt{\frac{1}{9}}|(\mathbf{70}_{cs},\mathbf{8}_c,\mathbf{4};\mathbf{8}_f),
(\overline{\mathbf{20}}_{cs},\mathbf{8}_c,\mathbf{2};\overline{\mathbf{10}}_f),\mathbf{1}_c,\mathbf{3};\mathbf{8}_f\otimes\overline{\mathbf{10}}_f\rangle\\
\nonumber&&-\sqrt{\frac{4}{9}}|(\mathbf{70}_{cs},\mathbf{8}_c,\mathbf{2};\mathbf{8}_f),
(\overline{\mathbf{20}}_{cs},\mathbf{8}_c,\mathbf{2};\overline{\mathbf{10}}_f),\mathbf{1}_c,\mathbf{3};\mathbf{8}_f\otimes\overline{\mathbf{10}}_f\rangle+\sqrt{\frac{4}{9}}|(\mathbf{70}_{cs},\mathbf{1}_c,\mathbf{2};\mathbf{8}_f),\\
\nonumber&&(\overline{\mathbf{20}}_{cs},\mathbf{1}_c,\mathbf{4};\overline{\mathbf{10}}_f),\mathbf{1}_c,\mathbf{3};\mathbf{8}_f\otimes\overline{\mathbf{10}}_f\rangle;\\
\nonumber&&|\mathbf{280}_{cs},\mathbf{1}_c,\mathbf{3};\mathbf{8}_f\otimes\overline{\mathbf{10}}_f\rangle=-\sqrt{\frac{4}{9}}|(\mathbf{70}_{cs},\mathbf{8}_c,\mathbf{4};\mathbf{8}_f),
(\overline{\mathbf{20}}_{cs},\mathbf{8}_c,\mathbf{2};\overline{\mathbf{10}}_f),\mathbf{1}_c,\mathbf{3};\mathbf{8}_f\otimes\overline{\mathbf{10}}_f\rangle\\
\nonumber&&-\sqrt{\frac{1}{9}}|(\mathbf{70}_{cs},\mathbf{8}_c,\mathbf{2};\mathbf{8}_f),
(\overline{\mathbf{20}}_{cs},\mathbf{8}_c,\mathbf{2};\overline{\mathbf{10}}_f),\mathbf{1}_c,\mathbf{3};\mathbf{8}_f\otimes\overline{\mathbf{10}}_f\rangle-\sqrt{\frac{4}{9}}|(\mathbf{70}_{cs},\mathbf{1}_c,\mathbf{2};\mathbf{8}_f),\\
\nonumber&&(\overline{\mathbf{20}}_{cs},\mathbf{1}_c,\mathbf{4};\overline{\mathbf{10}}_f),\mathbf{1}_c,\mathbf{3};\mathbf{8}_f\otimes\overline{\mathbf{10}}_f\rangle;\\
\nonumber&&|\mathbf{35}_{cs},\mathbf{1}_c,\mathbf{3};\mathbf{8}_f\otimes\overline{\mathbf{10}}_f\rangle=-\sqrt{\frac{4}{9}}|(\mathbf{70}_{cs},\mathbf{8}_c,\mathbf{4};\mathbf{8}_f),
(\overline{\mathbf{20}}_{cs},\mathbf{8}_c,\mathbf{2};\overline{\mathbf{10}}_f),\mathbf{1}_c,\mathbf{3};\mathbf{8}_f\otimes\overline{\mathbf{10}}_f\rangle\\
\nonumber&&+\sqrt{\frac{4}{9}}|(\mathbf{70}_{cs},\mathbf{8}_c,\mathbf{2};\mathbf{8}_f),
(\overline{\mathbf{20}}_{cs},\mathbf{8}_c,\mathbf{2};\overline{\mathbf{10}}_f),\mathbf{1}_c,\mathbf{3};\mathbf{8}_f\otimes\overline{\mathbf{10}}_f\rangle+\sqrt{\frac{1}{9}}|(\mathbf{70}_{cs},\mathbf{1}_c,\mathbf{2};\mathbf{8}_f),\\
\label{2.42}&&(\overline{\mathbf{20}}_{cs},\mathbf{1}_c,\mathbf{4};\overline{\mathbf{10}}_f),\mathbf{1}_c,\mathbf{3};\mathbf{8}_f\otimes\overline{\mathbf{10}}_f\rangle
\end{eqnarray}
These states are mixed by colormagnetic interaction $H^{\prime}$,
from above we can calculate the eigenstates
\begin{eqnarray}
\nonumber&&|\mathbf{3},\mathbf{8}_f\otimes\overline{\mathbf{10}}_f\rangle_1=0.841|\mathbf{896}_{cs},\mathbf{1}_c,\mathbf{3};\mathbf{8}_f\otimes\overline{\mathbf{10}}_f\rangle+0.536|\mathbf{280}_{cs},\mathbf{1}_c,\mathbf{3};\mathbf{8}_f\otimes\overline{\mathbf{10}}_f\rangle\\
\nonumber&&+0.069|\mathbf{35}_{cs},\mathbf{1}_c,\mathbf{3};\mathbf{8}_f\otimes\overline{\mathbf{10}}_f\rangle=-0.684|(\mathbf{70}_{cs},\mathbf{8}_c,\mathbf{4};\mathbf{8}_f),
(\overline{\mathbf{20}}_{cs},\mathbf{8}_c,\mathbf{2};\overline{\mathbf{10}}_f),\mathbf{1}_c,\mathbf{3};\mathbf{8}_f\otimes\overline{\mathbf{10}}_f\rangle\\
\nonumber&&-0.693|(\mathbf{70}_{cs},\mathbf{8}_c,\mathbf{2};\mathbf{8}_f),(\overline{\mathbf{20}}_{cs},\mathbf{8}_c,\mathbf{2};\overline{\mathbf{10}}_f),
\mathbf{1}_c,\mathbf{3};\mathbf{8}_f\otimes\overline{\mathbf{10}}_f\rangle+0.226|(\mathbf{70}_{cs},\mathbf{1}_c,\mathbf{2};\mathbf{8}_f),\\
\label{2.43}&&(\overline{\mathbf{20}}_{cs},\mathbf{1}_c,\mathbf{4};\overline{\mathbf{10}}_f),\mathbf{1}_c,\mathbf{3};\mathbf{8}_f\otimes\overline{\mathbf{10}}_f\rangle,\\
\nonumber&&|\mathbf{3},\mathbf{8}_f\otimes\overline{\mathbf{10}}_f\rangle_2=0.156|\mathbf{896}_{cs},\mathbf{1}_c,\mathbf{3};\mathbf{8}_f\otimes\overline{\mathbf{10}}_f\rangle-0.364|\mathbf{280}_{cs},\mathbf{1}_c,\mathbf{3};\mathbf{8}_f\otimes\overline{\mathbf{10}}_f\rangle\\
\nonumber&&+0.918|\mathbf{35}_{cs},\mathbf{1}_c,\mathbf{3};\mathbf{8}_f\otimes\overline{\mathbf{10}}_f\rangle=-0.422|(\mathbf{70}_{cs},\mathbf{8}_c,\mathbf{4};\mathbf{8}_f),
(\overline{\mathbf{20}}_{cs},\mathbf{8}_c,\mathbf{2};\overline{\mathbf{10}}_f),\mathbf{1}_c,\mathbf{3};\mathbf{8}_f\otimes\overline{\mathbf{10}}_f\rangle\\
\nonumber&&+0.629|(\mathbf{70}_{cs},\mathbf{8}_c,\mathbf{2};\mathbf{8}_f),(\overline{\mathbf{20}}_{cs},\mathbf{8}_c,\mathbf{2};\overline{\mathbf{10}}_f),\mathbf{1}_c,\mathbf{3};\mathbf{8}_f\otimes\overline{\mathbf{10}}_f\rangle+0.653|(\mathbf{70}_{cs},\mathbf{1}_c,\mathbf{2};\mathbf{8}_f),\\
\label{2.44}&&(\overline{\mathbf{20}}_{cs},\mathbf{1}_c,\mathbf{4};\overline{\mathbf{10}}_f),\mathbf{1}_c,\mathbf{3};\mathbf{8}_f\otimes\overline{\mathbf{10}}_f\rangle,\\
\nonumber&&|\mathbf{3},\mathbf{8}_f\otimes\overline{\mathbf{10}}_f\rangle_3=-0.518|\mathbf{896}_{cs},\mathbf{1}_c,\mathbf{3};\mathbf{8}_f\otimes\overline{\mathbf{10}}_f\rangle+0.762|\mathbf{280}_{cs},\mathbf{1}_c,\mathbf{3};\mathbf{8}_f\otimes\overline{\mathbf{10}}_f\rangle\\
\nonumber&&+0.389|\mathbf{35}_{cs},\mathbf{1}_c,\mathbf{3};\mathbf{8}_f\otimes\overline{\mathbf{10}}_f\rangle=-0.595|(\mathbf{70}_{cs},\mathbf{8}_c,\mathbf{4};\mathbf{8}_f),
(\overline{\mathbf{20}}_{cs},\mathbf{8}_c,\mathbf{2};\overline{\mathbf{10}}_f),\mathbf{1}_c,\mathbf{3};\mathbf{8}_f\otimes\overline{\mathbf{10}}_f\rangle\\
\nonumber&&+0.351|(\mathbf{70}_{cs},\mathbf{8}_c,\mathbf{2};\mathbf{8}_f),(\overline{\mathbf{20}}_{cs},\mathbf{8}_c,\mathbf{2};\overline{\mathbf{10}}_f),\mathbf{1}_c,\mathbf{3};\mathbf{8}_f\otimes\overline{\mathbf{10}}_f\rangle-0.723|(\mathbf{70}_{cs},\mathbf{1}_c,\mathbf{2};\mathbf{8}_f),\\
\label{2.45}&&(\overline{\mathbf{20}}_{cs},\mathbf{1}_c,\mathbf{4};\overline{\mathbf{10}}_f),\mathbf{1}_c,\mathbf{3};\mathbf{8}_f\otimes\overline{\mathbf{10}}_f\rangle
\end{eqnarray}
The colormagnetic energy respectively are 24.634C, -12.007C, 7.373C.
\par
$\mathbf{C)}$, For the two spin-$\mathbf{2}$ states
$|\mathbf{189}_{cs},\mathbf{1}_c,\mathbf{5};\mathbf{8}_f\otimes\overline{\mathbf{10}}_f\rangle$
and
$|\mathbf{896}_{cs},\mathbf{1}_c,\mathbf{5};\mathbf{8}_f\otimes\overline{\mathbf{10}}_f\rangle$(
or
$|(\mathbf{70}_{cs},\mathbf{8}_c,\mathbf{4};\mathbf{8}_f),(\overline{\mathbf{20}}_{cs},\mathbf{8}_c,\mathbf{2};\overline{\mathbf{10}}_f),\mathbf{1}_c,\mathbf{5};\mathbf{8}_f\otimes\overline{\mathbf{10}}_f\rangle$,
$|(\mathbf{70}_{cs},\mathbf{1}_c,\mathbf{2};\mathbf{8}_f),(\overline{\mathbf{20}}_{cs},\mathbf{1}_c,\mathbf{4};\overline{\mathbf{10}}_f),\mathbf{1}_c,\mathbf{5};\mathbf{8}_f\otimes\overline{\mathbf{10}}_f\rangle$),
similar to calculating the mixing of the 3 spin-$\mathbf{1}$ states
in the above, we can obtain the following eigenstates,
\begin{eqnarray}
\nonumber&&|\mathbf{5};\mathbf{8}_f\otimes\overline{\mathbf{10}}_f\rangle_1=\frac{3}{5}|\mathbf{896}_{cs},\mathbf{1}_c,\mathbf{5};\mathbf{8}_f\otimes\overline{\mathbf{10}}_f\rangle
+\frac{4}{5}|\mathbf{189}_{cs},\mathbf{1}_c,\mathbf{5};\mathbf{8}_f\otimes\overline{\mathbf{10}}_f\rangle\\
\nonumber&&=-0.447|(\mathbf{70}_{cs},\mathbf{8}_c,\mathbf{4};\mathbf{8}_f),(\overline{\mathbf{20}}_{cs},\mathbf{8}_c,\mathbf{2};\overline{\mathbf{10}}_f),\mathbf{1}_c,\mathbf{5};\mathbf{8}_f\otimes\overline{\mathbf{10}}_f\rangle+0.894|(\mathbf{70}_{cs},\mathbf{1}_c,\mathbf{2};\mathbf{8}_f),\\
\label{2.46}&&(\overline{\mathbf{20}}_{cs},\mathbf{1}_c,\mathbf{4};\overline{\mathbf{10}}_f),\mathbf{1}_c,\mathbf{5};\mathbf{8}_f\otimes\overline{\mathbf{10}}_f\rangle,\\
\nonumber&& \\
\nonumber&&|\mathbf{5};\mathbf{8}_f\otimes\overline{\mathbf{10}}_f\rangle_2=-\frac{4}{5}|\mathbf{896}_{cs},\mathbf{1}_c,\mathbf{5};\mathbf{8}_f\otimes\overline{\mathbf{10}}_f\rangle
+\frac{3}{5}|\mathbf{189}_{cs},\mathbf{1}_c,\mathbf{5};\mathbf{8}_f\otimes\overline{\mathbf{10}}_f\rangle\\
\nonumber&&=-0.894|(\mathbf{70}_{cs},\mathbf{8}_c,\mathbf{4};\mathbf{8}_f),(\overline{\mathbf{20}}_{cs},\mathbf{8}_c,\mathbf{2};\overline{\mathbf{10}}_f),\mathbf{1}_c,\mathbf{5};\mathbf{8}_f\otimes\overline{\mathbf{10}}_f\rangle-0.447|(\mathbf{70}_{cs},\mathbf{1}_c,\mathbf{2};\mathbf{8}_f),\\
\label{2.47}&&(\overline{\mathbf{20}}_{cs},\mathbf{1}_c,\mathbf{4};\overline{\mathbf{10}}_f),\mathbf{1}_c,\mathbf{5};\mathbf{8}_f\otimes\overline{\mathbf{10}}_f\rangle
\end{eqnarray}
The colormagnetic energy are respectively -4C, 16C. The wavefunction
for $\mathbf{20}_{cs}\otimes\overline{\mathbf{70}}_{cs}$ can be
obtained by change quarks and antiquarks in the wavefunction
Eq.(\ref{2.41}), Eq.(\ref{2.43}-\ref{2.47}), and the corresponding
colormagnetic energy is the same.

\item$\mathbf{70}_{cs}\otimes\overline{\mathbf{56}}_{cs}$ and  $\mathbf{56}_{cs}\otimes\overline{\mathbf{70}}_{cs}$.

The multiplets are related by charge conjugation, so we only discuss
$\mathbf{70}_{cs}\otimes\overline{\mathbf{56}}_{cs}$ and the results
for $\mathbf{56}_{cs}\otimes\overline{\mathbf{70}}_{cs}$ can be
obtained by exchange quarks and antiquarks. Its flavor
representation is $\mathbf{8}_f\otimes\mathbf{1}_f=\mathbf{8}_f$,
The calculation is analogous to that of
$\mathbf{20}_{cs}\otimes\overline{\mathbf{20}}_{cs}$, and
\begin{equation}
\label{2.31}\mathbf{70}_{cs}\otimes\overline{\mathbf{56}}_{cs}=\mathbf{3200}_{cs}\oplus\mathbf{405}_{cs}\oplus\overline{\mathbf{280}}_{cs}\oplus\mathbf{35}_{cs}
\end{equation}
with color singlets as followings from  Table VII and Table VIII,
\begin{eqnarray}
\nonumber && (\mathbf{1}_{c},\mathbf{3})\subset\mathbf{35}_{cs} \\
\nonumber && (\mathbf{1}_{c},\mathbf{3})\subset\overline{\mathbf{280}}_{cs} \\
\nonumber && (\mathbf{1}_{c},\mathbf{1})\;\; \rm{and}\;\; (\mathbf{1}_{c},\mathbf{5})\subset\mathbf{405}_{cs} \\
\label{2.32} && (\mathbf{1}_{c},\mathbf{3})\;\; \rm{and}\;\;
(\mathbf{1}_{c},\mathbf{5})\subset\mathbf{3200}_{cs}
\end{eqnarray}
We study each spin states respectively:

$\bf{A}$), The spin-$\mathbf{0}$ state is simple, Its wave function
is
\begin{equation}
\label{2.33}|\mathbf{405}_{cs},\mathbf{1}_c,\mathbf{1};\mathbf{8}_f\otimes\mathbf{1}_f\rangle=|(\mathbf{70}_{cs},\mathbf{8}_c,\mathbf{2};\mathbf{8}_f),
(\overline{\mathbf{56}}_{cs},\mathbf{8}_c,\mathbf{2};\mathbf{1}_f),
\mathbf{1}_c,\mathbf{1};\mathbf{8}_f\otimes\mathbf{1}_f\rangle
\end{equation}
The colormagnetic energy is $-24C$.

$\mathbf{B}$), The three spin-$\mathbf{1}$ states
$|\mathbf{35}_{cs},\mathbf{1}_c,\mathbf{3};\mathbf{8}_f\hskip-0.15in\otimes\hskip-0.15in\mathbf{1}_f\rangle$,
$|\overline{\mathbf{280}}_{cs},\mathbf{1}_c,\mathbf{3};\mathbf{8}_f\hskip-0.1in\otimes\hskip-0.1in\mathbf{1}_f\rangle$
and
$|\mathbf{3200}_{cs},\mathbf{1}_c,\mathbf{3};\mathbf{8}_f\hskip-0.1in\otimes\hskip-0.1in\mathbf{1}_f\rangle$
( or
$|(\mathbf{70}_{cs},\mathbf{10}_c,\mathbf{2};\mathbf{8}_f),(\overline{\mathbf{56}}_{cs},
\overline{\mathbf{10}}_c,\mathbf{4};\mathbf{1}_f),\mathbf{1}_c,\mathbf{3};\mathbf{8}_f\hskip-0.1in\otimes\hskip-0.1in\mathbf{1}_f\rangle$,
$|(\mathbf{70}_{cs},\mathbf{8}_c,\mathbf{4};\mathbf{8}_f),(\overline{\mathbf{56}}_{cs},\mathbf{8}_c,\mathbf{2};
\mathbf{1}_f),\mathbf{1}_c,\mathbf{3};\mathbf{8}_f\otimes\mathbf{1}_f\rangle$,
$|(\mathbf{70}_{cs},\mathbf{8}_c,\mathbf{2};\mathbf{8}_f),(\overline{\mathbf{56}}_{cs},\mathbf{8}_c,\mathbf{2};
\mathbf{1}_f),\mathbf{1}_c,\mathbf{3};\mathbf{8}_f\otimes\mathbf{1}_f\rangle$)
are mixed by the interaction $H^{\prime}$, and the matrix form of
$H'$ is Eq.(\ref{B4}). The eigenstates are
\begin{eqnarray}
\nonumber&&|\mathbf{3},\mathbf{8}_f\otimes\mathbf{1}_f\rangle_1=0.416|(\mathbf{70}_{cs},\mathbf{10}_c,\mathbf{2};\mathbf{8}_f),(\overline{\mathbf{56}}_{cs},\overline{\mathbf{10}}_c,\mathbf{4};\mathbf{1}_f),\mathbf{1}_c,\mathbf{3};\mathbf{8}_f\otimes\mathbf{1}_f\rangle\\
\nonumber&&-0.670|(\mathbf{70}_{cs},\mathbf{8}_c,\mathbf{4};\mathbf{8}_f),(\overline{\mathbf{56}}_{cs},\mathbf{8}_c,\mathbf{2};\mathbf{1}_f),\mathbf{1}_c,\mathbf{3};\mathbf{8}_f\otimes\mathbf{1}_f\rangle\\
\label{2.34}&&-0.615|(\mathbf{70}_{cs},\mathbf{8}_c,\mathbf{2};\mathbf{8}_f),(\overline{\mathbf{56}}_{cs},\mathbf{8}_c,\mathbf{2};\mathbf{1}_f),\mathbf{1}_c,\mathbf{3};\mathbf{8}_f\otimes\mathbf{1}_f\rangle,\\
\nonumber&& \\
\nonumber&&|\mathbf{3},\mathbf{8}_f\otimes\mathbf{1}_f\rangle_2=0.732|(\mathbf{70}_{cs},\mathbf{10}_c,\mathbf{2};\mathbf{8}_f),(\overline{\mathbf{56}}_{cs},\overline{\mathbf{10}}_c,\mathbf{4};\mathbf{1}_f),\mathbf{1}_c,\mathbf{3};\mathbf{8}_f\otimes\mathbf{1}_f\rangle\\
\nonumber&&+0.648|(\mathbf{70}_{cs},\mathbf{8}_c,\mathbf{4};\mathbf{8}_f),(\overline{\mathbf{56}}_{cs},\mathbf{8}_c,\mathbf{2};\mathbf{1}_f),\mathbf{1}_c,\mathbf{3};\mathbf{8}_f\otimes\mathbf{1}_f\rangle\\
\label{2.35}&&-0.212|(\mathbf{70}_{cs},\mathbf{8}_c,\mathbf{2};\mathbf{8}_f),(\overline{\mathbf{56}}_{cs},\mathbf{8}_c,\mathbf{2};\mathbf{1}_f),\mathbf{1}_c,\mathbf{3};\mathbf{8}_f\otimes\mathbf{1}_f\rangle,\\
\nonumber&& \\
\nonumber&&|\mathbf{3},\mathbf{8}_f\otimes\mathbf{1}_f\rangle_3=0.540|(\mathbf{70}_{cs},\mathbf{10}_c,\mathbf{2};\mathbf{8}_f),(\overline{\mathbf{56}}_{cs},\overline{\mathbf{10}}_c,\mathbf{4};\mathbf{1}_f),\mathbf{1}_c,\mathbf{3};\mathbf{8}_f\otimes\mathbf{1}_f\rangle\\
\nonumber&&-0.362|(\mathbf{70}_{cs},\mathbf{8}_c,\mathbf{4};\mathbf{8}_f),(\overline{\mathbf{56}}_{cs},\mathbf{8}_c,\mathbf{2};\mathbf{1}_f),\mathbf{1}_c,\mathbf{3};\mathbf{8}_f\otimes\mathbf{1}_f\rangle\\
\label{2.36}&&+0.760|(\mathbf{70}_{cs},\mathbf{8}_c,\mathbf{2};\mathbf{8}_f),(\overline{\mathbf{56}}_{cs},\mathbf{8}_c,\mathbf{2};\mathbf{1}_f),\mathbf{1}_c,\mathbf{3};\mathbf{8}_f\otimes\mathbf{1}_f\rangle
\end{eqnarray}
The corresponding eigenvalues are $-45.078C,\; -14.477C,\; 7.555C $.

$\mathbf{C}$), The two spin-$\mathbf{2}$ states
$|\mathbf{405}_{cs},\mathbf{1}_c,\mathbf{5};\mathbf{8}_f\otimes\mathbf{1}_f\rangle$
and
$|\mathbf{3200}_{cs},\mathbf{1}_c,\mathbf{5};\mathbf{8}_f\otimes\mathbf{1}_f\rangle$
are linear combinations of the states
$|(\mathbf{70}_{cs},\mathbf{10}_c,\mathbf{2};\mathbf{8}_f),(\overline{\mathbf{56}}_{cs},\overline{\mathbf{10}}_c,\mathbf{4};\mathbf{1}_f),\mathbf{1}_c,\mathbf{5};\mathbf{8}_f\otimes\mathbf{1}_f\rangle$
and
$|(\mathbf{70}_{cs},\mathbf{8}_c,\mathbf{4};\mathbf{8}_f),(\overline{\mathbf{56}}_{cs},\mathbf{8}_c,\mathbf{2};\mathbf{1}_f),\mathbf{1}_c,\mathbf{5};\mathbf{8}_f\otimes\mathbf{1}_f\rangle$.
$H'$'s matrix form is Eq.(\ref{B5}), and the eigenstates are
\begin{eqnarray}
\nonumber&&|\mathbf{5},\mathbf{8}_f\otimes\mathbf{1}_f\rangle_1=0.845|(\mathbf{70}_{cs},\mathbf{10}_c,\mathbf{2};\mathbf{8}_f),(\overline{\mathbf{56}}_{cs},\overline{\mathbf{10}}_c,\mathbf{4};\mathbf{1}_f),\mathbf{1}_c,\mathbf{5};\mathbf{8}_f\otimes\mathbf{1}_f\rangle\\
\label{2.37}&&+0.535|(\mathbf{70}_{cs},\mathbf{8}_c,\mathbf{4};\mathbf{8}_f),(\overline{\mathbf{56}}_{cs},\mathbf{8}_c,\mathbf{2};\mathbf{1}_f),\mathbf{1}_c,\mathbf{5};\mathbf{8}_f\otimes\mathbf{1}_f\rangle,\\
\nonumber&& \\
\nonumber&&|\mathbf{5},\mathbf{8}_f\otimes\mathbf{1}_f\rangle_2=-0.535|(\mathbf{70}_{cs},\mathbf{10}_c,\mathbf{2};\mathbf{8}_f),(\overline{\mathbf{56}}_{cs},\overline{\mathbf{10}}_c,\mathbf{4};\mathbf{1}_f),\mathbf{1}_c,\mathbf{5};\mathbf{8}_f\otimes\mathbf{1}_f\rangle\\
\label{2.38}&&+0.845|(\mathbf{70}_{cs},\mathbf{8}_c,\mathbf{4};\mathbf{8}_f),(\overline{\mathbf{56}}_{cs},\mathbf{8}_c,\mathbf{2};\mathbf{1}_f),\mathbf{1}_c,\mathbf{5};\mathbf{8}_f\otimes\mathbf{1}_f\rangle
\end{eqnarray}
The corresponding eigenvalues are $16C$ and $-12C$. The eigenstates
for $\mathbf{56}_{cs}\otimes\overline{\mathbf{70}}_{cs}$ can be
obtained by interchanging the quarks and the antiquarks, and the
corresponding eigenvalues are the same as those of
$\mathbf{70}_{cs}\otimes\overline{\mathbf{56}}_{cs}$.

\item $\mathbf{70}_{cs}\otimes\overline{\mathbf{70}}_{cs}$.

Since
$\mathbf{8}_f\otimes\mathbf{8}_f=\mathbf{1}_f\oplus\mathbf{8}_f\oplus\mathbf{8}_f
\oplus\mathbf{10}_f\oplus\overline{\mathbf{10}}_f\oplus\mathbf{27}_f$,
the multiplets can be in flavor $\mathbf{1}_f$-plet,
$\mathbf{8}_f$-plet, $\mathbf{10}_f$-plet,
$\overline{\mathbf{10}}_f$-plet or $\mathbf{27}_f$-plet. From
Fig.1(I) we can see
\begin{equation}
\label{2.13}\mathbf{70}_{cs}\otimes\overline{\mathbf{70}}_{cs}=\mathbf{3675}_{cs}\oplus\mathbf{405}_{cs}\oplus\mathbf{280}_{cs}
\oplus\overline{\mathbf{280}}_{cs}\oplus\mathbf{35}_{cs}\oplus\mathbf{189}_{cs}\oplus\mathbf{35}_{cs}\oplus\mathbf{1}_{cs}
\end{equation}
The $\rm{SU}(6)_{cs}$ decomposition with respect to
$\rm{SU(3)}_c\otimes \rm{SU(2)}_s$ which are listed in Table VII and
Table VIII shows the following color singlet,
\begin{eqnarray}
\nonumber && (\mathbf{1}_c,\mathbf{1})\subset\mathbf{1}_{cs}\\
\nonumber && (\mathbf{1}_c,\mathbf{3})\subset\mathbf{35}_{cs}\;\;(\rm{there}\;\; \rm{exist}\;\; \rm{two}\;\;\mathbf{35}_{cs}-\rm{plets} )\\
\nonumber && (\mathbf{1}_c,\mathbf{1})\;\; \rm{and} \;\;(\mathbf{1}_c,\mathbf{5})\subset\mathbf{189}_{cs}\\
\nonumber && (\mathbf{1}_c,\mathbf{3})\subset\mathbf{280}_{cs}\\
\nonumber && (\mathbf{1}_c,\mathbf{3})\subset\overline{\mathbf{280}}_{cs}\\
\nonumber && (\mathbf{1}_c,\mathbf{1})\;\; \rm{and}
\;\;(\mathbf{1}_c,\mathbf{5})\subset\mathbf{405}_{cs}\\
\label{2.14} &&
(\mathbf{1}_c,\mathbf{1}),2(\mathbf{1}_c,\mathbf{3}),(\mathbf{1}_c,\mathbf{5})\;\;\rm{and}\;\;
(\mathbf{1}_c,\mathbf{7})\subset\mathbf{3675}_{cs}
\end{eqnarray}

{\bf A}), the spin-$\mathbf{3}$ eigenstates, since the
SU$_c$(3)$\otimes$ SU$_s$(2) decomposition of $\mathbf{70}_{cs}$ is
$\mathbf{70}_{cs}=(\mathbf{10}_c,\mathbf{2})+(\mathbf{8}_c,\mathbf{4})+(\mathbf{8}_c,\mathbf{2})+(\mathbf{1}_c,\mathbf{2})$,
it is straightforward to get the wave function of the system with
spin-$\mathbf{3}$ :
\begin{equation}
\label{2.15}|\mathbf{3675}_{cs},\mathbf{1}_c,\mathbf{7};\mathbf{8}_f\otimes\mathbf{8}_f\rangle=|(\mathbf{70}_{cs},\mathbf{8}_c,\mathbf{4};\mathbf{8}_f),
(\overline{\mathbf{70}}_{cs},\mathbf{8}_c,\mathbf{4};\mathbf{8}_f),
\mathbf{1}_c,\mathbf{7};\mathbf{8}_f\otimes\mathbf{8}_f\rangle.
\end{equation}
By the Eq.(\ref{1.2}), the corresponding colormagnetic energy is
$16C$.

{\bf B}), the spin-$\mathbf{2}$ eigenstates:  The three
spin-$\mathbf{2}$ states
$|\mathbf{3675}_{cs},\mathbf{1}_c,\mathbf{5};\mathbf{8}_f\otimes\mathbf{8}_f\rangle$,
$|\mathbf{405}_{cs},\mathbf{1}_c,\mathbf{5};\mathbf{8}_f\otimes\mathbf{8}_f\rangle$,
and
$|\mathbf{189}_{cs},\mathbf{1}_c,\mathbf{5};\mathbf{8}_f\otimes\mathbf{8}_f\rangle$
are linear combinations of
$|(\mathbf{70}_{cs},\;\mathbf{8}_c,\mathbf{4};\mathbf{8}_f),(\overline{\mathbf{70}}_{cs},\mathbf{8}_c,\mathbf{4};\mathbf{8}_f),$$\mathbf{1}_c,\mathbf{5},\mathbf{8}_f\otimes\mathbf{8}_f\rangle$,
$|(\mathbf{70}_{cs},\;\mathbf{8}_c,\mathbf{4};\mathbf{8}_f),\\(\overline{\mathbf{70}}_{cs},\mathbf{8}_c,\mathbf{2};\mathbf{8}_f),$$\mathbf{1}_c,\mathbf{5},\mathbf{8}_f\otimes\mathbf{8}_f\rangle$
and $|(\mathbf{70}_{cs},\mathbf{8}_c,\mathbf{2};\mathbf{8}_f),
(\overline{\mathbf{70}}_{cs},\mathbf{8}_c,\mathbf{4};\mathbf{8}_f),$
$\mathbf{1}_c,\mathbf{5},\mathbf{8}_f\otimes\mathbf{8}_f\rangle$,
but they are mixed by the colormagnetic interaction hamiltonian
$H^{\prime}$, and $H^{\prime}$ can be expressed as the  matrix
Eq.(\ref{B1}) in the subspace expanded by the latter three states,
diagonalizing this matrix, we can obtain the eigenstates and
eigenvalues.
The eigenstates are
\begin{eqnarray}
\nonumber&&|\mathbf{5},\mathbf{8}_f\otimes\mathbf{8}_f\rangle_1=-0.707|(\mathbf{70}_{cs},\;\mathbf{8}_c,\mathbf{4};\mathbf{8}_f),(\overline{\mathbf{70}}_{cs},\mathbf{8}_c,\mathbf{2};\mathbf{8}_f),\mathbf{1}_c,\mathbf{5},\mathbf{8}_f\otimes\mathbf{8}_f\rangle\\
\label{2.17}&&+0.707|(\mathbf{70}_{cs},\mathbf{8}_c,\mathbf{2};\mathbf{8}_f),(\overline{\mathbf{70}}_{cs},\mathbf{8}_c,\mathbf{4};\mathbf{8}_f),\mathbf{1}_c,\mathbf{5},\mathbf{8}_f\otimes\mathbf{8}_f\rangle,\\
\nonumber&& \\
\nonumber&&|\mathbf{5},\mathbf{8}_f\otimes\mathbf{8}_f\rangle_2=-0.526|(\mathbf{70}_{cs},\;\mathbf{8}_c,\mathbf{4};\mathbf{8}_f),(\overline{\mathbf{70}}_{cs},\mathbf{8}_c,\mathbf{4};\mathbf{8}_f),\mathbf{1}_c,\mathbf{5},\mathbf{8}_f\otimes\mathbf{8}_f\rangle\\
\nonumber&&-0.602|(\mathbf{70}_{cs},\;\mathbf{8}_c,\mathbf{4};\mathbf{8}_f),(\overline{\mathbf{70}}_{cs},\mathbf{8}_c,\mathbf{2};\mathbf{8}_f),\mathbf{1}_c,\mathbf{5},\mathbf{8}_f\otimes\mathbf{8}_f\rangle\\
\label{2.18}&&-0.602|(\mathbf{70}_{cs},\mathbf{8}_c,\mathbf{2};\mathbf{8}_f),(\overline{\mathbf{70}}_{cs},\mathbf{8}_c,\mathbf{4};\mathbf{8}_f),\mathbf{1}_c,\mathbf{5},\mathbf{8}_f\otimes\mathbf{8}_f\rangle,\\
\nonumber&& \\
\nonumber&&|\mathbf{5},\mathbf{8}_f\otimes\mathbf{8}_f\rangle_3=0.851|(\mathbf{70}_{cs},\;\mathbf{8}_c,\mathbf{4};\mathbf{8}_f),(\overline{\mathbf{70}}_{cs},\mathbf{8}_c,\mathbf{4};\mathbf{8}_f),\mathbf{1}_c,\mathbf{5},\mathbf{8}_f\otimes\mathbf{8}_f\rangle\\
\nonumber&&-0.372|(\mathbf{70}_{cs},\;\mathbf{8}_c,\mathbf{4};\mathbf{8}_f),(\overline{\mathbf{70}}_{cs},\mathbf{8}_c,\mathbf{2};\mathbf{8}_f),\mathbf{1}_c,\mathbf{5},\mathbf{8}_f\otimes\mathbf{8}_f\rangle\\
\label{2.19}&&-0.372|(\mathbf{70}_{cs},\mathbf{8}_c,\mathbf{2};\mathbf{8}_f),(\overline{\mathbf{70}}_{cs},\mathbf{8}_c,\mathbf{4};\mathbf{8}_f),\mathbf{1}_c,\mathbf{5},\mathbf{8}_f\otimes\mathbf{8}_f\rangle
\end{eqnarray}
And the corresponding eigenvalues are $16C,\; -12.944C,\; 4.944C$.

{$\mathbf{C}$}), the spin-$\mathbf{1}$ eigenstates: The six
spin-$\mathbf{1}$ states
$|\mathbf{35}_{cs},\mathbf{1}_c,\mathbf{3};\mathbf{8}_f\otimes\mathbf{8}_f\rangle_1$,
$|\mathbf{35}_{cs},\mathbf{1}_c,\mathbf{3};\mathbf{8}_f\otimes\mathbf{8}_f\rangle_2$,
$|\mathbf{280}_{cs},\mathbf{1}_c,\mathbf{3};\mathbf{8}_f\otimes\mathbf{8}_f\rangle$,
$|\overline{\mathbf{280}}_{cs},\mathbf{1}_c,\mathbf{3};\mathbf{8}_f\otimes\mathbf{8}_f\rangle$,
$|\mathbf{3675}_{cs},\mathbf{1}_c,\mathbf{3};\mathbf{8}_f\otimes\mathbf{8}_f\rangle_1$
and
$|\mathbf{3675}_{cs},\mathbf{1}_c,\mathbf{3};\mathbf{8}_f\otimes\mathbf{8}_f\rangle_2$
are linear combination of the six direct product states
$|(\mathbf{70}_{cs},\mathbf{10}_c,\mathbf{2};\mathbf{8}_f),(\overline{\mathbf{70}}_{cs},\overline{\mathbf{10}}_c,\mathbf{2};\mathbf{8}_f),$$\mathbf{1}_c,\mathbf{3};\mathbf{8}_f\otimes\mathbf{8}_f\rangle$,
$|(\mathbf{70}_{cs},\mathbf{8}_c,\mathbf{4};\mathbf{8}_f),(\overline{\mathbf{70}}_{cs},\mathbf{8}_c,\mathbf{4};\mathbf{8}_f),$$\mathbf{1}_c,\mathbf{3};\mathbf{8}_f\otimes\mathbf{8}_f\rangle$,
$|(\mathbf{70}_{cs},\mathbf{8}_c,\mathbf{4};\mathbf{8}_f),(\overline{\mathbf{70}}_{cs},\mathbf{8}_c,\mathbf{2};\mathbf{8}_f),$$\mathbf{1}_c,\mathbf{3};\mathbf{8}_f\otimes\mathbf{8}_f\rangle$,
$|(\mathbf{70}_{cs},\mathbf{8}_c,\mathbf{2};\mathbf{8}_f),(\overline{\mathbf{70}}_{cs},\mathbf{8}_c,\mathbf{4};\mathbf{8}_f),$$\mathbf{1}_c,\mathbf{3};\mathbf{8}_f\otimes\mathbf{8}_f\rangle$,
$|(\mathbf{70}_{cs},\mathbf{8}_c,\mathbf{2};\mathbf{8}_f),(\overline{\mathbf{70}}_{cs},\mathbf{8}_c,\mathbf{2};\mathbf{8}_f),$$\mathbf{1}_c,\mathbf{3};\mathbf{8}_f\otimes\mathbf{8}_f\rangle$
and
$|(\mathbf{70}_{cs},\mathbf{1}_c,\mathbf{2};\mathbf{8}_f),(|\overline{\mathbf{70}}_{cs},\mathbf{1}_c,\mathbf{2};\mathbf{8}_f),$$\mathbf{1}_c,\mathbf{3};\mathbf{8}_f\otimes\mathbf{8}_f\rangle$.
Noting that there are two $\mathbf{35}_{cs}$ in the outer product of
$\mathbf{70}_{cs}\otimes\overline{\mathbf{70}}_{cs}$, the states
$|\mathbf{35}_{cs},\mathbf{1}_c,\mathbf{3};\mathbf{8}_f\otimes\mathbf{8}_f\rangle_1$
and
$|\mathbf{35}_{cs},\mathbf{1}_c,\mathbf{3};\mathbf{8}_f\otimes\mathbf{8}_f\rangle_2$
represent the two states respectively belonging to the two
$\mathbf{35}_{cs}$. ($\mathbf{1}_c,\mathbf{3}$) appears twice in the
decomposition of $\mathbf{3675}_{cs}$, then the states
$|\mathbf{3675}_{cs},\mathbf{1}_c,\mathbf{3};\mathbf{8}_f\otimes\mathbf{8}_f\rangle_1$
and
$|\mathbf{3675}_{cs},\mathbf{1}_c,\mathbf{3};\mathbf{8}_f\otimes\mathbf{8}_f\rangle_2$
denote the two states corresponding to the two
($\mathbf{1}_c,\mathbf{3}$) respectively. Once again the states of
definite spin and $\rm{SU(6)}_{cs}$ representation are mixed by
$H^{\prime}$. By using the matrix representation of $H^{\prime}$,
i.e., Eq.(\ref{B2}),
the eigenstates are
\begin{eqnarray}
\nonumber&&|\mathbf{3},\mathbf{8}_f\otimes\mathbf{8}_f\rangle_1=0.2|(\mathbf{70}_{cs},\mathbf{10}_c,\mathbf{2};\mathbf{8}_f),(\overline{\mathbf{70}}_{cs},\overline{\mathbf{10}}_c,\mathbf{2};\mathbf{8}_f),\mathbf{1}_c,\mathbf{3};\mathbf{8}_f\otimes\mathbf{8}_f\rangle\\
\nonumber&&-0.324|(\mathbf{70}_{cs},\mathbf{8}_c,\mathbf{4};\mathbf{8}_f),(\overline{\mathbf{70}}_{cs},\mathbf{8}_c,\mathbf{4};\mathbf{8}_f),\mathbf{1}_c,\mathbf{3};\mathbf{8}_f\otimes\mathbf{8}_f\rangle\\
\nonumber&&-0.461|(\mathbf{70}_{cs},\mathbf{8}_c,\mathbf{4};\mathbf{8}_f),(\overline{\mathbf{70}}_{cs},\mathbf{8}_c,\mathbf{2};\mathbf{8}_f),\mathbf{1}_c,\mathbf{3};\mathbf{8}_f\otimes\mathbf{8}_f\rangle\\
\nonumber&&-0.461|(\mathbf{70}_{cs},\mathbf{8}_c,\mathbf{2};\mathbf{8}_f),(\overline{\mathbf{70}}_{cs},\mathbf{8}_c,\mathbf{4};\mathbf{8}_f),\mathbf{1}_c,\mathbf{3};\mathbf{8}_f\otimes\mathbf{8}_f\rangle\\
\nonumber&&+0.013|(\mathbf{70}_{cs},\mathbf{8}_c,\mathbf{2};\mathbf{8}_f),(\overline{\mathbf{70}}_{cs},\mathbf{8}_c,\mathbf{2};\mathbf{8}_f),\mathbf{1}_c,\mathbf{3};\mathbf{8}_f\otimes\mathbf{8}_f\rangle\\
\label{2.21}&&+0.656|(\mathbf{70}_{cs},\mathbf{1}_c,\mathbf{2};\mathbf{8}_f),(|\overline{\mathbf{70}}_{cs},\mathbf{1}_c,\mathbf{2};\mathbf{8}_f),\mathbf{1}_c,\mathbf{3};\mathbf{8}_f\otimes\mathbf{8}_f\rangle,\\
\nonumber&& \\
\nonumber&&|\mathbf{3},\mathbf{8}_f\otimes\mathbf{8}_f\rangle_2=-0.750|(\mathbf{70}_{cs},\mathbf{10}_c,\mathbf{2};\mathbf{8}_f),(\overline{\mathbf{70}}_{cs},\overline{\mathbf{10}}_c,\mathbf{2};\mathbf{8}_f),\mathbf{1}_c,\mathbf{3};\mathbf{8}_f\otimes\mathbf{8}_f\rangle\\
\nonumber&&+0.363|(\mathbf{70}_{cs},\mathbf{8}_c,\mathbf{4};\mathbf{8}_f),(\overline{\mathbf{70}}_{cs},\mathbf{8}_c,\mathbf{4};\mathbf{8}_f),\mathbf{1}_c,\mathbf{3};\mathbf{8}_f\otimes\mathbf{8}_f\rangle\\
\nonumber&&-0.303|(\mathbf{70}_{cs},\mathbf{8}_c,\mathbf{4};\mathbf{8}_f),(\overline{\mathbf{70}}_{cs},\mathbf{8}_c,\mathbf{2};\mathbf{8}_f),\mathbf{1}_c,\mathbf{3};\mathbf{8}_f\otimes\mathbf{8}_f\rangle\\
\nonumber&&-0.303|(\mathbf{70}_{cs},\mathbf{8}_c,\mathbf{2};\mathbf{8}_f),(\overline{\mathbf{70}}_{cs},\mathbf{8}_c,\mathbf{4};\mathbf{8}_f),\mathbf{1}_c,\mathbf{3};\mathbf{8}_f\otimes\mathbf{8}_f\rangle\\
\nonumber&&+0.349|(\mathbf{70}_{cs},\mathbf{8}_c,\mathbf{2};\mathbf{8}_f),(\overline{\mathbf{70}}_{cs},\mathbf{8}_c,\mathbf{2};\mathbf{8}_f),\mathbf{1}_c,\mathbf{3};\mathbf{8}_f\otimes\mathbf{8}_f\rangle\\
\label{2.22}&&-0.025|(\mathbf{70}_{cs},\mathbf{1}_c,\mathbf{2};\mathbf{8}_f),(|\overline{\mathbf{70}}_{cs},\mathbf{1}_c,\mathbf{2};\mathbf{8}_f),\mathbf{1}_c,\mathbf{3};\mathbf{8}_f\otimes\mathbf{8}_f\rangle,\\
\nonumber&& \\
\nonumber&&|\mathbf{3},\mathbf{8}_f\otimes\mathbf{8}_f\rangle_3=-0.274|(\mathbf{70}_{cs},\mathbf{10}_c,\mathbf{2};\mathbf{8}_f),(\overline{\mathbf{70}}_{cs},\overline{\mathbf{10}}_c,\mathbf{2};\mathbf{8}_f),\mathbf{1}_c,\mathbf{3};\mathbf{8}_f\otimes\mathbf{8}_f\rangle\\
\nonumber&&-0.764|(\mathbf{70}_{cs},\mathbf{8}_c,\mathbf{4};\mathbf{8}_f),(\overline{\mathbf{70}}_{cs},\mathbf{8}_c,\mathbf{4};\mathbf{8}_f),\mathbf{1}_c,\mathbf{3};\mathbf{8}_f\otimes\mathbf{8}_f\rangle\\
\nonumber&&+0.184|(\mathbf{70}_{cs},\mathbf{8}_c,\mathbf{4};\mathbf{8}_f),(\overline{\mathbf{70}}_{cs},\mathbf{8}_c,\mathbf{2};\mathbf{8}_f),\mathbf{1}_c,\mathbf{3};\mathbf{8}_f\otimes\mathbf{8}_f\rangle\\
\nonumber&&+0.184|(\mathbf{70}_{cs},\mathbf{8}_c,\mathbf{2};\mathbf{8}_f),(\overline{\mathbf{70}}_{cs},\mathbf{8}_c,\mathbf{4};\mathbf{8}_f),\mathbf{1}_c,\mathbf{3};\mathbf{8}_f\otimes\mathbf{8}_f\rangle\\
\nonumber&&+0.521|(\mathbf{70}_{cs},\mathbf{8}_c,\mathbf{2};\mathbf{8}_f),(\overline{\mathbf{70}}_{cs},\mathbf{8}_c,\mathbf{2};\mathbf{8}_f),\mathbf{1}_c,\mathbf{3};\mathbf{8}_f\otimes\mathbf{8}_f\rangle\\
\label{2.23}&&-0.044|(\mathbf{70}_{cs},\mathbf{1}_c,\mathbf{2};\mathbf{8}_f),(|\overline{\mathbf{70}}_{cs},\mathbf{1}_c,\mathbf{2};\mathbf{8}_f),\mathbf{1}_c,\mathbf{3};\mathbf{8}_f\otimes\mathbf{8}_f\rangle,\\
\nonumber&& \\
\nonumber&&|\mathbf{3},\mathbf{8}_f\otimes\mathbf{8}_f\rangle_4=0.482|(\mathbf{70}_{cs},\mathbf{10}_c,\mathbf{2};\mathbf{8}_f),(\overline{\mathbf{70}}_{cs},\overline{\mathbf{10}}_c,\mathbf{2};\mathbf{8}_f),\mathbf{1}_c,\mathbf{3};\mathbf{8}_f\otimes\mathbf{8}_f\rangle\\
\nonumber&&-0.323|(\mathbf{70}_{cs},\mathbf{8}_c,\mathbf{4};\mathbf{8}_f),(\overline{\mathbf{70}}_{cs},\mathbf{8}_c,\mathbf{2};\mathbf{8}_f),\mathbf{1}_c,\mathbf{3};\mathbf{8}_f\otimes\mathbf{8}_f\rangle\\
\nonumber&&-0.323|(\mathbf{70}_{cs},\mathbf{8}_c,\mathbf{2};\mathbf{8}_f),(\overline{\mathbf{70}}_{cs},\mathbf{8}_c,\mathbf{4};\mathbf{8}_f),\mathbf{1}_c,\mathbf{3};\mathbf{8}_f\otimes\mathbf{8}_f\rangle\\
\nonumber&&+0.431|(\mathbf{70}_{cs},\mathbf{8}_c,\mathbf{2};\mathbf{8}_f),(\overline{\mathbf{70}}_{cs},\mathbf{8}_c,\mathbf{2};\mathbf{8}_f),\mathbf{1}_c,\mathbf{3};\mathbf{8}_f\otimes\mathbf{8}_f\rangle\\
\label{2.24}&&-0.610|(\mathbf{70}_{cs},\mathbf{1}_c,\mathbf{2};\mathbf{8}_f),(|\overline{\mathbf{70}}_{cs},\mathbf{1}_c,\mathbf{2};\mathbf{8}_f),\mathbf{1}_c,\mathbf{3};\mathbf{8}_f\otimes\mathbf{8}_f\rangle,\\
\nonumber&& \\
\nonumber&&|\mathbf{3},\mathbf{8}_f\otimes\mathbf{8}_f\rangle_5=-0.299|(\mathbf{70}_{cs},\mathbf{10}_c,\mathbf{2};\mathbf{8}_f),(\overline{\mathbf{70}}_{cs},\overline{\mathbf{10}}_c,\mathbf{2};\mathbf{8}_f),\mathbf{1}_c,\mathbf{3};\mathbf{8}_f\otimes\mathbf{8}_f\rangle\\
\nonumber&&-0.424|(\mathbf{70}_{cs},\mathbf{8}_c,\mathbf{4};\mathbf{8}_f),(\overline{\mathbf{70}}_{cs},\mathbf{8}_c,\mathbf{4};\mathbf{8}_f),\mathbf{1}_c,\mathbf{3};\mathbf{8}_f\otimes\mathbf{8}_f\rangle\\
\nonumber&&-0.239|(\mathbf{70}_{cs},\mathbf{8}_c,\mathbf{4};\mathbf{8}_f),(\overline{\mathbf{70}}_{cs},\mathbf{8}_c,\mathbf{2};\mathbf{8}_f),\mathbf{1}_c,\mathbf{3};\mathbf{8}_f\otimes\mathbf{8}_f\rangle\\
\nonumber&&-0.239|(\mathbf{70}_{cs},\mathbf{8}_c,\mathbf{2};\mathbf{8}_f),(\overline{\mathbf{70}}_{cs},\mathbf{8}_c,\mathbf{4};\mathbf{8}_f),\mathbf{1}_c,\mathbf{3};\mathbf{8}_f\otimes\mathbf{8}_f\rangle\\
\nonumber&&-0.649|(\mathbf{70}_{cs},\mathbf{8}_c,\mathbf{2};\mathbf{8}_f),(\overline{\mathbf{70}}_{cs},\mathbf{8}_c,\mathbf{2};\mathbf{8}_f),\mathbf{1}_c,\mathbf{3};\mathbf{8}_f\otimes\mathbf{8}_f\rangle\\
\label{2.25}&& -0.441|(\mathbf{70}_{cs},\mathbf{1}_c,\mathbf{2};\mathbf{8}_f),(|\overline{\mathbf{70}}_{cs},\mathbf{1}_c,\mathbf{2};\mathbf{8}_f),\mathbf{1}_c,\mathbf{3};\mathbf{8}_f\otimes\mathbf{8}_f\rangle,\\
\nonumber&& \\
\nonumber&&|\mathbf{3},\mathbf{8}_f\otimes\mathbf{8}_f\rangle_6=-0.707|(\mathbf{70}_{cs},\mathbf{8}_c,\mathbf{4};\mathbf{8}_f),(\overline{\mathbf{70}}_{cs},\mathbf{8}_c,\mathbf{2};\mathbf{8}_f),\mathbf{1}_c,\mathbf{3};\mathbf{8}_f\otimes\mathbf{8}_f\rangle\\
\label{2.26}&&+0.707|(\mathbf{70}_{cs},\mathbf{8}_c,\mathbf{2};\mathbf{8}_f),(\overline{\mathbf{70}}_{cs},\mathbf{8}_c,\mathbf{4};\mathbf{8}_f),\mathbf{1}_c,\mathbf{3};\mathbf{8}_f\otimes\mathbf{8}_f\rangle
\end{eqnarray}
Their corresponding eigenvalues are $-29.461C,\; 27.631C,\;
-20.293C,\; -10.667C,\; 3.456C,\\ -2.667C $ respectively.

${\mathbf{D}})$, the spin-$\mathbf{0}$ eigenstates:  similarly the
four spin-$\mathbf{0}$ states
$|\mathbf{1}_{cs},\mathbf{1}_c,\mathbf{1};\mathbf{8}_f\otimes\mathbf{8}_f\rangle$,
$|\mathbf{189}_{cs},\mathbf{1}_c,\mathbf{1};\mathbf{8}_f\otimes\mathbf{8}_f\rangle$,
$|\mathbf{405}_{cs},\mathbf{1}_c,\mathbf{1};\mathbf{8}_f\otimes\mathbf{8}_f\rangle$,
$|\mathbf{3675}_{cs},\mathbf{1}_c,\mathbf{1};\mathbf{8}_f\otimes\mathbf{8}_f\rangle$,
are linear combination of the four states
$|(\mathbf{70}_{cs},\mathbf{10}_c,\mathbf{2};\mathbf{8}_f),(\overline{\mathbf{70}}_{cs},\overline{\mathbf{10}}_c,\mathbf{2};\mathbf{8}_f),\mathbf{1}_c,\mathbf{1};\mathbf{8}_f\otimes\mathbf{8}_f\rangle$,
$|(\mathbf{70}_{cs},\mathbf{8}_c,\mathbf{4};\mathbf{8}_f),(\overline{\mathbf{70}}_{cs},\mathbf{8}_c,\mathbf{4};\mathbf{8}_f),\mathbf{1}_c,\mathbf{1};\mathbf{8}_f\otimes\mathbf{8}_f\rangle$,
$|(\mathbf{70}_{cs},\mathbf{8}_c,\mathbf{2};\mathbf{8}_f),(\overline{\mathbf{70}}_{cs},\mathbf{8}_c,\mathbf{2};\mathbf{8}_f),\mathbf{1}_c,\mathbf{1};\mathbf{8}_f\otimes\mathbf{8}_f\rangle$
and
$|(\mathbf{70}_{cs},\mathbf{1}_c,\mathbf{2};\mathbf{8}_f),(\overline{\mathbf{70}}_{cs},\mathbf{1}_c,\mathbf{2};\mathbf{8}_f),\mathbf{1}_c,\mathbf{1};\mathbf{8}_f\otimes\mathbf{8}_f\rangle$.
$H'$'s matrix form is Eq.(\ref{B3}), thus the eigenstates are
\begin{eqnarray}
\nonumber&&|\mathbf{1},\mathbf{8}_f\otimes\mathbf{8}_f\rangle_1=0.394|(\mathbf{70}_{cs},\mathbf{10}_c,\mathbf{2};\mathbf{8}_f),(\overline{\mathbf{70}}_{cs},\overline{\mathbf{10}}_c,\mathbf{2};\mathbf{8}_f),\mathbf{1}_c,\mathbf{1};\mathbf{8}_f\otimes\mathbf{8}_f\rangle\\
\nonumber&&+0.604|(\mathbf{70}_{cs},\mathbf{8}_c,\mathbf{4};\mathbf{8}_f),(\overline{\mathbf{70}}_{cs},\mathbf{8}_c,\mathbf{4};\mathbf{8}_f),\mathbf{1}_c,\mathbf{1};\mathbf{8}_f\otimes\mathbf{8}_f\rangle\\
\nonumber&&-0.576|(\mathbf{70}_{cs},\mathbf{8}_c,\mathbf{2};\mathbf{8}_f),(\overline{\mathbf{70}}_{cs},\mathbf{8}_c,\mathbf{2};\mathbf{8}_f),\mathbf{1}_c,\mathbf{1};\mathbf{8}_f\otimes\mathbf{8}_f\rangle\\
\label{2.27}&&-0.384|(\mathbf{70}_{cs},\mathbf{1}_c,\mathbf{2};\mathbf{8}_f),(\overline{\mathbf{70}}_{cs},\mathbf{1}_c,\mathbf{2};\mathbf{8}_f),\mathbf{1}_c,\mathbf{1};\mathbf{8}_f\otimes\mathbf{8}_f\rangle,\\
\nonumber&& \\
\nonumber&&|\mathbf{1},\mathbf{8}_f\otimes\mathbf{8}_f\rangle_2=-0.725|(\mathbf{70}_{cs},\mathbf{10}_c,\mathbf{2};\mathbf{8}_f),(\overline{\mathbf{70}}_{cs},\overline{\mathbf{10}}_c,\mathbf{2};\mathbf{8}_f),\mathbf{1}_c,\mathbf{1};\mathbf{8}_f\otimes\mathbf{8}_f\rangle\\
\nonumber&&-0.649|(\mathbf{70}_{cs},\mathbf{8}_c,\mathbf{2};\mathbf{8}_f),(\overline{\mathbf{70}}_{cs},\mathbf{8}_c,\mathbf{2};\mathbf{8}_f),\mathbf{1}_c,\mathbf{1};\mathbf{8}_f\otimes\mathbf{8}_f\rangle\\
\label{2.28}&&+0.229|(\mathbf{70}_{cs},\mathbf{1}_c,\mathbf{2};\mathbf{8}_f),(\overline{\mathbf{70}}_{cs},\mathbf{1}_c,\mathbf{2};\mathbf{8}_f),\mathbf{1}_c,\mathbf{1};\mathbf{8}_f\otimes\mathbf{8}_f\rangle,\\
\nonumber&& \\
\nonumber&&|\mathbf{1},\mathbf{8}_f\otimes\mathbf{8}_f\rangle_3=0.371|(\mathbf{70}_{cs},\mathbf{10}_c,\mathbf{2};\mathbf{8}_f),(\overline{\mathbf{70}}_{cs},\overline{\mathbf{10}}_c,\mathbf{2};\mathbf{8}_f),\mathbf{1}_c,\mathbf{1};\mathbf{8}_f\otimes\mathbf{8}_f\rangle\\
\nonumber&&+0.232|(\mathbf{70}_{cs},\mathbf{8}_c,\mathbf{4};\mathbf{8}_f),(\overline{\mathbf{70}}_{cs},\mathbf{8}_c,\mathbf{4};\mathbf{8}_f),\mathbf{1}_c,\mathbf{1};\mathbf{8}_f\otimes\mathbf{8}_f\rangle\\
\nonumber&&-0.099|(\mathbf{70}_{cs},\mathbf{8}_c,\mathbf{2};\mathbf{8}_f),(\overline{\mathbf{70}}_{cs},\mathbf{8}_c,\mathbf{2};\mathbf{8}_f),\mathbf{1}_c,\mathbf{1};\mathbf{8}_f\otimes\mathbf{8}_f\rangle\\
\label{2.29}&&+0.894|(\mathbf{70}_{cs},\mathbf{1}_c,\mathbf{2};\mathbf{8}_f),(\overline{\mathbf{70}}_{cs},\mathbf{1}_c,\mathbf{2};\mathbf{8}_f),\mathbf{1}_c,\mathbf{1};\mathbf{8}_f\otimes\mathbf{8}_f\rangle,\\
\nonumber&& \\
\nonumber&&|\mathbf{1},\mathbf{8}_f\otimes\mathbf{8}_f\rangle_4=0.425|(\mathbf{70}_{cs},\mathbf{10}_c,\mathbf{2};\mathbf{8}_f),(\overline{\mathbf{70}}_{cs},\overline{\mathbf{10}}_c,\mathbf{2};\mathbf{8}_f),\mathbf{1}_c,\mathbf{1};\mathbf{8}_f\otimes\mathbf{8}_f\rangle\\
\nonumber&&-0.763|(\mathbf{70}_{cs},\mathbf{8}_c,\mathbf{4};\mathbf{8}_f),(\overline{\mathbf{70}}_{cs},\mathbf{8}_c,\mathbf{4};\mathbf{8}_f),\mathbf{1}_c,\mathbf{1};\mathbf{8}_f\otimes\mathbf{8}_f\rangle\\
\nonumber&&-0.487|(\mathbf{70}_{cs},\mathbf{8}_c,\mathbf{2};\mathbf{8}_f),(\overline{\mathbf{70}}_{cs},\mathbf{8}_c,\mathbf{2};\mathbf{8}_f),\mathbf{1}_c,\mathbf{1};\mathbf{8}_f\otimes\mathbf{8}_f\rangle\\
\label{2.30}&&-0.033|(\mathbf{70}_{cs},\mathbf{1}_c,\mathbf{2};\mathbf{8}_f),(\overline{\mathbf{70}}_{cs},\mathbf{1}_c,\mathbf{2};\mathbf{8}_f),\mathbf{1}_c,\mathbf{1};\mathbf{8}_f\otimes\mathbf{8}_f\rangle
\end{eqnarray}
The eigenvalues are $-45.547C,\; 16C,\; -12.671C,\; 2.218C$
respectively.

\par
~~~Finally, as a matter of convenience, we list all the QCD-allowed
states (i.e., color singlets)  and their colormagnetic energy in the
Table I. From this Table, we learned at first that the
configurations corresponding to the
$\Lambda\overline{\Lambda},\;p\overline{\Lambda},\;p\overline{p},
\;\cdots$ enhancements can have rather large negative colormagnetic
energy. This indicates that these states should be visible in
experiments, and therefore these results are consistent with the
experiments reported in Ref.[2-8]. Secondly, we can see also that
the binding colormagnetic energy of spin-$\mathbf{0}$ and
spin-$\mathbf{1}$ states are larger than ones of the states with
spin-$\mathbf{2}$ and spin-$\mathbf{3}$.
\par
~~~Specifically, and more concretely,
$|\mathbf{1},\mathbf{1}_f\otimes\mathbf{1}_f\rangle_1$(the wave
function is Eq.(\ref{2.56})) has the largest colormagnetic energy
-82.533C, and it has the same quantum numbers with the
$\Lambda\overline{\Lambda}$ enhancement \cite{bell3}, so this state
possibly is just the candidate $\Lambda\overline{\Lambda}$
enhancement. And the state
$|\mathbf{3},\mathbf{1}_f\otimes\mathbf{1}_f\rangle_1$ with the wave
function Eq.(\ref{2.58}) has colormagnetic energy -48.331C, then
this state is also rather stable. This spin-$\mathbf{1}$
$\Lambda\overline{\Lambda}$ enhancement as a prediction is expected
to be observed by BES, Bell or CLEO in future. By similar analysis,
the colormagnetic interaction energy of the state
$|\mathbf{1},\mathbf{8}_f\otimes\mathbf{8}_f\rangle_1$(the wave
function is Eq.(\ref{2.27})) is $-45.547C$, and the $p\overline{p}$
enhancement\cite{Bes1,Bes2} should belong to these states. Since the
colormagnetic energy of the proton or antiproton is $-8C$ (please
see Table II), we can estimate that the mass defect of the
$p\overline{p}$ enhancement approximately is $29.547C$ which is
compatible with the experimental observation, that the
$p\overline{p}$ enhancement is bellow the threshold. Also we see
that the states
$|\mathbf{3},\mathbf{8}_f\otimes\mathbf{8}_f\rangle_1$ whose wave
function is Eq.(\ref{2.21}) has colormagnetic energy -29.461C, so
these states could also be seen experimentally. Therefore we
conjecture that both the spin-$\mathbf{0}$ and spin-$\mathbf{1}$
$p\overline{p}$ enhancement may have been observed by
BES\cite{Bes1,Bes2}. The states
$|\mathbf{1},\mathbf{8}_f\otimes\mathbf{1}_f\rangle$(the wave
function is given by Eq.(\ref{2.33})) has colormagnetic energy -24C,
and the $p\overline{\Lambda}$ enhancement\cite{Bes3,bell2} can
belong to these states. However the states
$|\mathbf{3},\mathbf{8}_f\otimes\mathbf{1}_f\rangle_1$ whose wave
function is given by Eq.(\ref{2.34}) has colormagnetic energy
-45.078C, which is larger than the colormagnetic energy of the state
$|\mathbf{1},\mathbf{8}_f\otimes\mathbf{1}_f\rangle$, so it
indicates that the vector $p\overline{\Lambda}$ enhancement may be
favored over the scalar one. Because the $p\overline{\Lambda}$
enhancement has been seen in the process $J/\psi\rightarrow
p\overline{\Lambda}K^{-}$ and $\psi^{\prime}\rightarrow
p\overline{\Lambda}K^{-}$, and $J/\psi$ and $\psi^{\prime}$ are
vector mesons, $K^{-}$ is a pseudoscalar meson, the relative angular
momentum between $K^-$ and the $p\overline{\Lambda}$ enhancement
$l=0$ is favored, if the $p\overline{\Lambda}$ enhancement is a
vector state. We also conjecture that both the spin-$\mathbf{0}$ and
spin-$\mathbf{1}$ $p\overline{\Lambda}$ enhancement has been
observed by BES but the main content is spin-$\mathbf{1}$
$p\overline{\Lambda}$ enhancement.
\par
\begin{table}
\begin{center}
\caption{Colormagnetic interaction energy of
$\rm{q}^3\overline{\rm{q}}^3$ eigenstates in common quark model, and
the conjugates states are not included. }
\begin{tabular}{|c|c|c|c|c|c|}\hline\hline
State& Wave Function& Colormagnetic  &State & Wave Function& Colormagnetic \\
 & &Interaction Energy & & &Interaction Energy\\\hline \hline
$|\mathbf{1},\mathbf{8}_f\otimes\mathbf{8}_f\rangle_1$&Eq.(\ref{2.27})&-45.547C&$|\mathbf{1},\mathbf{8}_f\otimes\mathbf{8}_f\rangle_2$&Eq.(\ref{2.28})&16C\\\hline
$|\mathbf{1},\mathbf{8}_f\otimes\mathbf{8}_f\rangle_3$&Eq.(\ref{2.29})&-12.671C&$|\mathbf{1},\mathbf{8}_f\otimes\mathbf{8}_f\rangle_4$&Eq.(\ref{2.30})&2.218C\\\hline
$|\mathbf{1},\mathbf{8}_f\otimes\mathbf{1}_f\rangle$&Eq.(\ref{2.33})&-24C&$|\mathbf{1},\mathbf{8}_f\otimes\overline{\mathbf{10}}_f\rangle$&Eq.(\ref{2.41})&8C\\\hline
$|\mathbf{1},\mathbf{1}_f\otimes\overline{\mathbf{10}}_f\rangle$&Eq.(\ref{2.50})&16C&$|\mathbf{1},\mathbf{1}_f\otimes\mathbf{1}_f\rangle_1$&Eq.(\ref{2.56})&-82.533C\\\hline
$|\mathbf{1},\mathbf{1}_f\otimes\mathbf{1}_f\rangle_2$&Eq.(\ref{2.57})&-29.467C&$|\mathbf{1},\mathbf{10}_f\otimes\overline{\mathbf{10}}_f\rangle_1$&Eq.(\ref{2.67})&25.889C\\\hline
$|\mathbf{1},\mathbf{10}_f\otimes\overline{\mathbf{10}}_f\rangle_2$&Eq.(\ref{2.68})&-9.889C&$|\mathbf{3},\mathbf{8}_f\otimes\mathbf{8}_f\rangle_1$&Eq.(\ref{2.21})&-29.461C\\\hline
$|\mathbf{3},\mathbf{8}_f\otimes\mathbf{8}_f\rangle_2$&Eq.(\ref{2.22})&27.631C&$|\mathbf{3},\mathbf{8}_f\otimes\mathbf{8}_f\rangle_3$&Eq.(\ref{2.23})&-20.293C\\\hline
$|\mathbf{3},\mathbf{8}_f\otimes\mathbf{8}_f\rangle_4$&Eq.(\ref{2.24})&-10.667C&$|\mathbf{3},\mathbf{8}_f\otimes\mathbf{8}_f\rangle_5$&Eq.(\ref{2.25})&3.456C\\\hline
$|\mathbf{3},\mathbf{8}_f\otimes\mathbf{8}_f\rangle_6$&Eq.(\ref{2.26})&-2.667C&$|\mathbf{3},\mathbf{8}_f\otimes\mathbf{1}_f\rangle_1$&Eq.(\ref{2.34})&-45.078C\\\hline
$|\mathbf{3},\mathbf{8}_f\otimes\mathbf{1}_f\rangle_2$&Eq.(\ref{2.35})&-14.477C&$|\mathbf{3},\mathbf{8}_f\otimes\mathbf{1}_f\rangle_3$&Eq.(\ref{2.36})&7.555C\\\hline
$|\mathbf{3},\mathbf{8}_f\otimes\overline{\mathbf{10}}_f\rangle_1$&Eq.(\ref{2.43})&24.634C&$|\mathbf{3},\mathbf{8}_f\otimes\overline{\mathbf{10}}_f\rangle_2$&Eq.(\ref{2.44})&-12.007C\\\hline
$|\mathbf{3},\mathbf{8}_f\otimes\overline{\mathbf{10}}_f\rangle_3$&Eq.(\ref{2.45})&7.373C&$|\mathbf{3},\mathbf{1}_f\otimes\overline{\mathbf{10}}_f\rangle$&Eq.(\ref{2.51})&$-\frac{32}{3}$C\\\hline
$|\mathbf{3},\mathbf{1}_f\otimes\mathbf{1}_f\rangle_1$&Eq.(\ref{2.58})&-48.331C&$|\mathbf{3},\mathbf{1}_f\otimes\mathbf{1}_f\rangle_2$&Eq.(\ref{2.59})&-5.003C\\\hline
$|\mathbf{3},\mathbf{10}_f\otimes\overline{\mathbf{10}}_f\rangle_1$&Eq.(\ref{2.65})&34.397C&$|\mathbf{3},\mathbf{10}_f\otimes\overline{\mathbf{10}}_f\rangle_2$&Eq.(\ref{2.66})&8.269C\\\hline
$|\mathbf{5},\mathbf{8}_f\otimes\mathbf{8}_f\rangle_1$&Eq.(\ref{2.17})&16C&$|\mathbf{5},\mathbf{8}_f\otimes\mathbf{8}_f\rangle_2$&Eq.(\ref{2.18})&-12.944C\\\hline
$|\mathbf{5},\mathbf{8}_f\otimes\mathbf{8}_f\rangle_3$&Eq.(\ref{2.19})&4.944C&$|\mathbf{5},\mathbf{8}_f\otimes\mathbf{1}_f\rangle_1$&Eq.(\ref{2.37})&16C\\\hline
$|\mathbf{5},\mathbf{8}_f\otimes\mathbf{1}_f\rangle_2$&Eq.(\ref{2.38})&-12C&$|\mathbf{5},\mathbf{8}_f\otimes\overline{\mathbf{10}}_f\rangle_1$&Eq.(\ref{2.46})&-4C\\\hline
$|\mathbf{5},\mathbf{8}_f\otimes\overline{\mathbf{10}}_f\rangle_2$&Eq.(\ref{2.47})&16C&$|\mathbf{5},\mathbf{1}_f\otimes\mathbf{1}_f\rangle$&Eq.(\ref{2.55})&-16C\\\hline
$|\mathbf{5},\mathbf{10}_f\otimes\overline{\mathbf{10}}_f\rangle$&Eq.(\ref{2.63})&16C&$|\mathbf{7},\mathbf{8}_f\otimes\mathbf{8}_f\rangle$&Eq.(\ref{2.15})&16C\\\hline
$|\mathbf{7},\mathbf{1}_f\otimes\mathbf{1}_f\rangle$&Eq.(\ref{2.54})&16C&$|\mathbf{7},\mathbf{10}_f\otimes\overline{\mathbf{10}}_f\rangle$&Eq.(\ref{2.62})&16C\\\hline\hline
\end{tabular}
\end{center}
\end{table}
\end{enumerate}

\section{the interesting configurations}
Quark correlation turns out to play important role in multiquark
states \cite{correlation,ding2}, so we consider the effect of quark
correlation in the $\rm{q}^3\overline{\rm{q}}^3$ system. Generally
in the $\rm{q}^{3}\overline{\rm{q}}^3$ system there are many
possible configuration: $(\rm{q})-(\rm{q}^{2}\overline{\rm{q}}^3)$,
$(\overline{\rm{q}})-(\rm{q}^3\overline{\rm{q}}^2)$,
$(\rm{q}^{2})-(\rm{q}\overline{\rm{q}}^3)$,
$(\rm{q}\overline{\rm{q}})-(\rm{q}^{2}\overline{\rm{q}}^2)$,
$(\overline{\rm{q}}^2)-(\rm{q}^3\overline{\rm{q}})$,
$(\rm{q}^{3})-(\overline{\rm{q}}^3)$,
$(\rm{q}^2\overline{\rm{q}})-(\rm{q}\overline{\rm{q}}^2)$ and so on.
In this paper, we only consider the symmetric configurations: $
(\rm{q}^{3})-(\overline{\rm{q}}^3)$,
$(\rm{q}^2\overline{\rm{q}})-(\rm{q}\overline{\rm{q}}^2)$, where
under the changes $\rm{q}\leftrightarrow \overline{\rm{q}}$, the
$1^{st}$-cluster and $2^{nd}$-cluster interchange each other.   The
reasons why we choose these configuration are both because triquark
correlation has been proved to be important in
pentaquark\cite{correlation}, and because the configurations are
more symmetric than others. The cluster structure is postulated
instead of being established by a dynamical calculation. But with
the development of lattice QCD, we could study which configurations
are more stable than others through lattice numerical
calculations\cite{jaffe2}. It is assumed the two clusters are
separated by a distance larger than the range of the colormagnetic
force which is of short range and are kept together by the
colorelectric force. Consequently, the colormagnetic hyperfine
interaction operates only within each cluster, but is not felt
between them.
\par
Each cluster can be in a variety of SU(3)$_{c}$ representations, and
eventually two clusters together form the color singlet physical
baryonium state. However, if two clusters are
 color singlets respectively, they will interact only via
the van der Waal's force of QCD and will dissociate immediately into
a baryon and a antibaryon, or be bound very weakly to form a very
unstable particle which can not be regarded as a narrow resonance of
baronium that responds to the enhancement.

If the two clusters are connected via the color triplet bonds, the
creation of a $\rm{q}\overline{\rm{q}}$ pair permits the system to
break up into two separate states which are color singlet. So the
lifetime of the state is comparable to that of the excited
$\rm{q}\overline{\rm{q}}$ meson states or excited qqq baryon states,
and the decay width is larger than 100 MeV, since the ordinary meson
or baryon states are also bound together by color triplet
bond\cite{petcolor,jaffe1}. In Jaffe's language\cite{jaffe1} the
process is superallowed if it is heavy enough. Therefore, for our
purpose, we only consider the configurations that the two clusters
are separately in higher dimensional SU(3)$_{c}$ representation,
such as the
$\mathbf{15}_{c}-\overline{\mathbf{15}}_c$,$\mathbf{10}_{c}-\overline{\mathbf{10}}_c$,
$\mathbf{8}_{c}-\mathbf{8}_c$,
$\mathbf{6}_{c}-\overline{\mathbf{6}}_c$ configurations.

\subsection{The $(\rm{q}^3)-(\overline{\rm{q}}^3)$ Configuration}
The configuration is schematically shown in Fig.4, from
Eq.(\ref{2.11}) we see that the allowed states of the $\rm{q}^{3}$
cluster are
\begin{equation}
\label{r2.1}|\mathbf{70}_{cs},\mathbf{8}_f\rangle,~~|\mathbf{50}_{cs},\mathbf{1}_f\rangle,~~|\mathbf{20}_{cs},\mathbf{10}_f\rangle
\end{equation}
The $\rm{SU(6)}_{cs}$ irreducible representation $\mathbf{70}_{cs}$,
$\mathbf{56}_{cs}$, $\mathbf{20}_{cs}$ are decomposed as follows
under $\rm{SU(3)}_{c}\otimes \rm{SU(2)}_{s}$
\begin{eqnarray}
\nonumber
\mathbf{70}_{cs}&=&(\mathbf{10}_c,\mathbf{2})+(\mathbf{8}_c,\mathbf{4})
+(\mathbf{8}_c,\mathbf{2})+(\mathbf{1}_c,\mathbf{2}),\\
\nonumber \mathbf{56}_{cs}&=&(\mathbf{10}_c,\mathbf{4})+(\mathbf{8}_c,\mathbf{2}),\\
\label{2.2}
\mathbf{20}_{cs}&=&(\mathbf{8}_c,\mathbf{2})+(\mathbf{1}_c,\mathbf{4}),
\end{eqnarray}
where the first and the second numbers in the bracket are
respectively the irreducible representations of $\rm{SU(3)}_{c}$ and
$\rm{SU(2)}_{s}$. By using the conventional notation $ |
D_{cs},D_c,2S+1;D_f \rangle$ to express the states, the possible
states of the $\rm{q}^3$ cluster are
\begin{eqnarray}
\nonumber
&&|\mathbf{70}_{cs},\mathbf{10}_c,\mathbf{2};\mathbf{8}_{f}\rangle,\;\;\;
|\mathbf{70}_{cs},\mathbf{8}_c,\mathbf{4};\mathbf{8}_{f}\rangle,
\;\;\;
|\mathbf{70}_{cs},\mathbf{8}_c,\mathbf{2};\mathbf{8}_{f}\rangle,
\;\;\;
|\mathbf{70}_{cs},\mathbf{1}_c,\mathbf{2};\mathbf{8}_{f}\rangle,
\;\;\;
|\mathbf{56}_{cs},\mathbf{10}_c,\mathbf{4};\mathbf{1}_{f}\rangle,\;\;\;\\
\label{2.3}&&
|\mathbf{56}_{cs},\mathbf{8}_c,\mathbf{2};\mathbf{1}_{f}\rangle,
\;\;\;
|\mathbf{20}_{cs},\mathbf{8}_c,\mathbf{2};\mathbf{10}_{f}\rangle,
\;\;\;
|\mathbf{20}_{cs},\mathbf{1}_c,\mathbf{4};\mathbf{10}_{f}\rangle.
\;\;\;
\end{eqnarray}
Noting the symmetry group of both color and flavor are SU(3), we use
a subscript c and f to distinguish them.
 Similarly the possible states of $\overline{\rm{q}}^3$ cluster are
\begin{eqnarray}
\nonumber
&&|\overline{\mathbf{70}}_{cs},\overline{\mathbf{10}}_c,\mathbf{2};\mathbf{8}_{f}\rangle,\;\;\;
|\overline{\mathbf{70}}_{cs},\mathbf{8}_c,\mathbf{4};\mathbf{8}_{f}\rangle,\;\;\;
|\overline{\mathbf{70}}_{cs},\mathbf{8}_c,\mathbf{2};\mathbf{8}_{f}\rangle,\;\;\;
|\overline{\mathbf{70}}_{cs},\mathbf{1}_c,\mathbf{2};\mathbf{8}_{f}\rangle,\;\;\;
|\overline{\mathbf{56}}_{cs},\overline{\mathbf{10}}_c,\mathbf{4};\mathbf{1}_{f}\rangle,\;\;\;\\
\label{2.4}&&
|\overline{\mathbf{56}}_{cs},\mathbf{8}_c,\mathbf{2};\mathbf{1}_{f}\rangle,\;\;\;
|\overline{\mathbf{20}}_{cs},\mathbf{8}_c,\mathbf{2};\overline{\mathbf{10}}_{f}\rangle,\;\;\;
|\overline{\mathbf{20}}_{cs},\mathbf{1}_c,\mathbf{4};\overline{\mathbf{10}}_{f}\rangle\;\;\;
\end{eqnarray}
where $\mathbf{20}_{cs}=\overline{\mathbf{20}}_{cs}$.

The above states are eigenstates of the colormagnetic interaction
hamiltonian $H^{\prime}$, and we can calculate the corresponding
eigenvalues, the results are listed in Table II.
\begin{table}[hptb]
\begin{center}
\begin{tabular}{|c|c|}\hline\hline
States & Eigenvalue of the Hamiltonian $H^{\prime}$\\\hline
$|\mathbf{70}_{cs},\mathbf{10}_c,\mathbf{2};\mathbf{8}_{f}\rangle
\;\; (\;\rm{or}
|\overline{\mathbf{70}}_{cs},\overline{\mathbf{10}}_c,\mathbf{2};\mathbf{8}_{f}\rangle
\;)\;\;$& 4C\\\hline
$|\mathbf{70}_{cs},\mathbf{8}_c,\mathbf{4};\mathbf{8}_{f}\rangle\;\;(\;\rm{or}
|\overline{\mathbf{70}}_{cs},\mathbf{8}_c,\mathbf{4};\mathbf{8}_{f}\rangle
\;)\;\;$& 2C\\\hline
$|\mathbf{70}_{cs},\mathbf{8}_c,\mathbf{2};\mathbf{8}_{f}\rangle\;\;(\;\rm{or}
|\overline{\mathbf{70}}_{cs},\mathbf{8}_c,\mathbf{2};\mathbf{8}_{f}\rangle
\;)\;\;$& -2C\\\hline
$|\mathbf{70}_{cs},\mathbf{1}_c,\mathbf{2};\mathbf{8}_{f}\rangle\;\;(\;\rm{or}
|\overline{\mathbf{70}}_{cs},\mathbf{1}_c,\mathbf{2};\mathbf{8}_{f}\rangle
\;)\;\;$& -8C\\\hline
$|\mathbf{56}_{cs},\mathbf{10}_c,\mathbf{4};\mathbf{1}_{f}\rangle\;\;(\;\rm{or}
|\overline{\mathbf{56}}_{cs},\overline{\mathbf{10}}_c,\mathbf{4};\mathbf{1}_{f}\rangle
\;)\;\;$& -4C\\\hline
$|\mathbf{56}_{cs},\mathbf{8}_c,\mathbf{2};\mathbf{1}_{f}\rangle\;\;(\;\rm{or}
|\overline{\mathbf{56}}_{cs},\mathbf{8}_c,\mathbf{2};\mathbf{1}_{f}\rangle
\;)\;\;$& -14C\\\hline
$|\mathbf{20}_{cs},\mathbf{8}_c,\mathbf{2};\mathbf{10}_{f}\rangle\;\;(\;\rm{or}
|\overline{\mathbf{20}}_{cs},\mathbf{8}_c,\mathbf{2};\overline{\mathbf{10}}_{f}\rangle
\;)\;\;$& 10C\\\hline
$|\mathbf{20}_{cs},\mathbf{1}_c,\mathbf{4};\mathbf{10}_{f}\rangle\;\;(\;\rm{or}
|\overline{\mathbf{20}}_{cs},\mathbf{1}_c,\mathbf{4};\overline{\mathbf{10}}_{f}\rangle
\;)\;\;$& 8C\\\hline\hline
\end{tabular}
\caption{The possible states of the cluster $\rm{q}^{3}$ and the
corresponding eigenvalues of the colormagnetic interaction hamiliton
$H^{\prime}$ } \label{t1}
\end{center}
\end{table}

From Eq.(\ref{2.3}) we can learn that the color singlet states
$|\mathbf{70}_{cs},\mathbf{1}_c,\mathbf{2};\mathbf{8}_{f}\rangle$
and
$|\mathbf{20}_{cs},\mathbf{1}_c,\mathbf{4};\mathbf{10}_{f}\rangle$
are the usual baryon octet and baryon decuplet respectively, and
from Table II, their mass difference is $16C$,
The states of the total system are the product of the subsystem
$\rm{q}^{3}$'s states Eq.(\ref{2.3}) and the subsystem
$\overline{\rm{q}}^3$'s states Eq.(\ref{2.4}). It is obvious that
the system can be in the states
\begin{eqnarray}
\nonumber
&&|\mathbf{70}_{cs},\mathbf{1}_c,\mathbf{2};\mathbf{8}_{f}\rangle|\overline{\mathbf{70}}_{cs},\mathbf{1}_c,\mathbf{2};\mathbf{8}_{f}\rangle,\;\;\;
|\mathbf{70}_{cs},\mathbf{1}_c,\mathbf{2};\mathbf{8}_{f}\rangle|\overline{\mathbf{20}}_{cs},\mathbf{1}_c,\mathbf{4};\overline{\mathbf{10}}_{f}\rangle\\
\label{2.6}&&
|\mathbf{20}_{cs},\mathbf{1}_c,\mathbf{4};\mathbf{10}_{f}\rangle|\overline{\mathbf{70}}_{cs},\mathbf{1}_c,\mathbf{2};\mathbf{8}_{f}\rangle,\;\;\;
|\mathbf{20}_{cs},\mathbf{1}_c,\mathbf{4};\mathbf{10}_{f}\rangle|\overline{\mathbf{20}}_{cs},\mathbf{1}_c,\mathbf{4};\overline{\mathbf{10}}_{f}\rangle
\end{eqnarray}
Their colormagnetic energy is $( -16C, 0, 0, 16C )$. In the above
four states the two clusters are connected by color singlet bond.
They easily dissociate into a baryon and an antibaryon just as we
mentioned in the above, so we do not discuss these states further.
We are interested in the states in which the two clusters are
connected by color $\mathbf{10}$-plet bond or color
$\mathbf{8}$-plet bond. It is easy to write out the possible states
of this system, to reveal their flavor structure and the spin
structure, and to calculate colormagnetic interaction energy. The
results are listed in the Table III. We address that the clusters in
these states are with color, and can not be observed individually,
but the combines of them may be colorless and emerge as baryoniums.
Using the results in the Table III, the discussions on enhancement
are in order:

\begin{table}[hptb]
\begin{center}
\begin{tabular}{|c|c|c|c|}\hline\hline
States of the Total System & SU$_{f}$(3)  Structure & the Spin
Structure& the Mass Splitting \\\hline
$|\mathbf{70}_{cs},\mathbf{10}_c,\mathbf{2};\mathbf{8}_{f}\rangle|\overline{\mathbf{70}}_{cs},\overline{\mathbf{10}}_c,\mathbf{2};\mathbf{8}_{f}\rangle$&$\mathbf{1}_f\oplus
\mathbf{8}_f\oplus\mathbf{8}_f\oplus\mathbf{10}_f\oplus\overline{\mathbf{10}}_f\oplus\mathbf{27}_f$&$\mathbf{0},\mathbf{1}$&8C\\\hline
$|\mathbf{70}_{cs},\mathbf{10}_c,\mathbf{2};\mathbf{8}_{f}\rangle|\overline{\mathbf{56}}_{cs},\overline{\mathbf{10}}_c,\mathbf{4};\mathbf{1}_{f}\rangle$&$\mathbf{8}_f$&$\mathbf{1},\mathbf{2}$&0\\\hline
$|\mathbf{70}_{cs},\mathbf{8}_c,\mathbf{4};\mathbf{8}_{f}\rangle|\overline{\mathbf{70}}_{cs},\mathbf{8}_c,\mathbf{4};\mathbf{8}_{f}\rangle$&$\mathbf{1}_f\oplus
\mathbf{8}_f\oplus\mathbf{8}_f\oplus\mathbf{10}_f\oplus\overline{\mathbf{10}}_f\oplus\mathbf{27}_f$&$\mathbf{0},\mathbf{1},\mathbf{2},\mathbf{3}$&4C\\\hline
$|\mathbf{70}_{cs},\mathbf{8}_c,\mathbf{4};\mathbf{8}_{f}\rangle|\overline{\mathbf{70}}_{cs},\mathbf{8}_c,\mathbf{2};\mathbf{8}_{f}\rangle$&$\mathbf{1}_f\oplus
\mathbf{8}_f\oplus\mathbf{8}_f\oplus\mathbf{10}_f\oplus\overline{\mathbf{10}}_f\oplus\mathbf{27}_f$&$\mathbf{1},\mathbf{2}$&0\\\hline
$|\mathbf{70}_{cs},\mathbf{8}_c,\mathbf{4};\mathbf{8}_{f}\rangle|\overline{\mathbf{56}}_{cs},\mathbf{8}_c,\mathbf{2};\mathbf{1}_{f}\rangle$&$\mathbf{8}_f$&$\mathbf{1},\mathbf{2}$&-12C\\\hline
$|\mathbf{70}_{cs},\mathbf{8}_c,\mathbf{4};\mathbf{8}_{f}\rangle|\overline{\mathbf{20}}_{cs},\mathbf{8}_c,\mathbf{2};\overline{\mathbf{10}}_{f}\rangle$&$\mathbf{8}_f\oplus\overline{\mathbf{10}}_f\oplus\mathbf{27}_f\oplus\overline{\mathbf{35}}_f$&$\mathbf{1},\mathbf{2}$&12C\\\hline
$|\mathbf{70}_{cs},\mathbf{8}_c,\mathbf{2};\mathbf{8}_{f}\rangle|\overline{\mathbf{70}}_{cs},\mathbf{8}_c,\mathbf{4};\mathbf{8}_{f}\rangle$&$\mathbf{1}_f\oplus
\mathbf{8}_f\oplus\mathbf{8}_f\oplus\mathbf{10}_f\oplus\overline{\mathbf{10}}_f\oplus\mathbf{27}_f$&$\mathbf{1},\mathbf{2}$&0\\\hline
$|\mathbf{70}_{cs},\mathbf{8}_c,\mathbf{2};\mathbf{8}_{f}\rangle|\overline{\mathbf{70}}_{cs},\mathbf{8}_c,\mathbf{2};\mathbf{8}_{f}\rangle$&$\mathbf{1}_f\oplus
\mathbf{8}_f\oplus\mathbf{8}_f\oplus\mathbf{10}_f\oplus\overline{\mathbf{10}}_f\oplus\mathbf{27}_f$&$\mathbf{0},\mathbf{1}$&-4C\\\hline
$|\mathbf{70}_{cs},\mathbf{8}_c,\mathbf{2};\mathbf{8}_{f}\rangle|\overline{\mathbf{56}}_{cs},\mathbf{8}_c,\mathbf{2};\mathbf{1}_{f}\rangle$&$\mathbf{8}_f$&$\mathbf{0},\mathbf{1}$&-16C\\\hline
$|\mathbf{70}_{cs},\mathbf{8}_c,\mathbf{2};\mathbf{8}_{f}\rangle|\overline{\mathbf{20}}_{cs},\mathbf{8}_c,\mathbf{2};\overline{\mathbf{10}}_{f}\rangle$&$\mathbf{8}_f\oplus\overline{\mathbf{10}}_f\oplus\mathbf{27}_f\oplus\overline{\mathbf{35}}_f$&$\mathbf{0},\mathbf{1}$&8C\\\hline
$|\mathbf{56}_{cs},\mathbf{10}_c,\mathbf{4};\mathbf{1}_{f}\rangle|\overline{\mathbf{70}}_{cs},\overline{\mathbf{10}}_c,\mathbf{2};\mathbf{8}_{f}\rangle$&$\mathbf{8}_f$&$\mathbf{1},\mathbf{2}$&0\\\hline
$|\mathbf{56}_{cs},\mathbf{10}_c,\mathbf{4};\mathbf{1}_{f}\rangle|\overline{\mathbf{56}}_{cs},\overline{\mathbf{10}}_c,\mathbf{4};\mathbf{1}_{f}\rangle$&$\mathbf{1}_f$&$\mathbf{0},\mathbf{1},\mathbf{2},\mathbf{3}$&-8C\\\hline
$|\mathbf{56}_{cs},\mathbf{8}_c,\mathbf{2};\mathbf{1}_{f}\rangle|\overline{\mathbf{70}}_{cs},\mathbf{8}_c,\mathbf{4};\mathbf{8}_{f}\rangle$&$\mathbf{8}_{f}$&$\mathbf{1},\mathbf{2}$&-12C\\\hline
$|\mathbf{56}_{cs},\mathbf{8}_c,\mathbf{2};\mathbf{1}_{f}\rangle|\overline{\mathbf{70}}_{cs},\mathbf{8}_c,\mathbf{2};\mathbf{8}_{f}\rangle$&$\mathbf{8}_{f}$&$\mathbf{0},\mathbf{1}$&-16C\\\hline
$|\mathbf{56}_{cs},\mathbf{8}_c,\mathbf{2};\mathbf{1}_{f}\rangle|\overline{\mathbf{56}}_{cs},\mathbf{8}_c,\mathbf{2};\mathbf{1}_{f}\rangle$&$\mathbf{1}_{f}$&$\mathbf{0},\mathbf{1}$&-28C\\\hline
$|\mathbf{56}_{cs},\mathbf{8}_c,\mathbf{2};\mathbf{1}_{f}\rangle|\overline{\mathbf{20}}_{cs},\mathbf{8}_c,\mathbf{2};\overline{\mathbf{10}}_{f}\rangle$&$\overline{\mathbf{10}}_{f}$&$\mathbf{0},\mathbf{1}$&-4C\\\hline
$|\mathbf{20}_{cs},\mathbf{8}_c,\mathbf{2};\mathbf{10}_{f}\rangle|\overline{\mathbf{70}}_{cs},\mathbf{8}_c,\mathbf{4};\mathbf{8}_{f}\rangle$&$\mathbf{8}_f\oplus\mathbf{10}_f+\mathbf{27}_f\oplus\mathbf{35}_f$&$\mathbf{1},\mathbf{2}$&12C\\\hline
$|\mathbf{20}_{cs},\mathbf{8}_c,\mathbf{2};\mathbf{10}_{f}\rangle|\overline{\mathbf{70}}_{cs},\mathbf{8}_c,\mathbf{2};\mathbf{8}_{f}\rangle$&$\mathbf{8}_f\oplus\mathbf{10}_f\oplus\mathbf{27}_f\oplus\mathbf{35}_f$&$\mathbf{0},\mathbf{1}$&8C\\\hline
$|\mathbf{20}_{cs},\mathbf{8}_c,\mathbf{2};\mathbf{10}_{f}\rangle|\overline{\mathbf{56}}_{cs},\mathbf{8}_c,\mathbf{2};\mathbf{1}_{f}\rangle$&$\mathbf{10}_f$&$\mathbf{0},\mathbf{1}$&-4C\\\hline
$|\mathbf{20}_{cs},\mathbf{8}_c,\mathbf{2};\mathbf{10}_{f}\rangle|\overline{\mathbf{20}}_{cs},\mathbf{8}_c,\mathbf{2};\overline{\mathbf{10}}_{f}\rangle$&$\mathbf{1}_f\oplus\mathbf{8}_f\oplus\mathbf{27}_f\oplus\mathbf{64}_f$&$\mathbf{0},\mathbf{1}$&20C\\\hline\hline
\end{tabular}
\caption{The allowed states of the total system in the configuration
$(\rm{q}^{3})-(\overline{\rm{q}}^3)$ and the mass splitting.}
\label{t2}
\end{center}
\end{table}

\begin{enumerate}

\item
$\Lambda_1\overline{\Lambda}_1 $enhancement: From the Table III, we
can see that the state
$|\mathbf{56}_{cs},\mathbf{8}_c,\mathbf{2};\mathbf{1}_{f}\rangle|\overline{\mathbf{56}}_{cs},\mathbf{8}_c,\mathbf{2};\mathbf{1}_{f}\rangle$
with flavor $\mathbf{1}_f$ and spin equaling $\mathbf{0}$ or
$\mathbf{1}$ have the largest colormagnetic energy $-28C$, so this
quark configuration is rather stable. Since the two clusters are all
in flavor $\mathbf{1}_f$, they are of flavor $\Lambda_1$ ,
$\overline{\Lambda}_1$ with color octet. It is well known that in
the color singlet ground state multiplet, the SU$_{f}$(3) flavor
singlet $\Lambda_1$ is forbidden by Fermi statistics\cite{PDG}. But
the above composite boson state of $(\Lambda_1\overline{\Lambda}_1)$
can belong to color singlet, and is not forbidden. And its quantum
numbers are same as ones of the state $(\Lambda
\overline{\Lambda})$. This $\Lambda_1\overline{\Lambda}_1$ (or
$\Lambda\overline{\Lambda}$) enhancement has been reported in
Ref.\cite{bell3}.

\item
$p\overline{\Lambda}_1$ enhancement: Both the states
$|\mathbf{70}_{cs},\mathbf{8}_c,\mathbf{2};\mathbf{8}_{f}\rangle|\overline{\mathbf{56}}_{cs},\mathbf{8}_c,\mathbf{2};\mathbf{1}_{f}\rangle$
and
$|\mathbf{56}_{cs},\mathbf{8}_c,\mathbf{2};\mathbf{1}_{f}\rangle|\overline{\mathbf{70}}_{cs},\mathbf{8}_c,\mathbf{2};\mathbf{8}_{f}\rangle$
have colormagnetic energy -16C, they are in the flavor
$\mathbf{8}_{f}$-plet with spin $\mathbf{0}$ or $\mathbf{1}$. The
total system can be in the $p\overline{\Lambda}$ like state because
one of the clusters is in flavor representation $\mathbf{1}_{f}$.
One of these states has also been reported by BES\cite{Bes3}and
Belle\cite{bell2}, but its spin can not be fixed.

\item
$p\overline{p}$ enhancement: It is obvious that the $p\overline{p}$
enhancement\cite{Bes1,Bes2} possibly belongs to the states
$|\mathbf{70}_{cs},\mathbf{10}_c,\mathbf{2};\mathbf{8}_{f}\rangle|\overline{\mathbf{70}}_{cs},\overline{\mathbf{10}}_c,\mathbf{2};\mathbf{8}_{f}\rangle$,
$|\mathbf{70}_{cs},\mathbf{8}_c,\mathbf{4};\mathbf{8}_{f}\rangle|\overline{\mathbf{70}}_{cs},\mathbf{8}_c,\mathbf{4};\mathbf{8}_{f}\rangle$,
$|\mathbf{70}_{cs},\mathbf{8}_c,\mathbf{2};\mathbf{8}_{f}\rangle|\overline{\mathbf{70}}_{cs},\mathbf{8}_c,\mathbf{2};\mathbf{8}_{f}\rangle$,
$|\mathbf{70}_{cs},\mathbf{8}_c,\mathbf{2};\mathbf{8}_{f}\rangle|\overline{\mathbf{20}}_{cs},
\mathbf{8}_c,\mathbf{2};\overline{\mathbf{10}}_{f}\rangle$,
$|\mathbf{20}_{cs},\mathbf{8}_c,\mathbf{2};\mathbf{10}_{f}\rangle|\overline{\mathbf{70}}_{cs},
\mathbf{8}_c,\mathbf{2};\mathbf{8}_{f}\rangle$, and the
corresponding colormagnetic energy  are $8C,\; 4C,\;-4C,\; 8C,\; 8C$
respectively. This indicates that among them, the quark
configuration
$|\mathbf{70}_{cs},\mathbf{8}_c,\mathbf{2};\mathbf{8}_{f}
\rangle|\overline{\mathbf{70}}_{cs},\mathbf{8}_c,\mathbf{2};\mathbf{8}_{f}\rangle$
is more stable. Namely, in the $(\rm{q}^{3})-(\overline{\rm{q}}^3$)
configuration, the the $p\overline{p}$ enhancement most possibly
belongs to the states
$|\mathbf{70}_{cs},\mathbf{8}_c,\mathbf{2};\mathbf{8}_{f}\rangle|\overline{\mathbf{70}}_{cs},
\mathbf{8}_c,\mathbf{2};\mathbf{8}_{f}\rangle$.

\end{enumerate}
Noting that both spin-$\mathbf{0}$, spin-$\mathbf{1}$
$p\overline{p}$ enhancements and spin-$\mathbf{0}$,
spin-$\mathbf{1}$ $p\overline{\Lambda}$ enhancements are predicted.
\subsection{The ($\rm{q}^{2}\overline{\rm{q}})-(\rm{q}\overline{\rm{q}}^2$) Configuration}
The configuration is schematically shown in Fig.5. Due to Fermi
statistics, the two quarks in the cluster  belong to the
$(\mathbf{6}_{cs}\otimes\mathbf{6}_{cs})_{S}=\mathbf{21}_{cs}$
dimensional representation of $\rm{SU(6)}_{cs}$ and irreducible
$\rm{SU(3)}_f$ representation $\overline{\mathbf{3}}_f$, or to
$(\mathbf{6}_{cs}\otimes\mathbf{6}_{cs})_{A}=\mathbf{15}_{cs}$
dimensional representation of $\rm{SU(6)}_{cs}$ and $\rm{SU(3)}_f$
representation $\mathbf{6}_f$. Performing the outer product of
$\rm{SU(6)}_{cs}$ representations
\begin{equation}
\label{2.7}\mathbf{21}_{cs}\otimes\overline{\mathbf{6}}_{cs}=\mathbf{6}_{cs}\oplus\mathbf{120}_{cs},\;\;\;
\mathbf{15}_{cs}\otimes\overline{\mathbf{6}}_{cs}=\mathbf{6}_{cs}\oplus\overline{\mathbf{84}}_{cs},
\end{equation}
we find out that the cluster $\rm{q}^2\overline{\rm{q}}$ can be in
the $\rm{SU(6)}_{cs}$ representation $\mathbf{6}_{cs}$,
$\overline{\mathbf{84}}_{cs}$ or $\mathbf{120}_{cs}$, and the
corresponding $\rm{SU(3)}_c\otimes \rm{SU(2)}_s$ decompositions are
as followings
\begin{eqnarray}
\nonumber &&
\mathbf{6}_{cs}=(\mathbf{3}_c,\mathbf{2}),\;\;\mathbf{15}_{cs}=(\mathbf{6}_c,\mathbf{1})+(\overline{\mathbf{3}}_c,\mathbf{3}),\;\;\mathbf{21}_{cs}=(\mathbf{6}_{c},\mathbf{3})+(\overline{\mathbf{3}}_c,\mathbf{1})\\
\nonumber&&\overline{\mathbf{84}}_{cs}=(\mathbf{3}_{c},\mathbf{2})+(\mathbf{3}_{c},\mathbf{4})+(\overline{\mathbf{6}}_{c},\mathbf{2})+(\overline{\mathbf{6}}_{c},\mathbf{4})+(\mathbf{15}_{c},\mathbf{2})\\
\label{2.8}&&\mathbf{120}_{cs}=(\mathbf{3}_{c},\mathbf{2})+(\mathbf{3}_{c},\mathbf{4})
+(\mathbf{15}_{c},\mathbf{2})+(\mathbf{15}_{c},\mathbf{4})+(\overline{\mathbf{6}}_{c},\mathbf{2}).
\end{eqnarray}
And the cluster $\rm{q}^2\overline{\rm{q}}$ can be in the flavor
representation
$\overline{\mathbf{3}}_{f}\otimes\overline{\mathbf{3}}_f=\mathbf{3}_f\oplus\overline{\mathbf{6}}_f$
or in
$\mathbf{6}_f\otimes\overline{\mathbf{3}}_f=\mathbf{3}_f\oplus\mathbf{15}_f$.
We can easily see the possible states of the cluster
$\rm{q}^2\overline{\rm{q}}$, which are listed in the followings.
\begin{eqnarray}
\nonumber
&&|\mathbf{120}_{cs},\mathbf{3}_c,\mathbf{2},\mathbf{3}_f\oplus\overline{\mathbf{6}}_f\rangle,\;\;|\mathbf{120}_{cs},\mathbf{3}_c,\mathbf{4},\mathbf{3}_f\oplus\overline{\mathbf{6}}_f\rangle,\;\;
|\mathbf{120}_{cs},\mathbf{15}_c,\mathbf{2},\mathbf{3}_f\oplus\overline{\mathbf{6}}_f\rangle,\\
\nonumber&&|\mathbf{120}_{cs},\mathbf{15}_c,\mathbf{4},\mathbf{3}_f\oplus\overline{\mathbf{6}}_f\rangle,\;\;|\mathbf{120}_{cs},\overline{\mathbf{6}}_c,\mathbf{2},\mathbf{3}_f\oplus\overline{\mathbf{6}}_f\rangle,\;\;
|\mathbf{6}_{cs},\mathbf{3}_c,\mathbf{2},\mathbf{3}_f\oplus\overline{\mathbf{6}}_f\rangle,\\
\nonumber&&|\overline{\mathbf{84}}_{cs},\mathbf{3}_c,\mathbf{2},\mathbf{3}_f\oplus\mathbf{15}_f\rangle,\;\;
|\overline{\mathbf{84}}_{cs},\mathbf{3}_c,\mathbf{4},\mathbf{3}_f\oplus\mathbf{15}_f\rangle,\;\;
|\overline{\mathbf{84}}_{cs},\overline{\mathbf{6}}_c,\mathbf{2},\mathbf{3}_f\oplus\mathbf{15}_f\rangle\\
\label{2.9}&&|\overline{\mathbf{84}}_{cs},\overline{\mathbf{6}}_c,\mathbf{4},\mathbf{3}_f\oplus\mathbf{15}_f\rangle,\;\;
|\overline{\mathbf{84}}_{cs},\mathbf{15}_c,\mathbf{2},\mathbf{3}_f\oplus\mathbf{15}_f\rangle,\;\;|\mathbf{6}_{cs},\mathbf{3}_c,\mathbf{2},\mathbf{3}_f\oplus\mathbf{15}_f\rangle
.
\end{eqnarray}
Similarly the possible states of another cluster
$\rm{q}\overline{\rm{q}}^2$ are
\begin{eqnarray}
\nonumber
&&|\overline{\mathbf{120}}_{cs},\overline{\mathbf{3}}_c,\mathbf{2},\overline{\mathbf{3}}_f\oplus\mathbf{6}_f\rangle,\;\;|\overline{\mathbf{120}}_{cs},\overline{\mathbf{3}}_c,\mathbf{4},\overline{\mathbf{3}}_f\oplus\mathbf{6}_f\rangle,\;\;
|\overline{\mathbf{120}}_{cs},\overline{\mathbf{15}}_c,\mathbf{2},\overline{\mathbf{3}}_f\oplus\mathbf{6}_f\rangle,\\
\nonumber&&|\overline{\mathbf{120}}_{cs},\overline{\mathbf{15}}_c,\mathbf{4},\overline{\mathbf{3}}_f\oplus\mathbf{6}_f\rangle,\;\;|\overline{\mathbf{120}}_{cs},\mathbf{6}_c,\mathbf{2},\overline{\mathbf{3}}_f\oplus\mathbf{6}_f\rangle,\;\;
|\overline{\mathbf{6}}_{cs},\overline{\mathbf{3}}_c,\mathbf{2},\overline{\mathbf{3}}_f\oplus\mathbf{6}_f\rangle,\\
\nonumber&&|\mathbf{84}_{cs},\overline{\mathbf{3}}_c,\mathbf{2},\overline{\mathbf{3}}_f\oplus\overline{\mathbf{15}}_f\rangle,\;\;
|\mathbf{84}_{cs},\overline{\mathbf{3}}_c,\mathbf{4},\overline{\mathbf{3}}_f\oplus\overline{\mathbf{15}}_f\rangle,\;\;
|\mathbf{84}_{cs},\mathbf{6}_c,\mathbf{2},\overline{\mathbf{3}}_f\oplus\overline{\mathbf{15}}_f\rangle\\
\label{2.10}&&|\mathbf{84}_{cs},\mathbf{6}_c,\mathbf{4},\overline{\mathbf{3}}_f\oplus\overline{\mathbf{15}}_f\rangle,\;\;
|\mathbf{84}_{cs},\overline{\mathbf{15}}_c,\mathbf{2},\overline{\mathbf{3}}_f\oplus\overline{\mathbf{15}}_f\rangle,\;\;
|\overline{\mathbf{6}}_{cs},\overline{\mathbf{3}}_c,\mathbf{2},\overline{\mathbf{3}}_f\oplus\overline{\mathbf{15}}_f\rangle
\end{eqnarray}
We now consider the multiplets whose two subsystems are connected by
$\overline{\mathbf{6}}_c$($\mathbf{6}_c$) or
$\mathbf{15}_c$($\overline{\mathbf{15}}_c$) bond, neglecting these
with the clusters being in $\mathbf{3}_c$ or
$\overline{\mathbf{3}}_c$, since color triplet bond is easily
broken. The states in Eq.(\ref{2.9}) and Eq.(\ref{2.10}) are
eigenstates of the hamiltonian $H^{\prime}$, and the eigenvalues can
be  calculated straightforwardly. The results are listed in Table
IV.

\begin{table}[hptb]
\begin{center}
\begin{tabular}{|c|c|}\hline\hline
States& Eigenvalues of the hamiltonian $H^{\prime}$\\\hline
$|\mathbf{120}_{cs},\mathbf{15}_c,\mathbf{2},\mathbf{3}_f\oplus\overline{\mathbf{6}}_f\rangle\;\;
(\;\rm{or}\;|\overline{\mathbf{120}}_{cs},\overline{\mathbf{15}}_c,\mathbf{2},\overline{\mathbf{3}}_f\oplus\mathbf{6}_f\rangle\;)$&
$\frac{4}{3}C$\\\hline
$|\mathbf{120}_{cs},\mathbf{15}_c,\mathbf{4},\mathbf{3}_f\oplus\overline{\mathbf{6}}_f\rangle\;\;
(\;\rm{or}\;|\overline{\mathbf{120}}_{cs},\overline{\mathbf{15}}_c,\mathbf{4},\overline{\mathbf{3}}_f\oplus\mathbf{6}_f\rangle\;)$&
$-\frac{8}{3}C$\\\hline
$|\mathbf{120}_{cs},\overline{\mathbf{6}}_c,\mathbf{2},\mathbf{3}_f\oplus\overline{\mathbf{6}}_f\rangle\;\;(\;
\rm{or}\;|\overline{\mathbf{120}}_{cs},\mathbf{6}_c,\mathbf{2},\overline{\mathbf{3}}_f\oplus\mathbf{6}_f\rangle\;)$&-8C\\\hline
$|\overline{\mathbf{84}}_{cs},\mathbf{15}_c,\mathbf{2},\mathbf{3}_f\oplus\mathbf{15}_f\rangle\;\;(\;
\rm{or}\;|\mathbf{84}_{cs},\overline{\mathbf{15}}_c,\mathbf{2},\overline{\mathbf{3}}_f\oplus\overline{\mathbf{15}}_f\rangle\;)$&4C\\\hline
$|\overline{\mathbf{84}}_{cs},\overline{\mathbf{6}}_c,\mathbf{2},\mathbf{3}_f\oplus\mathbf{15}_f\rangle\;\;(\;\rm{or}
\;|\mathbf{84}_{cs},\mathbf{6}_c,\mathbf{2},\overline{\mathbf{3}}_f\oplus\overline{\mathbf{15}}_f\rangle\;)$&$\frac{16}{3}C$\\\hline
$|\overline{\mathbf{84}}_{cs},\overline{\mathbf{6}}_c,\mathbf{4},\mathbf{3}_f\oplus\mathbf{15}_f\rangle\;\;(\;\rm{or}
\;|\mathbf{84}_{cs},\mathbf{6}_c,\mathbf{4},\overline{\mathbf{3}}_f\oplus\overline{\mathbf{15}}_f\rangle\;)$&$\frac{4}{3}C$\\\hline\hline
\end{tabular}
\caption{The allowed states of the cluster
$\rm{q}^2\overline{\rm{q}}$ (or another cluster
$\rm{q}\overline{\rm{q}}^2$ ) except those states whose color are in
representation $\mathbf{3}_c$ or $\overline{\mathbf{3}}_c$, and the
corresponding eigenvalues of the colormagnetic interaction
hamiltonian $H^{\prime}$. } \label{t3}
\end{center}
\end{table}

The states of the total system are formed by combining the states of
the cluster $\rm{q}^2\overline{\rm{q}}$ in Eq.(\ref{2.9}) and that
of another cluster $\rm{q}\overline{\rm{q}}^2$ in Eq.(\ref{2.10}),
the allowed states of the system and the corresponding eigenvalues
of the hamiltonian $H^{\prime}$ are given in the Table V and Table
VI.
\begin{table}[hptb]
\begin{center}
\caption{The allowed states of the total system in the configuration
$(\rm{q}^2\overline{\rm{q}}$)-($\rm{q}\overline{\rm{q}}^2$) and the
mass splitting.}
\begin{tabular}{|c|l|c|c|}\hline\hline
States of the total system&$~~~~~~~\;\;\rm{SU}_{f}$(3) structure&
the Spin  & the Mass  \\
 & & Structure & Splitting\\\hline
$|\mathbf{120}_{cs},\mathbf{15}_c,\mathbf{2},\mathbf{3}_f\oplus\overline{\mathbf{6}}_f\rangle
|\overline{\mathbf{120}}_{cs},\overline{\mathbf{15}}_c,\mathbf{2},\overline{\mathbf{3}}_f\oplus\mathbf{6}_f\rangle$
&$\mathbf{1}_f(2)\oplus\mathbf{8}_f(4)\oplus\mathbf{10}_f\oplus\overline{\mathbf{10}}_f\oplus\mathbf{27}_f$
&$\mathbf{0},\mathbf{1}$&$\frac{8}{3}C$\\\hline
$|\mathbf{120}_{cs},\mathbf{15}_c,\mathbf{2},\mathbf{3}_f\oplus\overline{\mathbf{6}}_f\rangle
|\overline{\mathbf{120}}_{cs},\overline{\mathbf{15}}_c,\mathbf{4},\overline{\mathbf{3}}_f\oplus\mathbf{6}_f\rangle$
&$\mathbf{1}_f(2)\oplus\mathbf{8}_f(4)\oplus\mathbf{10}_f\oplus\overline{\mathbf{10}}_f\oplus\mathbf{27}_f$
&$\mathbf{1},\mathbf{2}$&$\frac{-4}{3}C$\\\hline
$|\mathbf{120}_{cs},\mathbf{15}_c,\mathbf{2},\mathbf{3}_f\oplus\overline{\mathbf{6}}_f\rangle
|\mathbf{84}_{cs},\overline{\mathbf{15}}_c,\mathbf{2},\overline{\mathbf{3}}_f\oplus\overline{\mathbf{15}}_f\rangle$
&$\mathbf{1}_f\oplus\mathbf{8}_f(4)\oplus\mathbf{10}_f\oplus\overline{\mathbf{10}}_f(3)$
&$\mathbf{0},\mathbf{1}$&$\frac{16}{3}C$\\
 &$\oplus\mathbf{27}_f(2)\oplus\overline{\mathbf{35}}_f$& & \\\hline
$|\mathbf{120}_{cs},\mathbf{15}_c,\mathbf{4},\mathbf{3}_f\oplus\overline{\mathbf{6}}_f\rangle
|\overline{\mathbf{120}}_{cs},\overline{\mathbf{15}}_c,\mathbf{2},\overline{\mathbf{3}}_f\oplus\mathbf{6}_f\rangle$
&$\mathbf{1}_f(2)\oplus\mathbf{8}_f(4)\oplus\mathbf{10}_f\oplus\overline{\mathbf{10}}_f\oplus\mathbf{27}_f$
&$\mathbf{1},\mathbf{2}$&$-\frac{4}{3}C$\\\hline
$|\mathbf{120}_{cs},\mathbf{15}_c,\mathbf{4},\mathbf{3}_f\oplus\overline{\mathbf{6}}_f\rangle
|\overline{\mathbf{120}}_{cs},\overline{\mathbf{15}}_c,\mathbf{4},\overline{\mathbf{3}}_f\oplus\mathbf{6}_f\rangle$
&$\mathbf{1}_f(2)\oplus\mathbf{8}_f(4)\oplus\mathbf{10}_f\oplus\overline{\mathbf{10}}_f\oplus\mathbf{27}_f$
&$\mathbf{0},\mathbf{1},\mathbf{2},\mathbf{3}$&$-\frac{16}{3}C$\\\hline
$|\mathbf{120}_{cs},\mathbf{15}_c,\mathbf{4},\mathbf{3}_f\oplus\overline{\mathbf{6}}_f\rangle
|\mathbf{84}_{cs},\overline{\mathbf{15}}_c,\mathbf{2},\overline{\mathbf{3}}_f\oplus\overline{\mathbf{15}}_f\rangle$
&$\mathbf{1}_f\oplus\mathbf{8}_f(4)\oplus\mathbf{10}_f\oplus\overline{\mathbf{10}}_f(3)$
&$\mathbf{1},\mathbf{2}$&$\frac{4}{3}C$\\
 &$\oplus\mathbf{27}_f(2)\oplus\overline{\mathbf{35}}_f$& & \\\hline
$|\overline{\mathbf{84}}_{cs},\mathbf{15}_c,\mathbf{2},\mathbf{3}_f\oplus\mathbf{15}_f\rangle
|\overline{\mathbf{120}}_{cs},\overline{\mathbf{15}}_c,\mathbf{2},\overline{\mathbf{3}}_f\oplus\mathbf{6}_f\rangle$
&$\mathbf{1}_f\oplus\mathbf{8}_f(4)\oplus\mathbf{10}_f(3)\oplus\overline{\mathbf{10}}_f$
&$\mathbf{0},\mathbf{1}$&$\frac{16}{3}C$\\
 &$\oplus\mathbf{27}_f(2)\oplus\mathbf{35}_f$& & \\\hline
$|\overline{\mathbf{84}}_{cs},\mathbf{15}_c,\mathbf{2},\mathbf{3}_f\oplus\mathbf{15}_f\rangle
|\overline{\mathbf{120}}_{cs},\overline{\mathbf{15}}_c,\mathbf{4},\overline{\mathbf{3}}_f\oplus\mathbf{6}_f\rangle$
&$\mathbf{1}_f\oplus\mathbf{8}_f(4)\oplus\mathbf{10}_f(3)\oplus\overline{\mathbf{10}}_f$
&$\mathbf{1},\mathbf{2}$&$\frac{4}{3}C$\\
 &$\oplus\mathbf{27}_f(2)\oplus\mathbf{35}_f$& & \\\hline
$|\overline{\mathbf{84}}_{cs},\mathbf{15}_c,\mathbf{2},\mathbf{3}_f\oplus\mathbf{15}_f\rangle
|\mathbf{84}_{cs},\overline{\mathbf{15}}_c,\mathbf{2},\overline{\mathbf{3}}_f\oplus\overline{\mathbf{15}}_f\rangle$
&$\mathbf{1}_f(2)\oplus\mathbf{8}_f(5)\oplus\mathbf{10}_f(2)\oplus\overline{\mathbf{10}}(2)$
&$\mathbf{0},\mathbf{1}$&$8C$\\
 &$\oplus\mathbf{27}_f(4)\oplus\mathbf{35}_f\oplus\overline{\mathbf{35}}_f\oplus\mathbf{64}_f$& & \\\hline
$|\mathbf{120}_{cs},\overline{\mathbf{6}}_c,\mathbf{2},\mathbf{3}_f\oplus\overline{\mathbf{6}}_f\rangle
|\overline{\mathbf{120}}_{cs},\mathbf{6}_c,\mathbf{2},\overline{\mathbf{3}}_f\oplus\mathbf{6}_f\rangle$
&$\mathbf{1}_f(2)\oplus\mathbf{8}_f(4)\oplus\mathbf{10}_f\oplus\overline{\mathbf{10}}_f\oplus\mathbf{27}_f$
&$\mathbf{0},\mathbf{1}$&-16C\\\hline
$|\mathbf{120}_{cs},\overline{\mathbf{6}}_c,\mathbf{2},\mathbf{3}_f\oplus\overline{\mathbf{6}}_f\rangle
|\mathbf{84}_{cs},\mathbf{6}_c,\mathbf{2},\overline{\mathbf{3}}_f\oplus\overline{\mathbf{15}}_f\rangle$
&$\mathbf{1}_f\oplus\mathbf{8}_f(4)\oplus\mathbf{10}_f\oplus\overline{\mathbf{10}}_f(3)$
&$\mathbf{0},\mathbf{1}$&$-\frac{8}{3}C$\\
 &$\oplus\mathbf{27}_f(2)\oplus\overline{\mathbf{35}}_f$& & \\\hline
$|\mathbf{120}_{cs},\overline{\mathbf{6}}_c,\mathbf{2},\mathbf{3}_f\oplus\overline{\mathbf{6}}_f\rangle
|\mathbf{84}_{cs},\mathbf{6}_c,\mathbf{4},\overline{\mathbf{3}}_f\oplus\overline{\mathbf{15}}_f\rangle$
&$\mathbf{1}_f\oplus\mathbf{8}_f(4)\oplus\mathbf{10}_f\oplus\overline{\mathbf{10}}_f(3)$
&$\mathbf{1},\mathbf{2}$&$-\frac{20}{3}C$\\
 &$\oplus\mathbf{27}_f(2)\oplus\overline{\mathbf{35}}_f$& & \\\hline
$|\overline{\mathbf{84}}_{cs},\overline{\mathbf{6}}_c,\mathbf{2},\mathbf{3}_f\oplus\mathbf{15}_f\rangle
|\overline{\mathbf{120}}_{cs},\mathbf{6}_c,\mathbf{2},\overline{\mathbf{3}}_f\oplus\mathbf{6}_f\rangle$
&$\mathbf{1}_f\oplus\mathbf{8}_f(4)\oplus\mathbf{10}_f(3)\oplus\overline{\mathbf{10}}_f$
&$\mathbf{0},\mathbf{1}$&$-\frac{8}{3}C$\\
 &$\oplus\mathbf{27}_f(2)\oplus\mathbf{35}_f$& & \\\hline
$|\overline{\mathbf{84}}_{cs},\overline{\mathbf{6}}_c,\mathbf{2},\mathbf{3}_f\oplus\mathbf{15}_f\rangle
|\mathbf{84}_{cs},\mathbf{6}_c,\mathbf{2},\overline{\mathbf{3}}_f\oplus\overline{\mathbf{15}}_f\rangle$
&$\mathbf{1}_f(2)\oplus\mathbf{8}_f(5)\oplus\mathbf{10}_f(2)\oplus\overline{\mathbf{10}}(2)$
&$\mathbf{0},\mathbf{1}$&$\frac{32}{3}C$\\
 &$\oplus\mathbf{27}_f(4)\oplus\mathbf{35}_f\oplus\overline{\mathbf{35}}_f\oplus\mathbf{64}_f$& & \\\hline
$|\overline{\mathbf{84}}_{cs},\overline{\mathbf{6}}_c,\mathbf{2},\mathbf{3}_f\oplus\mathbf{15}_f\rangle
|\mathbf{84}_{cs},\mathbf{6}_c,\mathbf{4},\overline{\mathbf{3}}_f\oplus\overline{\mathbf{15}}_f\rangle$
&$\mathbf{1}_f(2)\oplus\mathbf{8}_f(5)\oplus\mathbf{10}_f(2)\oplus\overline{\mathbf{10}}(2)$
&$\mathbf{1},\mathbf{2}$&$\frac{20}{3}C$\\
 &$\oplus\mathbf{27}_f(4)\oplus\mathbf{35}_f\oplus\overline{\mathbf{35}}_f\oplus\mathbf{64}_f$& &
 \\\hline\hline
\end{tabular}
\label{t4}
\end{center}
\end{table}
\begin{table}
\caption{The Continuing of Table V.}
\begin{center}
\begin{tabular}{|c|l|c|c|}\hline\hline
States of the total system&$~~~~~~~\;\;\rm{SU}_{f}$(3) structure&
the Spin  & the Mass  \\
 & & Structure & Splitting\\\hline
$|\overline{\mathbf{84}}_{cs},\overline{\mathbf{6}}_c,\mathbf{4},\mathbf{3}_f\oplus\mathbf{15}_f\rangle
|\overline{\mathbf{120}}_{cs},\mathbf{6}_c,\mathbf{2},\overline{\mathbf{3}}_f\oplus\mathbf{6}_f\rangle$
&$\mathbf{1}_f\oplus\mathbf{8}_f(4)\oplus\mathbf{10}_f(3)\oplus\overline{\mathbf{10}}_f$
&$\mathbf{1},\mathbf{2}$&$-\frac{20}{3}C$\\
 &$\oplus\mathbf{27}_f(2)\oplus\overline{\mathbf{35}}_f$& & \\\hline
$|\overline{\mathbf{84}}_{cs},\overline{\mathbf{6}}_c,\mathbf{4},\mathbf{3}_f\oplus\mathbf{15}_f\rangle
|\mathbf{84}_{cs},\mathbf{6}_c,\mathbf{2},\overline{\mathbf{3}}_f\oplus\overline{\mathbf{15}}_f\rangle$
&$\mathbf{1}_f(2)\oplus\mathbf{8}_f(5)\oplus\mathbf{10}_f(2)\oplus\overline{\mathbf{10}}(2)$
&$\mathbf{1},\mathbf{2}$&$\frac{20}{3}C$\\
 &$\oplus\mathbf{27}_f(4)\oplus\mathbf{35}_f\oplus\overline{\mathbf{35}}_f\oplus\mathbf{64}_f$& & \\\hline
$|\overline{\mathbf{84}}_{cs},\overline{\mathbf{6}}_c,\mathbf{4},\mathbf{3}_f\oplus\mathbf{15}_f\rangle
|\mathbf{84}_{cs},\mathbf{6}_c,\mathbf{4},\overline{\mathbf{3}}_f\oplus\overline{\mathbf{15}}_f\rangle$
&$\mathbf{1}_f(2)\oplus\mathbf{8}_f(5)\oplus\mathbf{10}_f(2)\oplus\overline{\mathbf{10}}(2)$
&$\mathbf{0},\mathbf{1},\mathbf{2},\mathbf{3}$&$\frac{8}{3}C$\\
 &$\oplus\mathbf{27}_f(4)\oplus\mathbf{35}_f\oplus\overline{\mathbf{35}}_f\oplus\mathbf{64}_f$& &
 \\\hline\hline
\end{tabular}
\end{center}
\end{table}
By means of Table V and Table VI, we discuss the enhancements in the
$(\rm{q}^2\overline{\rm{q}})-(\rm{q}\overline{\rm{q}}^2)$ cluster
configuration in order:

\begin{enumerate}

\item $\Lambda\overline{\Lambda}$,
$p\overline{\Lambda}$ and $p\overline{p}$ enhancements with spin
$\mathbf{0}$ or $\mathbf{1}$: From the Table V and Table VI, we can
see the multiplet state
$|\mathbf{120}_{cs},\overline{\mathbf{6}}_c,\mathbf{2},\mathbf{3}_f\oplus\overline{\mathbf{6}}_f\rangle
|\overline{\mathbf{120}}_{cs},\mathbf{6}_c,\mathbf{2},\overline{\mathbf{3}}_f\oplus\mathbf{6}_f\rangle$
has the largest colormagnetic energy $-16C$, and its flavors are in
$\mathbf{1}(2)_f\oplus\mathbf{8}_f(4)\oplus\mathbf{10}_f
\oplus\overline{\mathbf{10}}_f\oplus\mathbf{27}_f$. The spin of
these states equals 0 or 1, and with the baryon number being zero.
Therefore the multiplet state
$|\mathbf{120}_{cs},\overline{\mathbf{6}}_c,\mathbf{2},\mathbf{3}_f\oplus\overline{\mathbf{6}}_f\rangle
|\overline{\mathbf{120}}_{cs},\mathbf{6}_c,\mathbf{2},\overline{\mathbf{3}}_f\oplus\mathbf{6}_f\rangle$
can decay to (or couple to) baryon octet plus antibaryon octet.
Namely, $(\Lambda\overline{\Lambda}$, $p\overline{\Lambda}$ and
$p\overline{p}$ enhancements)$\in$
$|\mathbf{120}_{cs},\overline{\mathbf{6}}_c,\mathbf{2},\mathbf{3}_f\oplus\overline{\mathbf{6}}_f\rangle
|\overline{\mathbf{120}}_{cs},\mathbf{6}_c,\mathbf{2},\overline{\mathbf{3}}_f\oplus\mathbf{6}_f\rangle$.
These baryonium states may mix with the states in the
$(q^3)-(\overline{q}^3)$ cluster configurations illustrated in the
previous subsection. Namely, this mixing can be expressed as follows
\begin{equation}\label{mix}
|i\rangle =\alpha_i |(q^3)-(\overline{q}^3)\rangle_i
+\beta_i|(q^2\overline{q})-(q\overline{q}^2)\rangle_i,
\end{equation} where
$i=\{\Lambda\overline{\Lambda},\;p\overline{\Lambda}, \;
p\overline{p}, \;\cdots \}$, $\alpha_i$ and $\beta_i$ are constants,
$|i\rangle$ is $i^{\rm th}$ physical baryonium state, $
|(q^3)-(\overline{q}^3)\rangle_i$ is $i^{\rm th}$ baryonium state in
$(q^3)-(\overline{q}^3)$ cluster configurations, and $
|(q^2\overline{q})-(q\overline{q}^2)\rangle_i$ is $i^{\rm th}$
baryonium state in $(q^2\overline{q})-(q\overline{q}^2)$ cluster
configurations. We would like to argue that the physical baryonium
states may favor the cluster structure with larger mass defect over
that with smaller defect. Therefore, comparing the results of this
subsection with ones of the previous subsection, we have
\begin{eqnarray}\label{mixLL}
{\rm to}~ \Lambda\overline{\Lambda}\;\;{\rm
enhancement:}~~|\alpha_{i=\Lambda\overline{\Lambda}}| &>&
|\beta_{i=\Lambda\overline{\Lambda}}|, \\
\label{mixpL} {\rm to}~~ p\overline{\Lambda}\;\;{\rm
enhancement:}~~|\alpha_{i=p\overline{\Lambda}}| &\simeq&
|\beta_{i=p\overline{\Lambda}}|,\\
\label{mixpp} {\rm to}~~~ p\overline{p}\;\;{\rm
enhancement:}~~|\alpha_{i=p\overline{p}}| &<&
|\beta_{i=p\overline{p}}|.
 \end{eqnarray}
In some recent works in literature\cite{Datta,yan,ding}, the
$p\overline{p}$ enhancement has been treated as a bound state of two
clusters.  Actually, under the approximation of neglecting van der
Waal's force between two color singlet baryons, the clusters as
elements of the enhancement are objects with color. The potentials
used in \cite{Datta}\cite{yan}\cite{ding} should be due to  "color
chemistry bonds" that are caused by the unfrozen color degree of
freedom in the system in the framework of this paper. The
Eq.(\ref{mixpp}) indicates that the color, spin, flavor structures
of $p \overline{p}$ enhancement are mainly of
$|\bf{120}_{cs},\overline{\bf{6}}_c,\bf{2},\bf{3}_f\oplus\overline{\bf{6}}_f\rangle
|\overline{\bf{120}}_{cs},\bf{6}_c,\bf{2},\overline{\bf{3}}_f\oplus\bf{6}_f\rangle$
in  the configuration
$(\rm{q}^2\overline{\rm{q}})-(\rm{q}\overline{\rm{q}}^2)$, and mix
with few $|\bf{70}_{cs},\bf{8}_c,\bf{2},\bf{8}_f\rangle
|\overline{\bf{70}}_{cs},\bf{8}_c,\bf{2},\bf{8}_f\rangle$ in the
configuration $(\rm{q}^3)-(\overline{\rm{q}}^3)$.

~~Similarly, Eq.(\ref{mixLL}) indicates the color-spin-flavor
structures of $\Lambda\overline{\Lambda}$ enhancement are mainly
$|\mathbf{56}_{cs},\mathbf{8}_c,\mathbf{2};\mathbf{1}_f\rangle
|\overline{\mathbf{56}}_{cs},\mathbf{8}_c,\mathbf{2};\mathbf{1}_f\rangle$
in the configuration $(\rm{q}^3)-(\overline{\rm{q}}^3)$, and plus
few mixtures of the corresponding $\Lambda\overline{\Lambda}$
enhancement state in  the configuration
$(\rm{q}^2\overline{\rm{q}})-(\rm{q}\overline{\rm{q}}^2)$. Moreover,
by Eq.(\ref{mixpL}), the $p\overline{\Lambda}$ enhancement should be
half of the states in the $(\rm{q}^3)-(\overline{\rm{q}}^3)$
configuration plus half of that in the
$(\rm{q}^2\overline{\rm{q}})-(\rm{q}\overline{\rm{q}}^2)$
approximately.

\item The enhancements with
spin -$\mathbf{1}$ or -$\mathbf{2}$: The states
$|\mathbf{120}_{cs},\overline{\mathbf{6}}_c,\mathbf{2},\mathbf{3}_f\oplus\overline{\mathbf{6}}_f\rangle
|\mathbf{84}_{cs},\mathbf{6}_c,\mathbf{4},\overline{\mathbf{3}}_f\oplus\overline{\mathbf{15}}_f\rangle$
or
$|\overline{\mathbf{84}}_{cs},\overline{\mathbf{6}}_c,\mathbf{4},\mathbf{3}_f\oplus\mathbf{15}_f\rangle
|\overline{\mathbf{120}}_{cs},\mathbf{6}_c,\mathbf{2},\overline{\mathbf{3}}_f\oplus\mathbf{6}_f\rangle$
has colormagnetic energy $ -\frac{20}{3} C$, but their spins are
$\mathbf{1}$ or $\mathbf{2}$. We address that the spins of the
baryoniums come from the quark's intrinsic spin in this quark model
with $S$-wave, instead of coming from their relative  orbital
angular momentum. These color singlet states are allowed.


~~On all accounts, both the analysis from common quark model and
from the model with quark correlation imply that the $p\overline{p}$
enhancement, $p\overline{\Lambda}$ enhancement and the
$\Lambda\overline{\Lambda}$ enhancement are relatively easy to be
found and these enhancements which have been seen by BES and Bell
are not experiment artifacts. We predict that spin-$\mathbf{1}$
$\Lambda\overline{\Lambda}$ enhancement may be observed in future
and we conjecture that BES has observed both spin-$\mathbf{0}$,
spin-$\mathbf{1}$ $p\overline{p}$ enhancement and spin-$\mathbf{0}$,
spin-$\mathbf{1}$ $p\overline{\Lambda}$ enhancement in section III,
these predictions are compatible with the above results obtained in
subsection IV.A and subsection IV.B. These results should be further
checked by experiment.
\end{enumerate}

\section{conclusion and discussion}
In summary, we have studied the mass spectrum of the
$\rm{q}^3\overline{\rm{q}}^3$ mesons and the baryon antibaryon
enhancements in the quark model with the interaction between quarks
being the well-known colormagnetic interaction. First of all, We
study ${\rm q}^3\overline{\rm q}^3$ system using common quark model
without quark correlation with colormagnetic hyperfine interactions
between quarks. By this model, we firstly found that the
configurations corresponding to the
$\Lambda\overline{\Lambda},\;p\overline{\Lambda},\;p\overline{p},
\;\cdots$ enhancement can have rather large negative  colormagnetic
energy. This indicates that these states should be ones with rather
low energy, and hence be visible in experiments. These predictions
are consistent with the experiments reported in literatures.
Secondly, we found further that among those enhancements, the
binding colormagnetic energy of spin-$\mathbf{0}$ and
spin-$\mathbf{1}$ states are larger than ones of the states with
spin-$\mathbf{2}$ and spin-$\mathbf{3}$. Thirdly, the existence of
the spin-$\mathbf{1}$ $\Lambda\overline{\Lambda}$,
$p\overline{\Lambda}$ and $p\overline{p}$ enhancement is predicted.

We further study this system in the quark model with quark
correlation, we consider two interesting configurations
$(\rm{q}^3)-(\overline{\rm{q}}^3)$ and
$(\rm{q}^2\overline{\rm{q}})-(\rm{q}\overline{\rm{q}}^2)$, and the
two clusters are  connected by color $\mathbf{8}$-plet bond,
$\mathbf{10}$-plet bond and color $\mathbf{6}$-plet bond,
$\mathbf{15}$-plet bond respectively. We find in the spectrum of
both configurations there exist rather stable states which has the
same quantum number with the $p\overline{p}$ enhancement,
$p\overline{\Lambda}$ enhancement and $\Lambda\overline{\Lambda}$
enhancement, hence these enhancements are not experimental
artifacts.

To the $p\overline{p}$ enhancement, we found out that  under the
approximation of neglecting van der Waal's force between color
singlet baryon and color singlet antibaryon, the clusters as
elements of the enhancement are objects with color. The potentials
used in the studies on the $p\overline{p}$-baryonium should be due
to "color chemistry bonds" that are caused by the unfrozen color
degree of freedom in the system in the framework of this paper. It
has been shown that if there exist quark correlation in the
baryonium, the state of $p\overline{p}$ enhancement are mainly of
$|\bf{120}_{cs},\overline{\bf{6}}_c,\bf{2},\bf{3}_f\oplus\overline{\bf{6}}_f\rangle
|\overline{\bf{120}}_{cs},\bf{6}_c,\bf{2},\overline{\bf{3}}_f\oplus\bf{6}_f\rangle$
in  the configuration
$(\rm{q}^2\overline{\rm{q}})-(\rm{q}\overline{\rm{q}}^2)$, with
small mixing $|\bf{70}_{cs},\bf{8}_c,\bf{2},\bf{8}_f\rangle
|\overline{\bf{70}}_{cs},\bf{8}_c,\bf{2},\bf{8}_f\rangle$ in the
configuration $(\rm{q}^3)-(\overline{\rm{q}}^3)$. Similarly, the
state of $\Lambda\overline{\Lambda}$ enhancement are mainly
$|\mathbf{56}_{cs},\mathbf{8}_c,\mathbf{2};\mathbf{1}_f\rangle
|\overline{\mathbf{56}}_{cs},\mathbf{8}_c,\mathbf{2};\mathbf{1}_f\rangle$
in the configuration $(\rm{q}^3)-(\overline{\rm{q}}^3)$, and plus
few mixtures of the corresponding $\Lambda\overline{\Lambda}$
enhancement state in  the configuration
$(\rm{q}^2\overline{\rm{q}})-(\rm{q}\overline{\rm{q}}^2)$. Moreover,
the $p\overline{\Lambda}$ enhancement should be half of the states
in the $(\rm{q}^3)-(\overline{\rm{q}}^3)$ configuration plus half of
that in the
$(\rm{q}^2\overline{\rm{q}})-(\rm{q}\overline{\rm{q}}^2)$
approximately.

~~Considering  all facts  from common quark model and from the
models
 with quark correlations, we conclude  that the $p\overline{p}$ enhancement,
$p\overline{\Lambda}$ enhancement and the
$\Lambda\overline{\Lambda}$ enhancement are in existence, and the
signal corresponding them must be visible. Actually, they have been
shown experimentally by BES and Bell. In other words, those
experiments are certainly not experimental artifacts. We predict
that spin-$\mathbf{1}$ $\Lambda\overline{\Lambda}$ enhancement will
be observed in future and we conjecture that BES has observed both
spin-$\mathbf{0}$, spin-$\mathbf{1}$ $p\overline{p}$ enhancement and
spin-$\mathbf{0}$, spin-$\mathbf{1}$ $p\overline{\Lambda}$
enhancement.

The existed quark model with colormagnetic interaction only consider
one gluon exchange interaction between quarks(antiquarks) and
quarks(antiquarks), and  the corrections from the quark's
annihilation contribution( the so-called OZI violated graphs ) are
not taken into account in the model's framework. The latter will
bring small errors to the model, and  the problem how to count  the
contributions of the quark's annihilation diagrams into the model
still remains to be open so far. However, since many successes and
reasonable results have been achieved in the studies  of the
multiquark systems within this model framework (i.e., the model is
supported by the phenomenology),  and our results presented in this
paper are exact in the model,
 we argue that the results and predictions presented in this paper
are reliable, and hence should be useful and meaningful for
understanding the physics of both $\rm{q}^3\overline{\rm{q}}^3$
system and the recent corresponding experiments  reported by BES and
Bell qualitatively.

\par
The effect of $\rm{SU(3)}_f$ symmetry broken is not taken into
account in this paper, and the results are qualitatively correct. It
is necessary to consider the effect of strange quark mass in order
to make serious comparison with the experiments and predict exactly
the masses of the multiplets of the $\rm{q}^3\overline{\rm{q}}^3$
meson.
\par
\begin{center}
{\bf ACKNOWLEDGMENTS}
\end{center}
This work is partially supported by National Natural Science
Foundation of China under Grant Numbers 90403021, and by the PhD
Program Funds 20020358040 of the Education Ministry of China and
KJCX2-SW-N10 of the Chinese Academy.

\begin{appendix}
\section{The mixing of states with definite spin under the colormagnetic interaction hamiltonian $H^{\prime}$ }
In this appendix, as an example, we will illustrate that in the case
of $\mathbf{70}_{cs}\otimes\overline{\mathbf{20}}_{cs}$  the three
spin-$\mathbf{1}$ states
$|\mathbf{35}_{cs},\mathbf{1}_c,\mathbf{3};\mathbf{8}_f\otimes\overline{\mathbf{10}}_f\rangle$,
$|\mathbf{280}_{cs},\mathbf{1}_c,\mathbf{3};\mathbf{8}_f\otimes\overline{\mathbf{10}}_f\rangle$
and
$|\mathbf{896}_{cs},\mathbf{1}_c,\mathbf{3};\mathbf{8}_f\otimes\overline{\mathbf{10}}_f\rangle$
are mixed due to the colormagnetic interaction $H^{\prime}$, and are
linear combinations of the states
$|(\mathbf{70}_{cs},\mathbf{8}_c,\mathbf{4};\mathbf{8}_f),(\overline{\mathbf{20}}_{cs},\mathbf{8}_c,\mathbf{2};\overline{\mathbf{10}}_f),\mathbf{1}_c,\mathbf{3},\mathbf{8}_f\otimes\overline{\mathbf{10}}_f\rangle$,
$|(\mathbf{70}_{cs},\mathbf{8}_c,\mathbf{2};\mathbf{8}_f)(\overline{\mathbf{20}}_{cs},\mathbf{8}_c,\mathbf{2};\overline{\mathbf{10}}_f),\mathbf{1}_c,\mathbf{3},\mathbf{8}_f\otimes\overline{\mathbf{10}}_f\rangle$
and
$|(\mathbf{70}_{cs},\mathbf{1}_c,\mathbf{2};\mathbf{8}_f),(\overline{\mathbf{20}}_{cs},\mathbf{1}_c,\mathbf{4};\overline{\mathbf{10}}_f),,\mathbf{1}_c,\mathbf{3},\mathbf{8}_f\otimes\overline{\mathbf{10}}_f\rangle$.
The eigenstates and the corresponding colormagnetic interaction
energy are calculated, for the other cases, these quantities can
also be calculated following the same method.
\begin{eqnarray}
\nonumber&&|\mathbf{896}_{cs},\mathbf{1}_c,\mathbf{3};\mathbf{8}_f\otimes\overline{\mathbf{10}}_f\rangle=\left\{\begin{array}{c
c }\mathbf{70}_{cs} & \overline{\mathbf{20}}_{cs}
\\(\mathbf{8}_c,\mathbf{4}) & (\mathbf{8}_c,\mathbf{2})\end{array} \right|
\left.\begin{array}{c}\mathbf{896}_{cs}\\(\mathbf{1}_c,\mathbf{3})\end{array}\right\}
|(\mathbf{70}_{cs},\mathbf{8}_c,\mathbf{4};\mathbf{8}_f),(\overline{\mathbf{20}}_{cs},\mathbf{8}_c,\mathbf{2};\overline{\mathbf{10}}_f),\\
\nonumber&&\mathbf{1}_c,\mathbf{3};\mathbf{8}_f\otimes\overline{\mathbf{10}}_f\rangle+\left\{\begin{array}{c
c }\mathbf{70}_{cs} & \overline{\mathbf{20}}_{cs}
\\(\mathbf{8}_c,\mathbf{2}) & (\mathbf{8}_c,\mathbf{2})\end{array} \right|
\left.\begin{array}{c}\mathbf{896}_{cs}\\(\mathbf{1}_c,\mathbf{3})\end{array}\right\}
|(\mathbf{70}_{cs},\mathbf{8}_c,\mathbf{2};\mathbf{8}_f),(\overline{\mathbf{20}}_{cs},\mathbf{8}_c,\mathbf{2};\overline{\mathbf{10}}_f),\\
\nonumber&&\mathbf{1}_c,\mathbf{3};\mathbf{8}_f\otimes\overline{\mathbf{10}}_f\rangle+\left\{\begin{array}{c
c }\mathbf{70}_{cs} & \overline{\mathbf{20}}_{cs}
\\(\mathbf{1}_c,\mathbf{2}) & (\mathbf{1}_c,\mathbf{4})\end{array} \right|
\left.\begin{array}{c}\mathbf{896}_{cs}\\(\mathbf{1}_c,\mathbf{3})\end{array}\right\}
|(\mathbf{70}_{cs},\mathbf{1}_c,\mathbf{2};\mathbf{8}_f),(\overline{\mathbf{20}}_{cs},\mathbf{1}_c,\mathbf{4};\overline{\mathbf{10}}_f),\\
\nonumber&&\mathbf{1}_c,\mathbf{3};\mathbf{8}_f\otimes\overline{\mathbf{10}}_f\rangle;\\
\nonumber&&|\mathbf{280}_{cs},\mathbf{1}_c,\mathbf{3};\mathbf{8}_f\otimes\overline{\mathbf{10}}_f\rangle=\left\{\begin{array}{c
c }\mathbf{70}_{cs} & \overline{\mathbf{20}}_{cs}
\\(\mathbf{8}_c,\mathbf{4}) & (\mathbf{8}_c,\mathbf{2})\end{array} \right|
\left.\begin{array}{c}\mathbf{280}_{cs}\\(\mathbf{1}_c,\mathbf{3})\end{array}\right\}
|(\mathbf{70}_{cs},\mathbf{8}_c,\mathbf{4};\mathbf{8}_f),(\overline{\mathbf{20}}_{cs},\mathbf{8}_c,\mathbf{2};\overline{\mathbf{10}}_f),\\
\nonumber&&\mathbf{1}_c,\mathbf{3};\mathbf{8}_f\otimes\overline{\mathbf{10}}_f\rangle+\left\{\begin{array}{c
c }\mathbf{70}_{cs} & \overline{\mathbf{20}}_{cs}
\\(\mathbf{8}_c,\mathbf{2}) & (\mathbf{8}_c,\mathbf{2})\end{array} \right|
\left.\begin{array}{c}\mathbf{280}_{cs}\\(\mathbf{1}_c,\mathbf{3})\end{array}\right\}
|(\mathbf{70}_{cs},\mathbf{8}_c,\mathbf{2};\mathbf{8}_f),(\overline{\mathbf{20}}_{cs},\mathbf{8}_c,\mathbf{2};\overline{\mathbf{10}}_f),\\
\nonumber&&\mathbf{1}_c,\mathbf{3};\mathbf{8}_f\otimes\overline{\mathbf{10}}_f\rangle+\left\{\begin{array}{c
c }\mathbf{70}_{cs} & \overline{\mathbf{20}}_{cs}
\\(\mathbf{1}_c,\mathbf{2}) & (\mathbf{1}_c,\mathbf{4})\end{array} \right|
\left.\begin{array}{c}\mathbf{280}_{cs}\\(\mathbf{1}_c,\mathbf{3})\end{array}\right\}
|(\mathbf{70}_{cs},\mathbf{1}_c,\mathbf{2};\mathbf{8}_f),(\overline{\mathbf{20}}_{cs},\mathbf{1}_c,\mathbf{4};\overline{\mathbf{10}}_f),\\
\nonumber&&\mathbf{1}_c,\mathbf{3};\mathbf{8}_f\otimes\overline{\mathbf{10}}_f\rangle;\\
\nonumber&&|\mathbf{35}_{cs},\mathbf{1}_c,\mathbf{3};\mathbf{8}_f\otimes\overline{\mathbf{10}}_f\rangle=\left\{\begin{array}{c
c }\mathbf{70}_{cs} & \overline{\mathbf{20}}_{cs}
\\(\mathbf{8}_c,\mathbf{4}) & (\mathbf{8}_c,\mathbf{2})\end{array} \right|
\left.\begin{array}{c}\mathbf{35}_{cs}\\(\mathbf{1}_c,\mathbf{3})\end{array}\right\}
|(\mathbf{70}_{cs},\mathbf{8}_c,\mathbf{4};\mathbf{8}_f),(\overline{\mathbf{20}}_{cs},\mathbf{8}_c,\mathbf{2};\overline{\mathbf{10}}_f),\\
\nonumber&&\mathbf{1}_c,\mathbf{3};\mathbf{8}_f\otimes\overline{\mathbf{10}}_f\rangle+\left\{\begin{array}{c
c }\mathbf{70}_{cs} & \overline{\mathbf{20}}_{cs}
\\(\mathbf{8}_c,\mathbf{2}) & (\mathbf{8}_c,\mathbf{2})\end{array} \right|
\left.\begin{array}{c}\mathbf{35}_{cs}\\(\mathbf{1}_c,\mathbf{3})\end{array}\right\}
|(\mathbf{70}_{cs},\mathbf{8}_c,\mathbf{2};\mathbf{8}_f),(\overline{\mathbf{20}}_{cs},\mathbf{8}_c,\mathbf{2};\overline{\mathbf{10}}_f),\\
\nonumber&&\mathbf{1}_c,\mathbf{3};\mathbf{8}_f\otimes\overline{\mathbf{10}}_f\rangle+\left\{\begin{array}{c
c }\mathbf{70}_{cs} & \overline{\mathbf{20}}_{cs}
\\(\mathbf{1}_c,\mathbf{2}) & (\mathbf{1}_c,\mathbf{4})\end{array} \right|
\left.\begin{array}{c}\mathbf{35}_{cs}\\(\mathbf{1}_c,\mathbf{3})\end{array}\right\}
|(\mathbf{70}_{cs},\mathbf{1}_c,\mathbf{2};\mathbf{8}_f),(\overline{\mathbf{20}}_{cs},\mathbf{1}_c,\mathbf{4};\overline{\mathbf{10}}_f),\\
\label{a1}&&\mathbf{1}_c,\mathbf{3};\mathbf{8}_f\otimes\overline{\mathbf{10}}_f\rangle;
\end{eqnarray}
$\rm{SU}_{cs}(6)\supset \rm{SU}_{c}(3)\times \rm{SU}_{s}(2)$
coefficients of fractional parentage can be calculated following the
Ref.\cite{cpf}, and some results has been listed in the Ref.
\cite{chenjq}. From this book, we can find the
$\rm{SU}_{cs}(6)\supset \rm{SU}_{c}(3)\times \rm{SU}_{s}(2)$
coefficients of fractional parentage which appear in the above
formula,
\begin{eqnarray}
\nonumber&&\left\{\begin{array}{c c }\mathbf{70}_{cs} &
\overline{\mathbf{20}}_{cs} \\(\mathbf{8}_c,\mathbf{4}) &
(\mathbf{8}_c,\mathbf{2})\end{array} \right|
\left.\begin{array}{c}\mathbf{896}_{cs}\\(\mathbf{1}_c,\mathbf{3})\end{array}\right\}=-\sqrt{\frac{1}{9}},\;\;
\left\{\begin{array}{c c }\mathbf{70}_{cs} &
\overline{\mathbf{20}}_{cs}
\\(\mathbf{8}_c,\mathbf{2}) & (\mathbf{8}_c,\mathbf{2})\end{array} \right|
\left.\begin{array}{c}\mathbf{896}_{cs}\\(\mathbf{1}_c,\mathbf{3})\end{array}\right\}=-\sqrt{\frac{4}{9}}\\
\nonumber&&\left\{\begin{array}{c c }\mathbf{70}_{cs} &
\overline{\mathbf{20}}_{cs} \\(\mathbf{1}_c,\mathbf{2}) &
(\mathbf{1}_c,\mathbf{4})\end{array} \right|
\left.\begin{array}{c}\mathbf{896}_{cs}\\(\mathbf{1}_c,\mathbf{3})\end{array}\right\}=\sqrt{\frac{4}{9}},\;\;
\left\{\begin{array}{c c }\mathbf{70}_{cs} &
\overline{\mathbf{20}}_{cs}
\\(\mathbf{8}_c,\mathbf{4}) & (\mathbf{8}_c,\mathbf{2})\end{array} \right|
\left.\begin{array}{c}\mathbf{280}_{cs}\\(\mathbf{1}_c,\mathbf{3})\end{array}\right\}=-\sqrt{\frac{4}{9}}\\
\nonumber&&\left\{\begin{array}{c c }\mathbf{70}_{cs} &
\overline{\mathbf{20}}_{cs}
\\(\mathbf{8}_c,\mathbf{2}) & (\mathbf{8}_c,\mathbf{2})\end{array} \right|
\left.\begin{array}{c}\mathbf{280}_{cs}\\(\mathbf{1}_c,\mathbf{3})\end{array}\right\}=-\sqrt{\frac{1}{9}},\;\;
\left\{\begin{array}{c c }\mathbf{70}_{cs} &
\overline{\mathbf{20}}_{cs}
\\(\mathbf{1}_c,\mathbf{2}) & (\mathbf{1}_c,\mathbf{4})\end{array} \right|
\left.\begin{array}{c}\mathbf{280}_{cs}\\(\mathbf{1}_c,\mathbf{3})\end{array}\right\}=-\sqrt{\frac{4}{9}}\\
\nonumber&&\left\{\begin{array}{c c }\mathbf{70}_{cs} &
\overline{\mathbf{20}}_{cs}
\\(\mathbf{8}_c,\mathbf{4}) & (\mathbf{8}_c,\mathbf{2})\end{array} \right|
\left.\begin{array}{c}\mathbf{35}_{cs}\\(\mathbf{1}_c,\mathbf{3})\end{array}\right\}=-\sqrt{\frac{4}{9}},\;\;
\left\{\begin{array}{c c }\mathbf{70}_{cs} &
\overline{\mathbf{20}}_{cs}
\\(\mathbf{8}_c,\mathbf{2}) & (\mathbf{8}_c,\mathbf{2})\end{array} \right|
\left.\begin{array}{c}\mathbf{35}_{cs}\\(\mathbf{1}_c,\mathbf{3})\end{array}\right\}=\sqrt{\frac{4}{9}}\\
\label{a2}&&\left\{\begin{array}{c c }\mathbf{70}_{cs} &
\overline{\mathbf{20}}_{cs} \\(\mathbf{1}_c,\mathbf{2}) &
(\mathbf{1}_c,\mathbf{4})\end{array} \right|
\left.\begin{array}{c}\mathbf{35}_{cs}\\(\mathbf{1}_c,\mathbf{3})\end{array}\right\}=\sqrt{\frac{1}{9}}
\end{eqnarray}
Substituting Eq.(\ref{a2}) into Eq.(\ref{a1}) we can obtain
\begin{eqnarray}
\nonumber&&|\mathbf{896}_{cs},\mathbf{1}_c,\mathbf{3};\mathbf{8}_f\otimes\overline{\mathbf{10}}_f\rangle=-\sqrt{\frac{1}{9}}|(\mathbf{70}_{cs},\mathbf{8}_c,\mathbf{4};\mathbf{8}_f),
(\overline{\mathbf{20}}_{cs},\mathbf{8}_c,\mathbf{2};\overline{\mathbf{10}}_f),\mathbf{1}_c,\mathbf{3};\mathbf{8}_f\otimes\overline{\mathbf{10}}_f\rangle\\
\nonumber&&-\sqrt{\frac{4}{9}}|(\mathbf{70}_{cs},\mathbf{8}_c,\mathbf{2};\mathbf{8}_f),
(\overline{\mathbf{20}}_{cs},\mathbf{8}_c,\mathbf{2};\overline{\mathbf{10}}_f),\mathbf{1}_c,\mathbf{3};\mathbf{8}_f\otimes\overline{\mathbf{10}}_f\rangle+\sqrt{\frac{4}{9}}|(\mathbf{70}_{cs},\mathbf{1}_c,\mathbf{2};\mathbf{8}_f),\\
\nonumber&&(\overline{\mathbf{20}}_{cs},\mathbf{1}_c,\mathbf{4};\overline{\mathbf{10}}_f),\mathbf{1}_c,\mathbf{3};\mathbf{8}_f\otimes\overline{\mathbf{10}}_f\rangle;\\
\nonumber&&|\mathbf{280}_{cs},\mathbf{1}_c,\mathbf{3};\mathbf{8}_f\otimes\overline{\mathbf{10}}_f\rangle=-\sqrt{\frac{4}{9}}|(\mathbf{70}_{cs},\mathbf{8}_c,\mathbf{4};\mathbf{8}_f),
(\overline{\mathbf{20}}_{cs},\mathbf{8}_c,\mathbf{2};\overline{\mathbf{10}}_f),\mathbf{1}_c,\mathbf{3};\mathbf{8}_f\otimes\overline{\mathbf{10}}_f\rangle\\
\nonumber&&-\sqrt{\frac{1}{9}}|(\mathbf{70}_{cs},\mathbf{8}_c,\mathbf{2};\mathbf{8}_f),
(\overline{\mathbf{20}}_{cs},\mathbf{8}_c,\mathbf{2};\overline{\mathbf{10}}_f),\mathbf{1}_c,\mathbf{3};\mathbf{8}_f\otimes\overline{\mathbf{10}}_f\rangle-\sqrt{\frac{4}{9}}|(\mathbf{70}_{cs},\mathbf{1}_c,\mathbf{2};\mathbf{8}_f),\\
\nonumber&&(\overline{\mathbf{20}}_{cs},\mathbf{1}_c,\mathbf{4};\overline{\mathbf{10}}_f),\mathbf{1}_c,\mathbf{3};\mathbf{8}_f\otimes\overline{\mathbf{10}}_f\rangle;\\
\nonumber&&|\mathbf{35}_{cs},\mathbf{1}_c,\mathbf{3};\mathbf{8}_f\otimes\overline{\mathbf{10}}_f\rangle=-\sqrt{\frac{4}{9}}|(\mathbf{70}_{cs},\mathbf{8}_c,\mathbf{4};\mathbf{8}_f),
(\overline{\mathbf{20}}_{cs},\mathbf{8}_c,\mathbf{2};\overline{\mathbf{10}}_f),\mathbf{1}_c,\mathbf{3};\mathbf{8}_f\otimes\overline{\mathbf{10}}_f\rangle\\
\nonumber&&+\sqrt{\frac{4}{9}}|(\mathbf{70}_{cs},\mathbf{8}_c,\mathbf{2};\mathbf{8}_f),
(\overline{\mathbf{20}}_{cs},\mathbf{8}_c,\mathbf{2};\overline{\mathbf{10}}_f),\mathbf{1}_c,\mathbf{3};\mathbf{8}_f\otimes\overline{\mathbf{10}}_f\rangle+\sqrt{\frac{1}{9}}|(\mathbf{70}_{cs},\mathbf{1}_c,\mathbf{2};\mathbf{8}_f),\\
\label{A3}&&(\overline{\mathbf{20}}_{cs},\mathbf{1}_c,\mathbf{4};\overline{\mathbf{10}}_f),\mathbf{1}_c,\mathbf{3};\mathbf{8}_f\otimes\overline{\mathbf{10}}_f\rangle
\end{eqnarray}
From the above, we can learn that $\langle
\mathbf{896}_{cs},\mathbf{1}_c,\mathbf{3};\mathbf{8}_f\otimes\overline{\mathbf{10}}_f|H^{\prime}|\mathbf{896}_{cs},\mathbf{1}_c,\mathbf{3};\mathbf{8}_f\otimes\overline{\mathbf{10}}_{cs}\rangle=\frac{516}{27}C$,
$\langle
\mathbf{896}_{cs},\mathbf{1}_c,\mathbf{3};\mathbf{8}_f\otimes\overline{\mathbf{10}}_f|H^{\prime}|\mathbf{280}_{cs},\mathbf{1}_c,\mathbf{3};\mathbf{8}_f\otimes\overline{\mathbf{10}}_{cs}\rangle=\frac{90}{9}C$
etc., so $H^{\prime}$ can be expressed as the the following matrix
form in the space expanded by the three spin-$\mathbf{1}$ states
$|\mathbf{896}_{cs},\mathbf{1}_c,\mathbf{3};\mathbf{8}\otimes\overline{\mathbf{10}}_f\rangle$,
$|\mathbf{280}_{cs},\mathbf{1}_c,\mathbf{3};\mathbf{8}\otimes\overline{\mathbf{10}}_f\rangle$,
$|\mathbf{35}_{cs},\mathbf{1}_c,\mathbf{3};\mathbf{8}\otimes\overline{\mathbf{10}}_f\rangle$
\begin{equation}
\label{A4} \frac{1}{27}C\left(\begin{array}{ccc}
516&240&-48\\
240&264&192\\
-48&192&-240
\end{array}\right)
\end{equation}
Diagonalizing the above matrix, we can obtain the eigenstates and
the corresponding colormagnetic interaction energy,
\begin{eqnarray}
\nonumber&&|\mathbf{3},\mathbf{8}_f\otimes\overline{\mathbf{10}}_f\rangle_1=0.841|\mathbf{896}_{cs},\mathbf{1}_c,\mathbf{3};\mathbf{8}_f\otimes\overline{\mathbf{10}}_f\rangle+0.536|\mathbf{280}_{cs},\mathbf{1}_c,\mathbf{3};\mathbf{8}_f\otimes\overline{\mathbf{10}}_f\rangle\\
\nonumber&&+0.069|\mathbf{35}_{cs},\mathbf{1}_c,\mathbf{3};\mathbf{8}_f\otimes\overline{\mathbf{10}}_f\rangle=-0.684|(\mathbf{70}_{cs},\mathbf{8}_c,\mathbf{4};\mathbf{8}_f),
(\overline{\mathbf{20}}_{cs},\mathbf{8}_c,\mathbf{2};\overline{\mathbf{10}}_f),\mathbf{1}_c,\mathbf{3};\mathbf{8}_f\otimes\overline{\mathbf{10}}_f\rangle\\
\nonumber&&-0.693|(\mathbf{70}_{cs},\mathbf{8}_c,\mathbf{2};\mathbf{8}_f),(\overline{\mathbf{20}}_{cs},\mathbf{8}_c,\mathbf{2};\overline{\mathbf{10}}_f),\mathbf{1}_c,\mathbf{3};\mathbf{8}_f\otimes\overline{\mathbf{10}}_f\rangle+0.226|(\mathbf{70}_{cs},\mathbf{1}_c,\mathbf{2};\mathbf{8}_f),\\
\label{A5}&&(\overline{\mathbf{20}}_{cs},\mathbf{1}_c,\mathbf{4};\overline{\mathbf{10}}_f),\mathbf{1}_c,\mathbf{3};\mathbf{8}_f\otimes\overline{\mathbf{10}}_f\rangle,\\
\nonumber&&|\mathbf{3},\mathbf{8}_f\otimes\overline{\mathbf{10}}_f\rangle_2=0.156|\mathbf{896}_{cs},\mathbf{1}_c,\mathbf{3};\mathbf{8}_f\otimes\overline{\mathbf{10}}_f\rangle-0.364|\mathbf{280}_{cs},\mathbf{1}_c,\mathbf{3};\mathbf{8}_f\otimes\overline{\mathbf{10}}_f\rangle\\
\nonumber&&+0.918|\mathbf{35}_{cs},\mathbf{1}_c,\mathbf{3};\mathbf{8}_f\otimes\overline{\mathbf{10}}_f\rangle=-0.422|(\mathbf{70}_{cs},\mathbf{8}_c,\mathbf{4};\mathbf{8}_f),
(\overline{\mathbf{20}}_{cs},\mathbf{8}_c,\mathbf{2};\overline{\mathbf{10}}_f),\mathbf{1}_c,\mathbf{3};\mathbf{8}_f\otimes\overline{\mathbf{10}}_f\rangle\\
\nonumber&&+0.629|(\mathbf{70}_{cs},\mathbf{8}_c,\mathbf{2};\mathbf{8}_f),(\overline{\mathbf{20}}_{cs},\mathbf{8}_c,\mathbf{2};\overline{\mathbf{10}}_f),\mathbf{1}_c,\mathbf{3};\mathbf{8}_f\otimes\overline{\mathbf{10}}_f\rangle+0.653|(\mathbf{70}_{cs},\mathbf{1}_c,\mathbf{2};\mathbf{8}_f),\\
\label{A6}&&(\overline{\mathbf{20}}_{cs},\mathbf{1}_c,\mathbf{4};\overline{\mathbf{10}}_f),\mathbf{1}_c,\mathbf{3};\mathbf{8}_f\otimes\overline{\mathbf{10}}_f\rangle,\\
\nonumber&&|\mathbf{3},\mathbf{8}_f\otimes\overline{\mathbf{10}}_f\rangle_3=-0.518|\mathbf{896}_{cs},\mathbf{1}_c,\mathbf{3};\mathbf{8}_f\otimes\overline{\mathbf{10}}_f\rangle+0.762|\mathbf{280}_{cs},\mathbf{1}_c,\mathbf{3};\mathbf{8}_f\otimes\overline{\mathbf{10}}_f\rangle\\
\nonumber&&+0.389|\mathbf{35}_{cs},\mathbf{1}_c,\mathbf{3};\mathbf{8}_f\otimes\overline{\mathbf{10}}_f\rangle=-0.595|(\mathbf{70}_{cs},\mathbf{8}_c,\mathbf{4};\mathbf{8}_f),
(\overline{\mathbf{20}}_{cs},\mathbf{8}_c,\mathbf{2};\overline{\mathbf{10}}_f),\mathbf{1}_c,\mathbf{3};\mathbf{8}_f\otimes\overline{\mathbf{10}}_f\rangle\\
\nonumber&&+0.351|(\mathbf{70}_{cs},\mathbf{8}_c,\mathbf{2};\mathbf{8}_f),(\overline{\mathbf{20}}_{cs},\mathbf{8}_c,\mathbf{2};\overline{\mathbf{10}}_f),\mathbf{1}_c,\mathbf{3};\mathbf{8}_f\otimes\overline{\mathbf{10}}_f\rangle-0.723|(\mathbf{70}_{cs},\mathbf{1}_c,\mathbf{2};\mathbf{8}_f),\\
\label{A7}&&(\overline{\mathbf{20}}_{cs},\mathbf{1}_c,\mathbf{4};\overline{\mathbf{10}}_f),\mathbf{1}_c,\mathbf{3};\mathbf{8}_f\otimes\overline{\mathbf{10}}_f\rangle
\end{eqnarray}
The colormagnetic energy respectively are 24.634C, -12.007C, 7.373C.

\section{The matrix representation of the hamiltonian $H^{\prime}$ in the subspace of states with fixed spin}
In the following, we list the matrix form of the colormagnetic
interaction hamiltonian $H^{\prime}$ in the subspace expanded by the
states of fixed spin in each case, and the constant $C$ in front
each of matrix is omitted.
\begin{enumerate}
\item $\mathbf{56}_{cs}\otimes\overline{\mathbf{56}}_{cs}$

{\bf{A}}), the spin-$\mathbf{0}$ state, there exist two
spin-$\mathbf{0}$ states
$|(\mathbf{56}_{cs},\mathbf{10}_c,\mathbf{4};\mathbf{1}_f),
(\overline{\mathbf{56}}_{cs},\overline{\mathbf{10}}_c,\mathbf{4};\mathbf{1}_f),\mathbf{1}_c,\mathbf{1};\mathbf{1}_f\otimes\mathbf{1}_f\rangle$
and $|(\mathbf{56}_{cs},\mathbf{8}_c,\mathbf{2};\mathbf{1}_f),
(\overline{\mathbf{56}}_{cs},\mathbf{8}_c,\mathbf{2};\mathbf{1}_f),\mathbf{1}_c,\mathbf{1};\mathbf{1}_f\otimes\mathbf{1}_f\rangle$,
and the hamiltonian $H^{\prime}$ can be expressed as the following
matrix form in the subspace expanded by the latter state vectors ,
\begin{equation}
\label{B7}\left(\begin{array}{cc}
-48&-8\sqrt{10}\\
-8\sqrt{10}&-64
\end{array}\right)
\end{equation}
{\bf{B}}), the spin-$\mathbf{1}$ state, in the subspace expanded by
the two spin-$\mathbf{1}$ states
$|(\mathbf{56}_{cs},\mathbf{10}_c,\mathbf{4};\mathbf{1}_f),
(\overline{\mathbf{56}}_{cs},\overline{\mathbf{10}}_c,\mathbf{4};\mathbf{1}_f),\mathbf{1}_c,\mathbf{3};\mathbf{1}_f\otimes\mathbf{1}_f\rangle$
and $|(\mathbf{56}_{cs},\mathbf{8}_c,\mathbf{2};\mathbf{1}_f),
(\overline{\mathbf{56}}_{cs},\mathbf{8}_c,\mathbf{2};\mathbf{1}_f),\mathbf{1}_c,\mathbf{3};\mathbf{1}_f\otimes\mathbf{1}_f\rangle,$
the matrix representation of $H^{\prime}$ is
\begin{equation}
\label{B8}\left(\begin{array}{cc}
-112/3&-40\sqrt{2}/3\\
-40\sqrt{2}/3&-16
\end{array}\right)
\end{equation}



\item $\mathbf{70}_{cs}\otimes\overline{\mathbf{56}}_{cs}$ and  $\mathbf{56}_{cs}\otimes\overline{\mathbf{70}}_{cs}$.

{\bf{A}}), the spin-$\mathbf{1}$ state, the three spin-$\mathbf{1}$
states
$|(\mathbf{70}_{cs},\mathbf{10}_c,\mathbf{2};\mathbf{8}_f),(\overline{\mathbf{56}}_{cs},
\overline{\mathbf{10}}_c,\mathbf{4};\mathbf{1}_f),\mathbf{1}_c,\mathbf{3};\mathbf{8}_f\hskip-0.0in\otimes\hskip-0.0in\mathbf{1}_f\rangle$,
$|(\mathbf{70}_{cs},\mathbf{8}_c,\mathbf{4};\mathbf{8}_f),(\overline{\mathbf{56}}_{cs},\mathbf{8}_c,\mathbf{2};
\mathbf{1}_f),\mathbf{1}_c,\mathbf{3};\mathbf{8}_f\otimes\mathbf{1}_f\rangle$,
and
$|(\mathbf{70}_{cs},\mathbf{8}_c,\mathbf{2};\mathbf{8}_f),(\overline{\mathbf{56}}_{cs},\mathbf{8}_c,\mathbf{2};
\mathbf{1}_f),\mathbf{1}_c,\mathbf{3};\mathbf{8}_f\otimes\mathbf{1}_f\rangle$
expand the Hilbert space, and the matrix form of $H^{\prime}$ is
\begin{equation}
\label{B4}\left(\begin{array}{ccc}
-40/3&4\sqrt{10}/3&16\sqrt{10}/3\\
4\sqrt{10}/3&-76/3&-56/3\\
16\sqrt{10}/3&-56/3&-40/3
\end{array}\right)
\end{equation}
{\bf{B}}), the spin-$\mathbf{2}$ state, there are two
spin-$\mathbf{2}$ base vectors
$|(\mathbf{70}_{cs},\mathbf{10}_c,\mathbf{2};\mathbf{8}_f),(\overline{\mathbf{56}}_{cs},\overline{\mathbf{10}}_c,\mathbf{4};\mathbf{1}_f),\mathbf{1}_c,\mathbf{5};\mathbf{8}_f\otimes\mathbf{1}_f\rangle$
and
$|(\mathbf{70}_{cs},\mathbf{8}_c,\mathbf{4};\mathbf{8}_f),(\overline{\mathbf{56}}_{cs},\mathbf{8}_c,\mathbf{2};\mathbf{1}_f),\mathbf{1}_c,\mathbf{5};\mathbf{8}_f\otimes\mathbf{1}_f\rangle$,
and $H^{\prime}$'s matrix form is  the following form:
\begin{equation}
\label{B5}\left(\begin{array}{cc}
8&4\sqrt{10}\\
4\sqrt{10}&-4
\end{array}\right)
\end{equation}

\item $\mathbf{70}_{cs}\otimes\overline{\mathbf{70}}_{cs}$

{\bf{A}}), the spin-$\mathbf{2}$ states, there are three
spin-$\mathbf{2}$ states
$|(\mathbf{70}_{cs},\;\mathbf{8}_c,\mathbf{4};\mathbf{8}_f),(\overline{\mathbf{70}}_{cs},\mathbf{8}_c,\mathbf{4};\mathbf{8}_f),$$\mathbf{1}_c,\mathbf{5},\mathbf{8}_f\otimes\mathbf{8}_f\rangle$,
$|(\mathbf{70}_{cs},\;\mathbf{8}_c,\mathbf{4};\mathbf{8}_f),\\(\overline{\mathbf{70}}_{cs},\mathbf{8}_c,\mathbf{2};\mathbf{8}_f),$$\mathbf{1}_c,\mathbf{5},\mathbf{8}_f\otimes\mathbf{8}_f\rangle$
and $|(\mathbf{70}_{cs},\mathbf{8}_c,\mathbf{2};\mathbf{8}_f),
(\overline{\mathbf{70}}_{cs},\mathbf{8}_c,\mathbf{4};\mathbf{8}_f),$
$\mathbf{1}_c,\mathbf{5},\mathbf{8}_f\otimes\mathbf{8}_f\rangle$,
and the hamiltonian $H^{\prime}$ can be expressed as the following
matrix form in the subspace expanded by them,
\begin{equation}
\label{B1} \left(\begin{array}{ccc}
0&-4\sqrt{2}&\;\;-4\sqrt{2}\\
-4\sqrt{2}&4&-12\\
-4\sqrt{2}&-12&4
\end{array}\right)
\end{equation}
{\bf{B}}), the spin-$\mathbf{1}$ states, in the subspace of six
spin-$\mathbf{1}$ states $|(\mathbf{70}_{cs},\mathbf{10}_c,\mathbf{2};\mathbf{8}_f),(\overline{\mathbf{70}}_{cs},\overline{\mathbf{10}}_c,\mathbf{2};\mathbf{8}_f),$$\mathbf{1}_c,\mathbf{3};\mathbf{8}_f\otimes\mathbf{8}_f\rangle$,\\
$|(\mathbf{70}_{cs},\mathbf{8}_c,\mathbf{4};\mathbf{8}_f),(\overline{\mathbf{70}}_{cs},\mathbf{8}_c,\mathbf{4};\mathbf{8}_f),$$\mathbf{1}_c,\mathbf{3};\mathbf{8}_f\otimes\mathbf{8}_f\rangle$,
$|(\mathbf{70}_{cs},\mathbf{8}_c,\mathbf{4};\mathbf{8}_f),(\overline{\mathbf{70}}_{cs},\mathbf{8}_c,\mathbf{2};\mathbf{8}_f),$$\mathbf{1}_c,\mathbf{3};\mathbf{8}_f\otimes\mathbf{8}_f\rangle$,
$|(\mathbf{70}_{cs},\mathbf{8}_c,\mathbf{2};\mathbf{8}_f),(\overline{\mathbf{70}}_{cs},\mathbf{8}_c,\mathbf{4};\mathbf{8}_f),$$\mathbf{1}_c,\mathbf{3};\mathbf{8}_f\otimes\mathbf{8}_f\rangle$,
$|(\mathbf{70}_{cs},\mathbf{8}_c,\mathbf{2};\mathbf{8}_f),(\overline{\mathbf{70}}_{cs},\mathbf{8}_c,\mathbf{2};\mathbf{8}_f),$$\mathbf{1}_c,\mathbf{3};\mathbf{8}_f\otimes\mathbf{8}_f\rangle$
and
$|(\mathbf{70}_{cs},\mathbf{1}_c,\mathbf{2};\mathbf{8}_f),(|\overline{\mathbf{70}}_{cs},\mathbf{1}_c,\mathbf{2};\mathbf{8}_f),$$\mathbf{1}_c,\mathbf{3};\mathbf{8}_f\otimes\mathbf{8}_f\rangle$,
the matrix representation of $H^{\prime}$ is
\begin{equation}
\label{B2} \left(\begin{array}{cccccc}
\frac{32}{3}&-\frac{20\sqrt{2}}{3}&\frac{16\sqrt{5}}{3}&\frac{16\sqrt{5}}{3}&-\frac{8\sqrt{5}}{3}&0\\
-\frac{20\sqrt{2}}{3}&-\frac{32}{3}&-\frac{4\sqrt{10}}{3}&-\frac{4\sqrt{10}}{3}& 4\sqrt{10}&\frac{8\sqrt{5}}{3}\\
\frac{16\sqrt{5}}{3}&-\frac{4\sqrt{10}}{3}&-\frac{20}{3}&-4&-\frac{8}{3}&\frac{16\sqrt{2}}{3}\\
\frac{16\sqrt{5}}{3}&-\frac{4\sqrt{10}}{3}&-4&-\frac{20}{3}&-\frac{8}{3}&\frac{16\sqrt{2}}{3}\\
-\frac{8\sqrt{5}}{3}&4\sqrt{10}&-\frac{8}{3}&-\frac{8}{3}&-\frac{8}{3}&\frac{8\sqrt{2}}{3}\\
0&\frac{8\sqrt{5}}{3}&\frac{16\sqrt{2}}{3}&\frac{16\sqrt{2}}{3}&\frac{8\sqrt{2}}{3}&-16
\end{array}\right)
\end{equation}
{\bf{C}}), the spin-$\mathbf{0}$ state, the four spin-$\mathbf{0}$
states
$|(\mathbf{70}_{cs},\mathbf{10}_c,\mathbf{2};\mathbf{8}_f),(\overline{\mathbf{70}}_{cs},\overline{\mathbf{10}}_c,\mathbf{2};\mathbf{8}_f),\mathbf{1}_c,\mathbf{1};\mathbf{8}_f\otimes\mathbf{8}_f\rangle$,
$|(\mathbf{70}_{cs},\mathbf{8}_c,\mathbf{4};\mathbf{8}_f),(\overline{\mathbf{70}}_{cs},\mathbf{8}_c,\mathbf{4};\mathbf{8}_f),\mathbf{1}_c,\mathbf{1};\mathbf{8}_f\otimes\mathbf{8}_f\rangle$,
$|(\mathbf{70}_{cs},\mathbf{8}_c,\mathbf{2};\mathbf{8}_f),(\overline{\mathbf{70}}_{cs},\mathbf{8}_c,\mathbf{2};\mathbf{8}_f),\mathbf{1}_c,\mathbf{1};\mathbf{8}_f\otimes\mathbf{8}_f\rangle$
and
$|(\mathbf{70}_{cs},\mathbf{1}_c,\mathbf{2};\mathbf{8}_f),(\overline{\mathbf{70}}_{cs},\mathbf{1}_c,\mathbf{2};\mathbf{8}_f),\mathbf{1}_c,\mathbf{1};\mathbf{8}_f\otimes\mathbf{8}_f\rangle$
expand the Hilbert space. And the hamiltonian $H^{\prime}$ can be
expressed as
\begin{equation}
\label{B3} \left(\begin{array}{cccc}
0&-4\sqrt{10}&8\sqrt{5}&0\\
-4\sqrt{10}&-16& 12\sqrt{2}&8\\
8\sqrt{5}&12\sqrt{2}&-8&-8\sqrt{2}\\
0&8&-8\sqrt{2}&-16
\end{array}\right)
\end{equation}

\end{enumerate}
\end{appendix}
\newpage
\begin{table}[hptb]
\begin{center}
\begin{tabular}{|c|c|l|c|}\hline\hline
Young tabular&Dimension&$\parbox{10cm}{\begin{center}
SU$_{c}$(3)$\otimes$ SU$_{s}$(2) content\end{center}}$&
Eigenvalue of \\
 & & &Casimir operator\\\hline
 & 3675&$(\mathbf{1}_c,\mathbf{1}),2(\mathbf{1}_c,\mathbf{3}),(\mathbf{1}_c,\mathbf{5}),(\mathbf{1}_c,\mathbf{7}),3(\mathbf{8}_c,\mathbf{1}),7(\mathbf{8}_c,\mathbf{3}),6(\mathbf{8}_c,\mathbf{5}),$&36\\
\includegraphics[width=1cm]{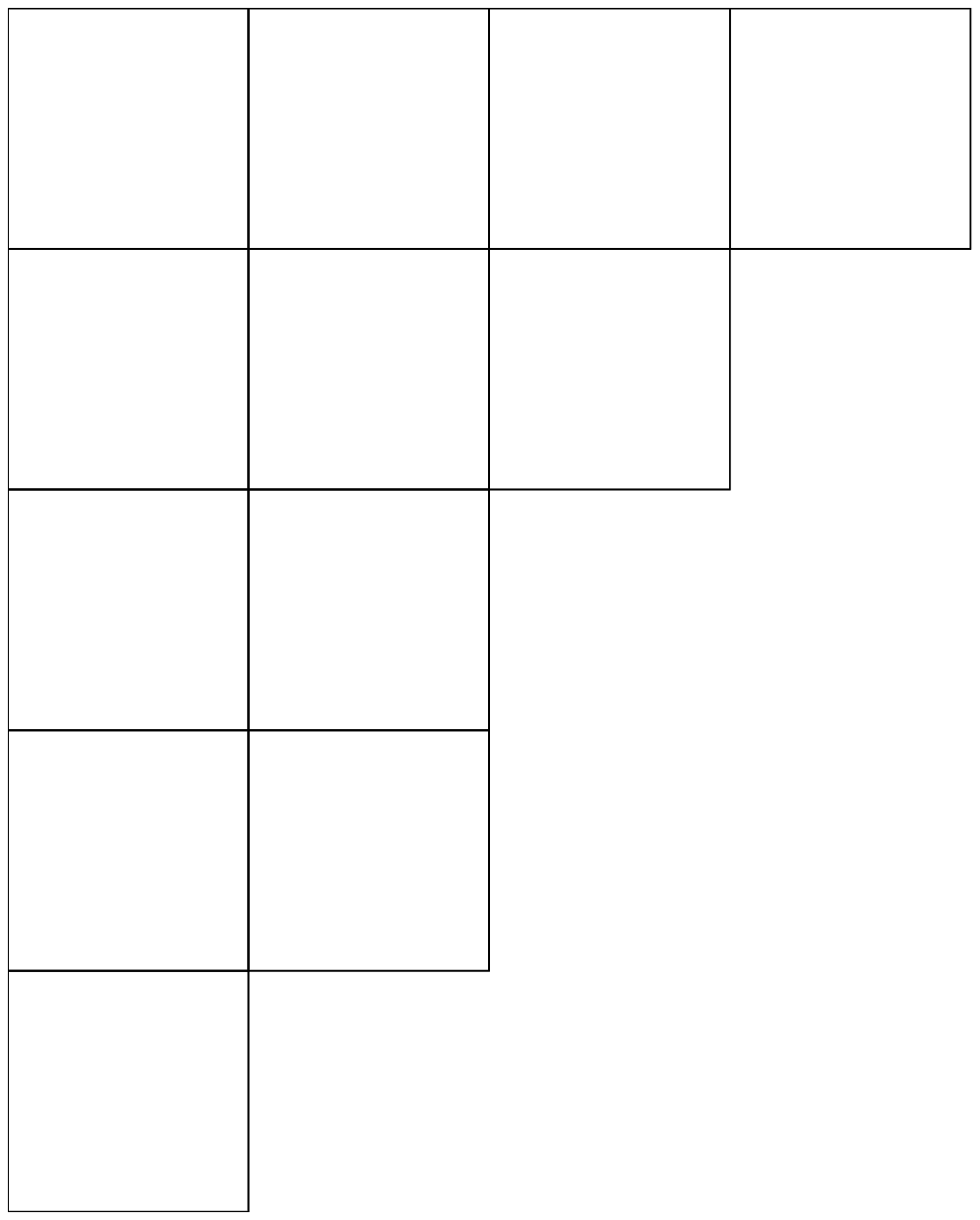}& & \parbox{10cm}{\vskip-55pt $2(\mathbf{8}_c,\mathbf{7}),2(\mathbf{10}_c,\mathbf{1}),4(\mathbf{10}_c,\mathbf{3}),3(\mathbf{10}_c,\mathbf{5}),(\mathbf{10}_c,\mathbf{7}),2(\overline{\mathbf{10}}_c,\mathbf{1}),$}& \\
 & &\parbox{10cm}{\vskip-55pt$4(\overline{\mathbf{10}}_c,\mathbf{3}),3(\overline{\mathbf{10}}_c,\mathbf{5}),(\overline{\mathbf{10}}_c,\mathbf{7}),3(\mathbf{27}_c,\mathbf{1}),6(\mathbf{27}_c,\mathbf{3}),4(\mathbf{27}_c,\mathbf{5}),$}& \\
 & &\parbox{10cm}{\vskip-55pt$(\mathbf{27}_c,\mathbf{7}),(\mathbf{35}_c,\mathbf{1}),2(\mathbf{35}_c,\mathbf{3}),(\mathbf{35}_c,\mathbf{5}),(\overline{\mathbf{35}}_c,\mathbf{1}),2(\overline{\mathbf{35}}_c,\mathbf{3}),$}& \\
 & &\parbox{10cm}{\vskip-55pt$(\overline{\mathbf{35}}_c,\mathbf{5}),(\mathbf{64}_c,\mathbf{1}),(\mathbf{64}_c,\mathbf{3})$}&  \\\hline
 &3200 &$(\mathbf{1}_c,\mathbf{3}),(\mathbf{1}_c,\mathbf{5}),2(\mathbf{8}_c,\mathbf{1}),4(\mathbf{8}_c,\mathbf{3}),3(\mathbf{8}_c,\mathbf{5}),(\mathbf{8}_c,\mathbf{7}),(\mathbf{10}_c,\mathbf{1}),$& 42\\
\includegraphics[width=1.25cm]{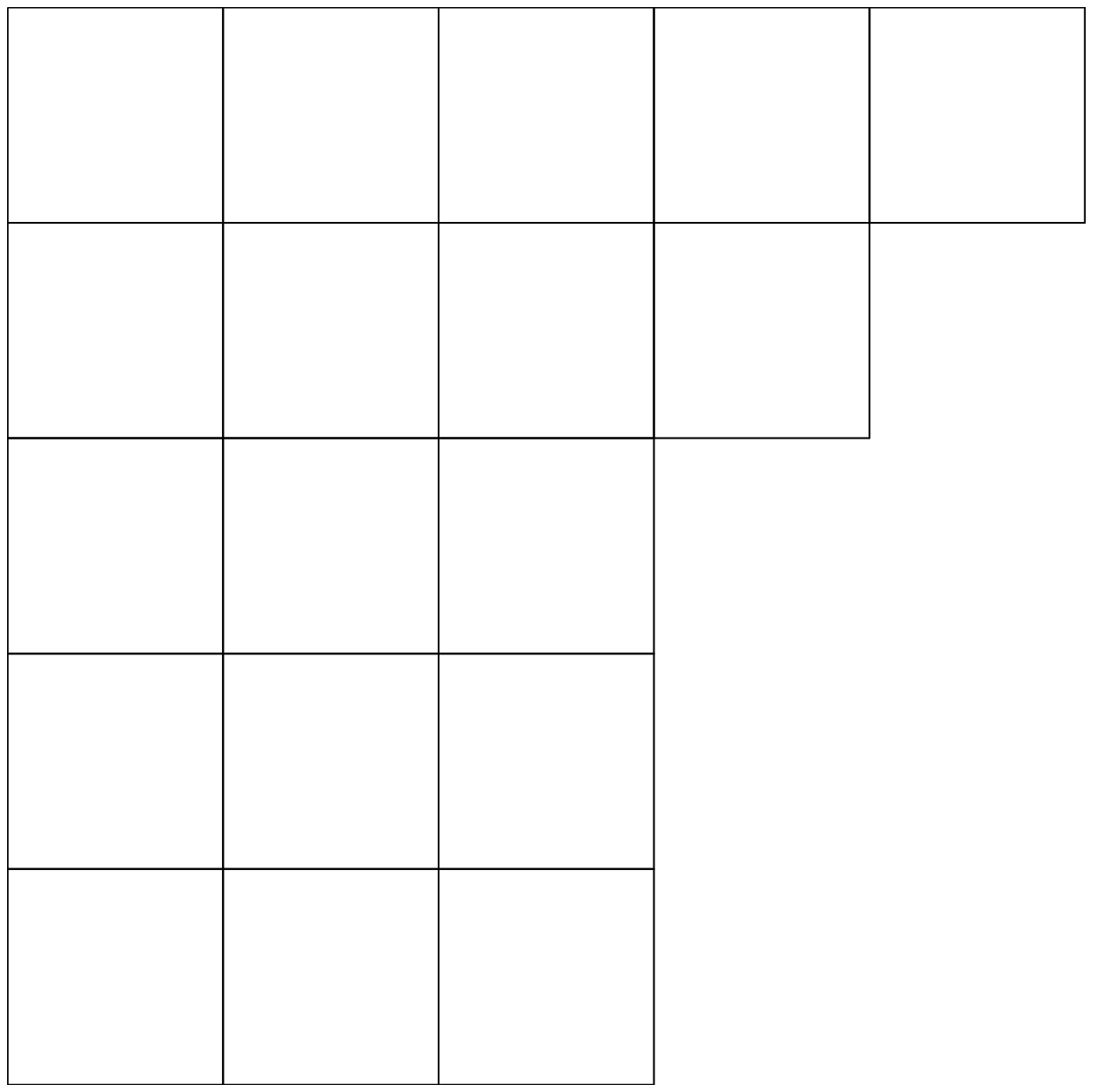} & &\parbox{10cm}{\vskip-35pt$2(\mathbf{10}_c,\mathbf{3}),(\mathbf{10}_c,\mathbf{5}),(\overline{\mathbf{10}}_c,\mathbf{1}),3(\overline{\mathbf{10}}_c,\mathbf{3}),(\overline{\mathbf{10}}_c,\mathbf{7}),2(\mathbf{27}_c,\mathbf{1}),$}& \\
 & &\parbox{10cm}{\vskip-45pt$4(\mathbf{27}_c,\mathbf{3}),3(\mathbf{27}_c,\mathbf{5}),(\mathbf{27}_c,\mathbf{7}),(\mathbf{35}_c,\mathbf{1}),(\mathbf{35}_c,\mathbf{3}),(\overline{\mathbf{35}}_c,\mathbf{1}),$}& \\
 & &\parbox{10cm}{\vskip-55pt$2(\overline{\mathbf{35}}_c,\mathbf{3}),2(\overline{\mathbf{35}}_c,\mathbf{5}),(\overline{\mathbf{35}}_c,\mathbf{7}),(\mathbf{64}_c,\mathbf{3}),(\mathbf{65}_c,\mathbf{5})$}& \\\hline
 & 2695&$(\mathbf{1}_c,\mathbf{3}),(\mathbf{1}_c,\mathbf{7}),(\mathbf{8}_c,\mathbf{1}),2(\mathbf{8}_c,\mathbf{3}),2(\mathbf{8}_c,\mathbf{5}),(\mathbf{8}_c,\mathbf{7}),(\mathbf{10}_c,\mathbf{1}),$&48\\
\includegraphics[width=1.5cm]{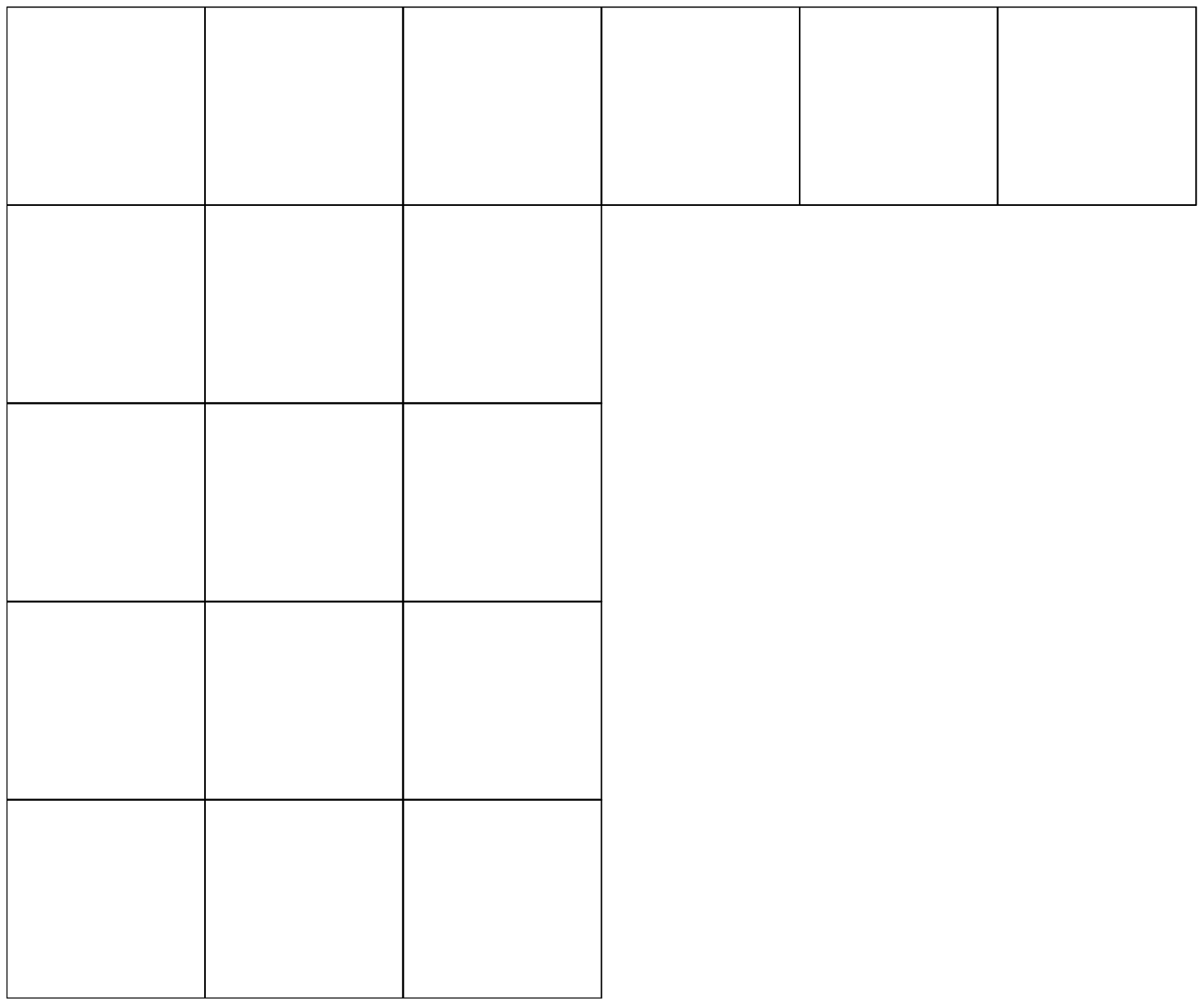} & &\parbox{10cm}{\vskip-40pt $(\mathbf{10}_c,\mathbf{3}),(\mathbf{10}_c,\mathbf{5}),(\overline{\mathbf{10}}_c,\mathbf{1}),(\overline{\mathbf{10}}_c,\mathbf{3}),(\overline{\mathbf{10}}_c,\mathbf{5}),(\mathbf{27}_c,\mathbf{1}),$}& \\
 & &\parbox{10cm}{\vskip-40pt$3(\mathbf{27}_c,\mathbf{3}),2(\mathbf{27}_c,\mathbf{5}),(\mathbf{27}_c,\mathbf{7}),(\mathbf{35}_c,\mathbf{3}),(\mathbf{35}_c,\mathbf{5}),(\overline{\mathbf{35}}_c,\mathbf{3}),$}& \\
 & &\parbox{10cm}{\vskip-40pt$(\overline{\mathbf{35}}_c,\mathbf{5}),(\mathbf{64}_c,\mathbf{1}),(\mathbf{64}_c,\mathbf{3}),(\mathbf{64}_c,\mathbf{5}),(\mathbf{64}_c,\mathbf{7})$}& \\\hline
 & 896&$(\mathbf{1}_c,\mathbf{3}),(\mathbf{1}_c,\mathbf{5}),2(\mathbf{8}_c,\mathbf{1}),3(\mathbf{8}_c,\mathbf{3}),2(\mathbf{8}_c,\mathbf{5}),(\mathbf{8}_c,\mathbf{7}),(\mathbf{10}_c,\mathbf{1}),$&30\\
\includegraphics[width=0.75cm]{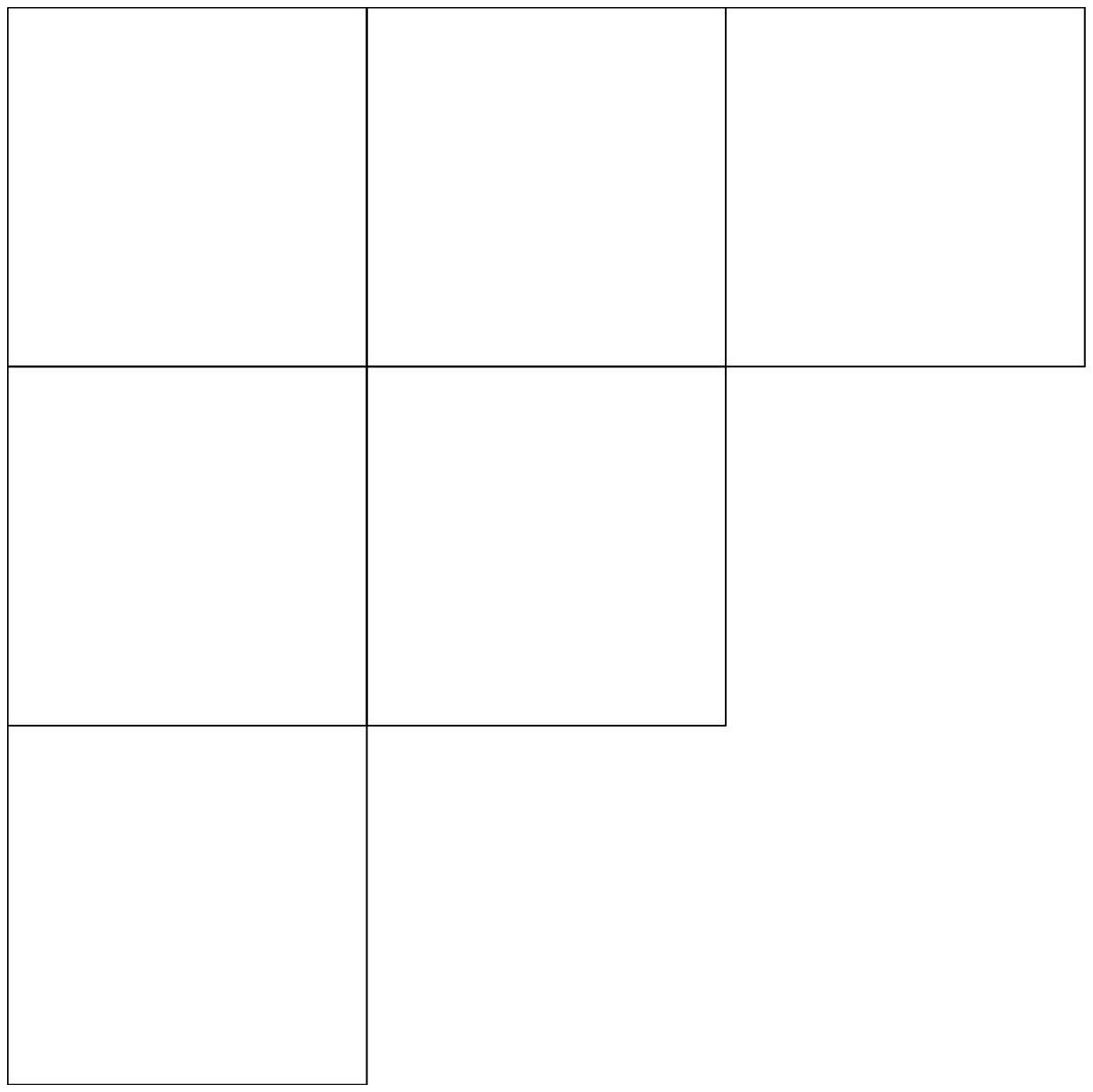}& &\parbox{10cm}{\vskip-20pt$2(\mathbf{10}_c,\mathbf{3}),(\mathbf{10}_c,\mathbf{5}),(\overline{\mathbf{10}}_c,\mathbf{3}),(\overline{\mathbf{10}}_c,\mathbf{5}),(\mathbf{27}_c,\mathbf{1}),2(\mathbf{27}_c,\mathbf{3}),$}& \\
 & &\parbox{10cm}{\vskip-20pt$(\mathbf{27}_c,\mathbf{5}),(\mathbf{35}_c,\mathbf{1}),(\mathbf{35}_c,\mathbf{3})$}& \\\hline
 & 840&$(\mathbf{1}_c,\mathbf{1}),(\mathbf{8}_c,\mathbf{1}),2(\mathbf{8}_c,\mathbf{3}),(\mathbf{8}_c,\mathbf{5}),(\mathbf{10}_c,\mathbf{1}),2(\mathbf{10}_c,\mathbf{3}),(\mathbf{10}_c,\mathbf{5}),$&36\\
\includegraphics[width=1cm]{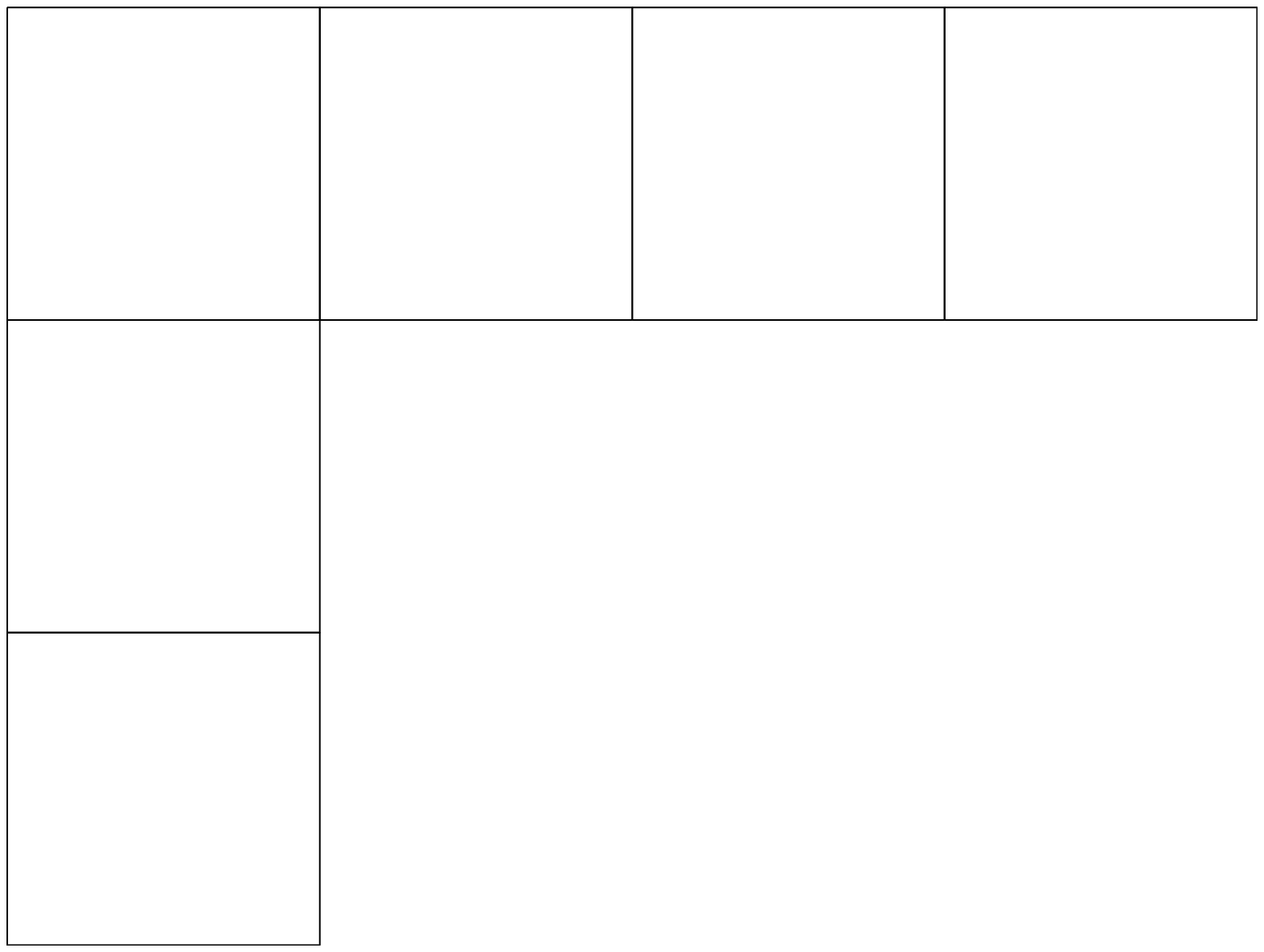} & &\parbox{10cm}{\vskip-20pt$(\mathbf{10}_c,\mathbf{7}),(\overline{\mathbf{10}}_c,\mathbf{3}),(\mathbf{27}_c,\mathbf{1}),(\mathbf{27}_c,\mathbf{3}),(\mathbf{27}_c,\mathbf{5}),(\mathbf{35}_c,\mathbf{3}),$}& \\
 & &\parbox{10cm}{\vskip-20pt$(\mathbf{35}_c,\mathbf{5})$}& \\\hline
& 405&$(\mathbf{1}_c,\mathbf{1}),(\mathbf{1}_c,\mathbf{5}),(\mathbf{8}_c,\mathbf{1}),2(\mathbf{8}_c,\mathbf{3}),(\mathbf{8}_c,\mathbf{5}),(\mathbf{10}_c,\mathbf{3}),(\overline{\mathbf{10}}_c,\mathbf{3}),$&28 \\
\includegraphics[width=1cm]{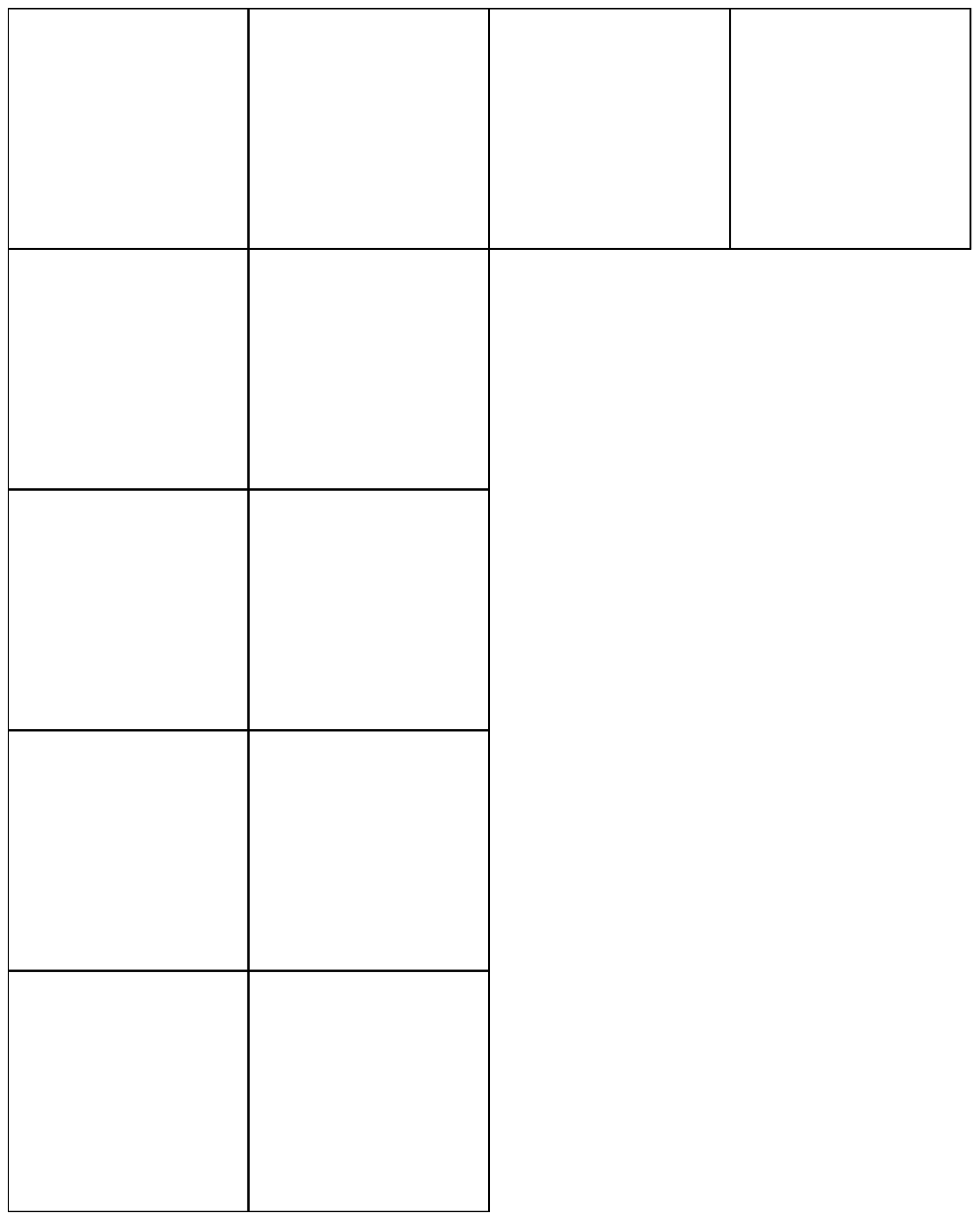} & &\parbox{10cm}{\vskip-35pt$(\overline{\mathbf{10}}_c,\mathbf{3}),(\mathbf{27}_c,\mathbf{1}),(\mathbf{27}_c,\mathbf{3}),(\mathbf{27}_c,\mathbf{5})$}&
\\\hline\hline
\end{tabular}
\caption{The possible SU(6)$_{cs}$ representation of the total
system, its SU$_{c}$(3)$\otimes$ SU$_{s}$(2) content and the
corresponding eigenvalue of Casimir operator. } \label{t5}
\end{center}
\end{table}
\begin{table}
\caption{The Continuing of Table VII.}
\begin{center}
\begin{tabular}{|c|c|l|c|}\hline\hline
Young tabular&Dimension&$\parbox{10cm}{\begin{center}
SU$_{c}$(3)$\otimes$ SU$_{s}$(2) content\end{center}}$&
Eigenvalue of \\
 & & &Casimir operator\\\hline
 & 280&$(\mathbf{1}_c,\mathbf{3}),(\mathbf{8}_c,\mathbf{1}),2(\mathbf{8}_c,\mathbf{3}),(\mathbf{8}_c,\mathbf{5}),(\mathbf{10}_c,\mathbf{1}),(\mathbf{10}_c,\mathbf{3}),(\mathbf{10}_c,\mathbf{5}),$&24\\
\includegraphics[width=0.75cm]{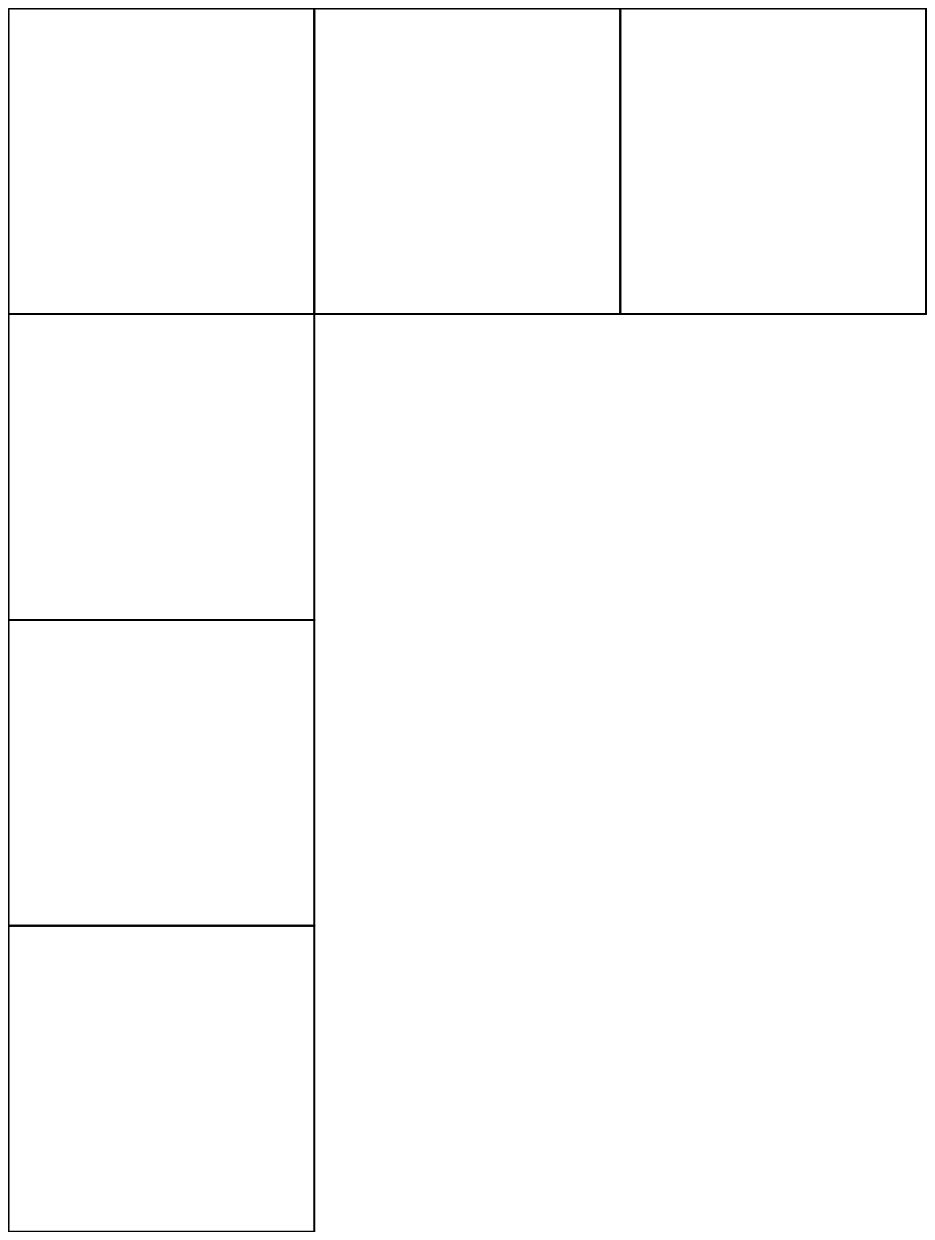} & &\parbox{10cm}{\vskip-40pt$(\overline{\mathbf{10}}_c,\mathbf{1}),(\mathbf{27}_c,\mathbf{3})$}& \\\hline
 & 189&$(\mathbf{1}_c,\mathbf{1}),(\mathbf{1}_c,\mathbf{5}),(\mathbf{8}_c,\mathbf{1}),2(\mathbf{8}_c,\mathbf{3}),(\mathbf{8}_c,\mathbf{5}),(\mathbf{10}_c,\mathbf{3}),(\overline{\mathbf{10}}_c,\mathbf{3}),$&20\\
\includegraphics[width=0.5cm]{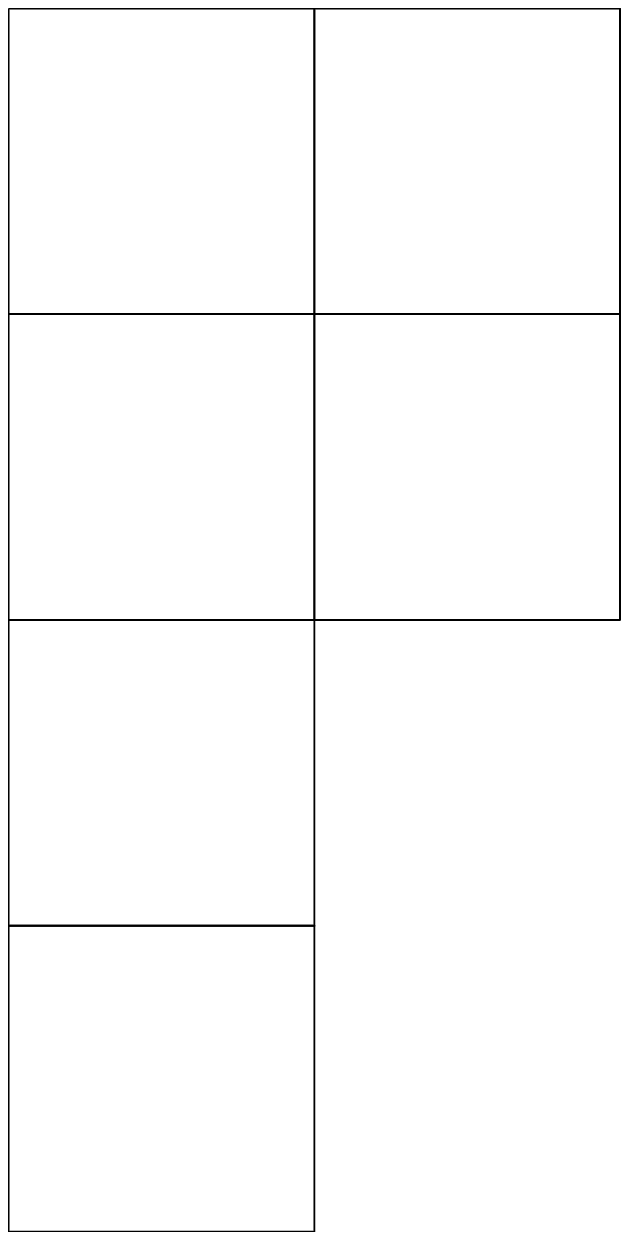} & &\parbox{10cm}{\vskip-30pt$(\mathbf{27}_c,\mathbf{1})$}& \\\hline
 & & & \\
\includegraphics[width=0.5cm]{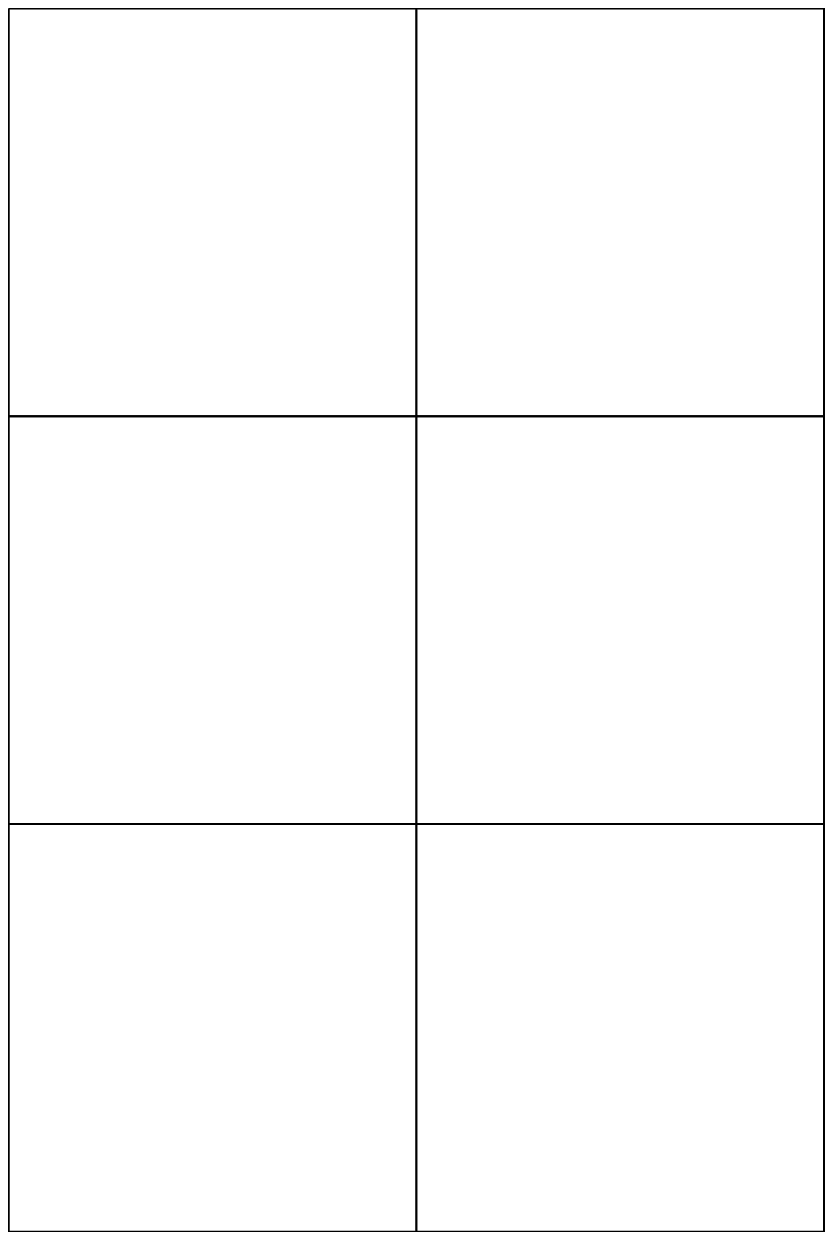}  & \parbox{1cm}{\vskip-50pt175}&\parbox{10cm}{\vskip-50pt$(\mathbf{1}_c,\mathbf{3}),(\mathbf{1}_c,\mathbf{7}),(\mathbf{8}_c,\mathbf{3}),(\mathbf{8}_c,\mathbf{5}),(\mathbf{10}_c,\mathbf{1}),(\overline{\mathbf{10}}_c,\mathbf{1}),(\mathbf{27}_c,\mathbf{3})$}&\parbox{1cm}{\vskip-50pt24}\\\hline
 & & & \\
\includegraphics[width=0.5cm]{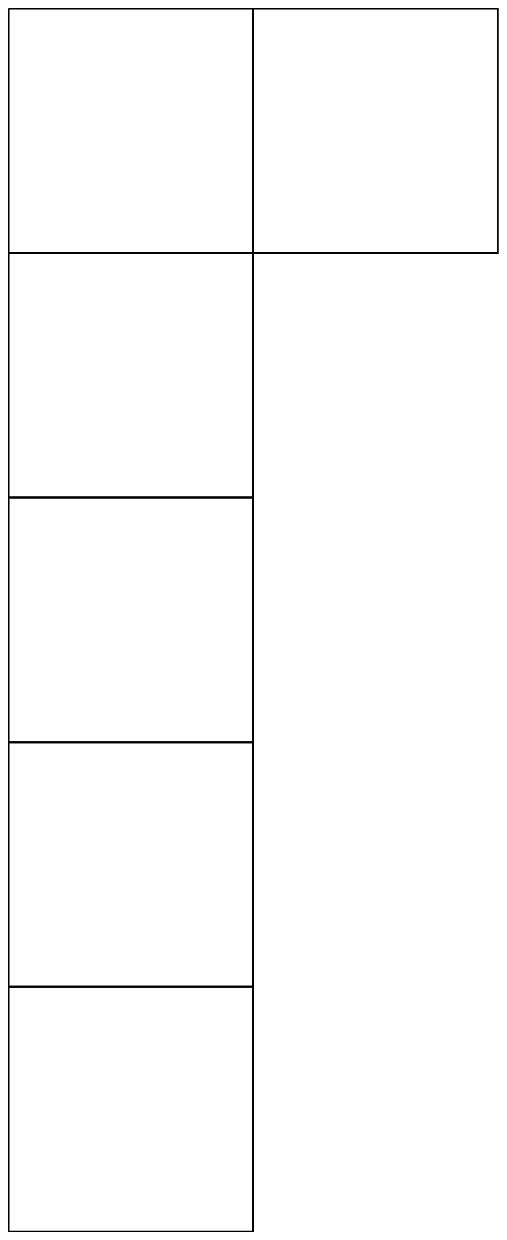}
&\parbox{1cm}{\vskip-70pt35}&\parbox{10cm}{\vskip-70pt$(\mathbf{1}_c,\mathbf{3}),(\mathbf{8}_c,\mathbf{1}),(\mathbf{8}_c,\mathbf{3})$}&\parbox{1cm}{\vskip-70pt12}\\\hline\hline
\end{tabular}
\end{center}
\end{table}
\newpage
\begin{figure}[hptb]
\begin{center}
\includegraphics[width=17.5cm]{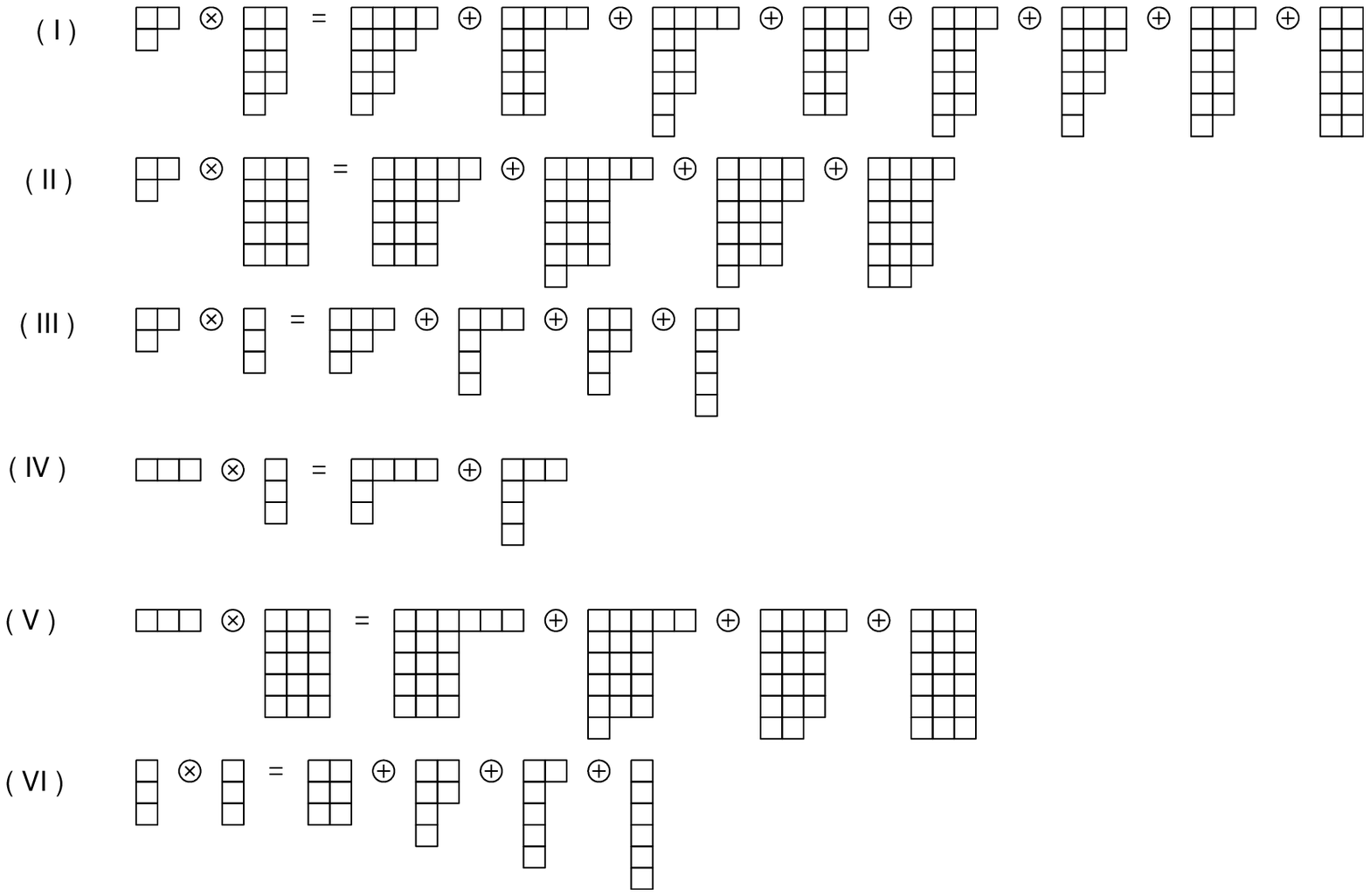}
\caption{The SU(6)$_{cs}$ outer product for $\rm{q}^3$ and
$\overline{\rm{q}}^3$ ~(I)~
$\mathbf{70}_{cs}\otimes\overline{\mathbf{70}}_{cs}=\mathbf{3675}_{cs}\oplus\mathbf{405}_{cs}\oplus\mathbf{280}_{cs}
\oplus\overline{\mathbf{280}}_{cs}\oplus\mathbf{35}_{cs}\oplus\mathbf{189}_{cs}\oplus\mathbf{35}_{cs}\oplus\mathbf{1}_{cs};$~(II)~$\mathbf{70}_{cs}\otimes\overline{\mathbf{56}}_{cs}=
\mathbf{3200}_{cs}\oplus\mathbf{405}_{cs}\oplus\overline{\mathbf{280}}_{cs}\oplus\mathbf{35}_{cs};\;
(\rm{III})\;\mathbf{70}_{cs}\otimes\overline{\mathbf{20}}_{cs}=\mathbf{896}_{cs}\oplus\mathbf{280}_{cs}\oplus
\mathbf{189}_{cs}\oplus\mathbf{35}_{cs};~(IV)~\mathbf{56}_{cs}\otimes\overline{\mathbf{20}}_{cs}=\mathbf{840}_{cs}\oplus\mathbf{280}_{cs};~(V)~\mathbf{56}_{cs}\otimes\overline{\mathbf{56}}_{cs}=\mathbf{2695}_{cs}\oplus
\mathbf{405}_{cs}\oplus\mathbf{35}_{cs}\oplus\mathbf{1}_{cs};~(VI)~\mathbf{20}_{cs}\otimes\overline{\mathbf{20}}=\mathbf{175}_{cs}\oplus\mathbf{189}_{cs}\oplus\mathbf{35}_{cs}\oplus\mathbf{1}_{cs}.$}
\end{center}
\vspace{0.5in}
\includegraphics[width=17.5cm]{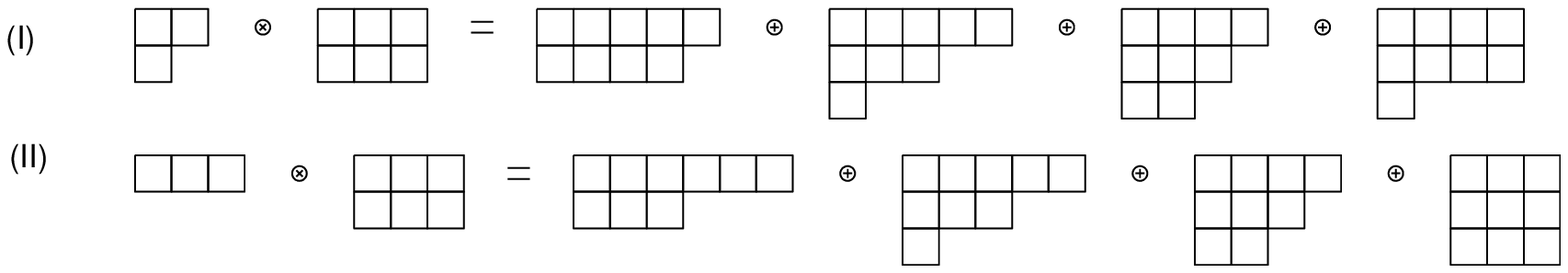}
\caption{The SU(3)$_f$ direct product for  $\rm{q}^{3}$ and
$\overline{\rm{q}}^3$ (I)
$\mathbf{8}_{f}\otimes\overline{\mathbf{10}}_f=\overline{\mathbf{35}}_f\oplus\mathbf{27}_f\oplus\mathbf{8}_{f}\oplus\overline{\mathbf{10}}_f;
\;(\rm{II})\;
\mathbf{10}_{f}\otimes\overline{\mathbf{10}}_f=\mathbf{64}_f\oplus\mathbf{27}_f\oplus\mathbf{8}_f\oplus\mathbf
{1}_f$}.
\end{figure}
\begin{figure}
\begin{center}
\includegraphics*[5pt,580pt][400pt,800pt]{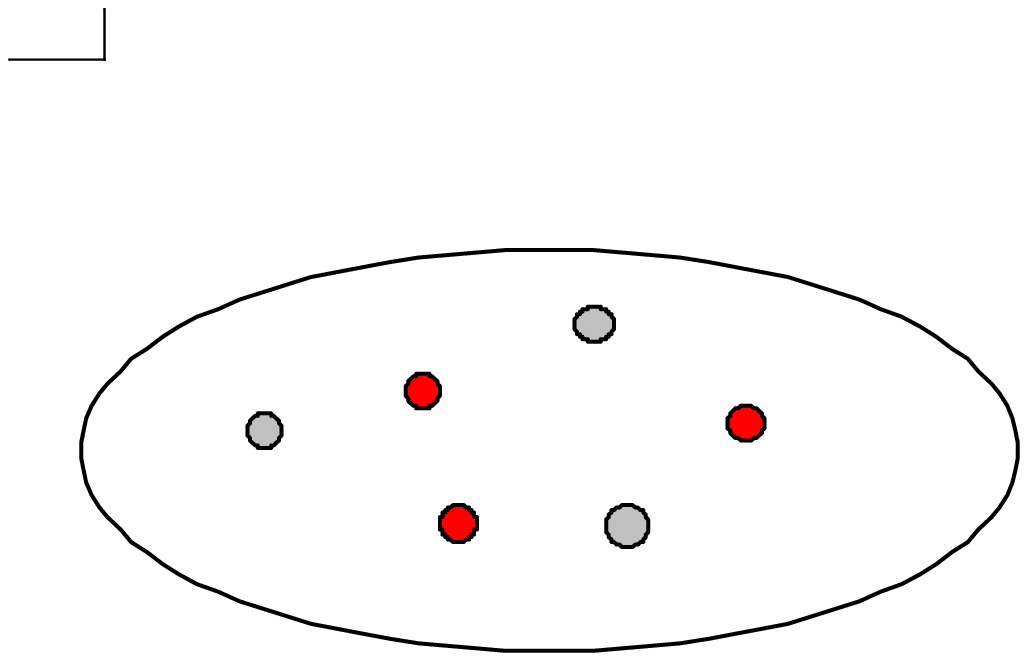}
\caption{The configuration in common quark model, where gray circles
denote quarks and red circles denote antiquarks}.
\end{center}
\end{figure}
\begin{figure}[hptb]
\begin{center}
\includegraphics*[5pt,580pt][400pt,800pt]{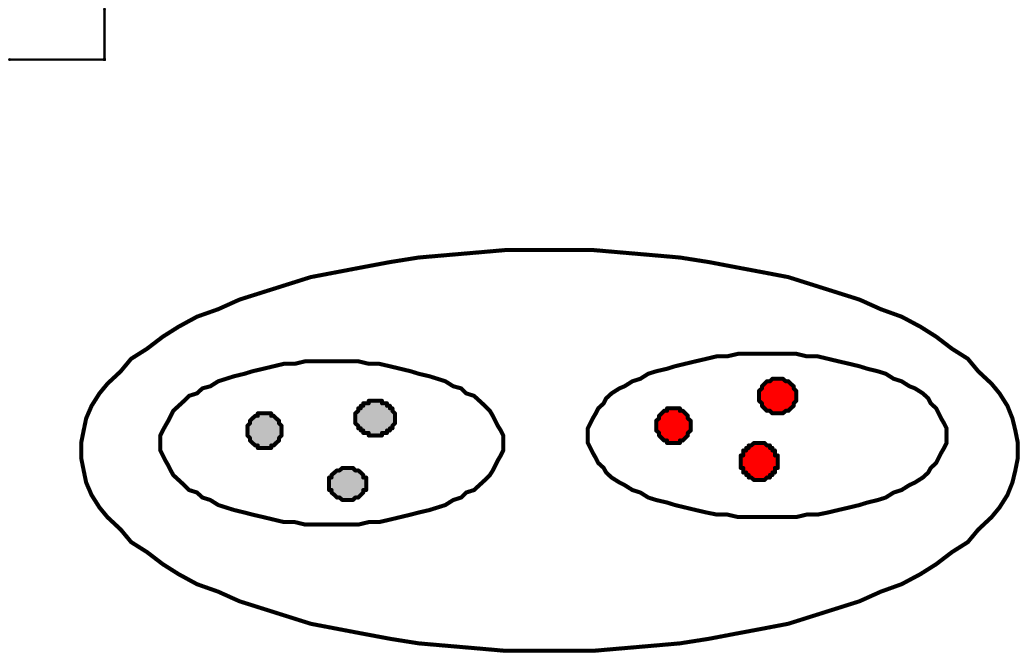}
\caption{The configuration $(\rm{q}^{3})-(\overline{\rm{q}}^3)$,
where gray circles denote quarks and red circles denote antiquarks.
}.
\end{center}
\end{figure}
\begin{figure}[hptb]
\begin{center}
\includegraphics*[5pt,580pt][400pt,800pt]{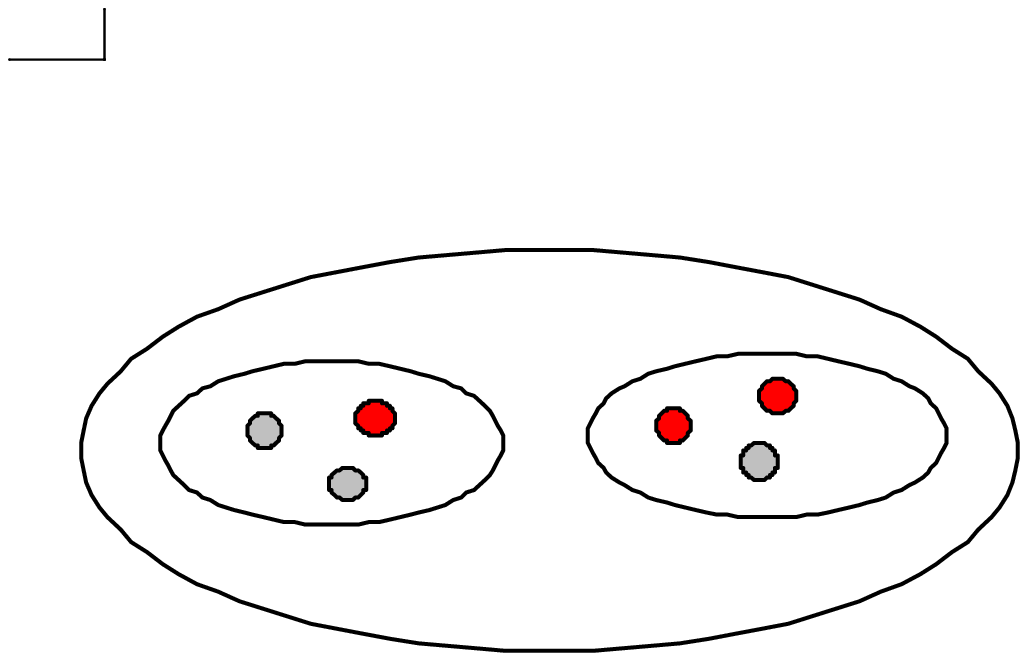}
\caption{The configuration
$(\rm{q}^{2}\overline{\rm{q}})-(\rm{q}\overline {\rm{q}}^2$), where
gray circles denote quarks and red circles denote antiquarks}.
\end{center}
\end{figure}

\begin{thebibliography}{99}
\bibitem{exotic} T.Nakano et al(The LEPS
collaboration),Phys.Rev.Lett.91(2003)012002,hep-ex/0301020; BABAR
Collaboration, B.Aubert et al., Phys.Rev.Lett.90,242001(2003); Belle
Collaboration, K.Abe et al., Phys.Rev.Lett.92,012002(2004).
\bibitem{Bes1}BES Collaboration,J.Z.Bai et
al.,Phys.Rev.Lett.91,022001(2003).
\bibitem{Bes}BES Collaboration, M. Ablikim et al.,Phys. Rev. Lett. 95 (2005) 262001;
hep-ex/0508025.
\bibitem{Bes2}BES Collaboration, M. Ablikim et al.,Phys.Rev. D71 (2005)
072006,hep-ex/0503030.
\bibitem{Bes3}BES Collaboration, M. Ablikim et
al.,Phys.Rev.Lett.93,112002(2004).
\bibitem{bell1}Belle Collaboration, K.Abe et al.,
Phys.Rev.Lett.88,181803(2002).
\bibitem{bell2}Belle Collaboration, M.-Z.Wang et al.,
Phys.Rev.Lett.90,201802(2003).
\bibitem{bell3}Belle Collaboration, Y.-J.Lee et al.,
Phys.Rev.Lett.93.211801(2004).
\bibitem{Datta} A.Datta and P.J.O'Donnell, Phys.Lett.B
567,273(2003), hep-ph/0306097.
\bibitem{interpretation1} B.S.Zou and H.C.Chiang,
Phys.Rev.D69.034004(2004); Xiao-Gang He, Xue-Qian Li and J.P.Ma,
Phys.Rev.D71(2005)014031, hep-ph/0407083; Shi-Lin Zhu and Chong-Shou
Gao, hep-ph/0507050; C.H.Chang and H.R.Pang, Commun.Theor.Phys.43,
275 (2005), hep-ph/ 0407188; Xiao-Gang He, Xue-Qian Li et al.,
hep-ph/0509140 ; N.Kochelev and Dong-Pil Min, hep-ph/0508288.

\bibitem{yan} Mu-Lin Yan, Si Li, Bin Wu and Bo-Qiang Ma, Phys. Rev.
{\bf D72}, 034027 (2005).
\bibitem{ding} Gui-Jun Ding and Mu-Lin Yan, Phys.Rev.C72:015208,2005,
hep-ph/0502127.

\bibitem{interpretation2}J.L.Rosner, Phys.Rev. D68 (2003)
014004, hep-ph/0303079; C.Z.Yuan, X.H.Mo and P.Wang, Phys.Lett.B626
(2005)95-100, hep-ph/0506019.
\bibitem{colorch} Hong-Mo Chan,
Nucl.Phys.B136:401,1978; Phys.Lett.B76:634-640,1978.
\bibitem{correlation}R.L.Jaffe and F.Wilcek, Phys.Rev.Lett 91,232003(2003),
hep-ph/0307341; M.Karliner and H.J.Lipkin,
Phys.Lett.B575:249-255,2003, hep-ph/0402260.

\bibitem{ding2} Gui-Jun Ding and Mu-Lin Yan,Phys.Rev.{\bf D72}, 034014,(2005);
hep-ph/0506119.
\bibitem{petcolor} H.Hogaasen and P.Sorba,
Nucl.Phys.B145:119,1978; Mod.Phys.Lett.A19:2403-2410,2004,
hep-ph/0406078; Fl. Stancu,Phys.Lett. B595 (2004) 269-276;
Erratum-ibid. B598 (2004) 295-295,hep-ph/0402044; Carl E.Carlson et
al.,Phys.Lett. B573 (2003) 101-108,hep-ph/0307396; Jialun Ping et
al., Phys.Lett. B602 (2004) 197-204, hep-ph/0408176.
\bibitem{colormag} A.De Rujula, H.Georgi and S.L.Glashow,
Phys,Rev.D.12,147(1975).
\bibitem{jaffe1}R.L.Jaffe and K.Johnson, Phys.Lett.B60,201(1976);
R.L.Jaffe, Phys.Rev.D15:267,1977; R.L.Jaffe, Phys.Rev.D15:281,1977.
\bibitem{q6}P.~J.~Mulders, A.~T.~M.~Aerts and J.~J.~De Swart,Phys.Rev.D21, 2653 (1980).
\bibitem{strottman}D.Strottman, Phys.Rev.D20,748 (1979).
\bibitem{jaffe2}R.L.Jaffe, hep-ph/0507149.
\bibitem{PDG}Partcle Data Group, {\it 14. Quark Model}, in Phys. Lett. {\bf B592}, 1
(2004), {\it pp154-159}.

\bibitem{group}C.Itzykson, M.Nauenberg, Rev.Mod.Phys.38:95-120,1966.
\bibitem{cpf} S. I. So and D. Strottman, J.Math.Phys.20:153-176, 1979;
D. Strottman, J.Math.Phys.20:1643-1647,1979.
\bibitem{chenjq}Jin-Quan Chen et al, Tables of the
SU(mn)$\supset$SU(m)$\times$SU(n) Coefficients of Fractional
Parentage(World Scientific,1991).
\bibitem{notation}W.G.McKay and J.Patera, Tables of Dimensions, Indices, and Branching Rules for
Representations of Simple Lie Algebras, (Marcel Dekker, New York,
1981) p.98.
\end{thebibliography}
\end{document}